\documentclass{article}
\usepackage[utf8]{inputenc}
\usepackage{float}
\usepackage{amsbsy}
\usepackage{amsmath}
\usepackage{amsthm}
\usepackage{amssymb}
\usepackage{epsfig}
\usepackage{multicol}
\usepackage{multirow}
\usepackage{adjustbox}
\usepackage{color}
\usepackage{bm}
\usepackage{bbm}
\usepackage{threeparttable}
\usepackage{subfigure}
\usepackage{rotating}
\usepackage{array}
\usepackage{verbatim}
\usepackage{fullpage}
\usepackage{setspace}
\usepackage{makecell}
\usepackage{hyperref}
\usepackage[ruled]{algorithm2e}
\usepackage{natbib}
\usepackage{graphicx}
\usepackage{booktabs}
\usepackage{authblk}
\usepackage{xr}
\usepackage{geometry}

\newcommand{\tr}{\mbox{tr}}
\newcommand{\mvec}{\mbox{vec}}
\newcommand{\btheta}{\boldsymbol{\Theta}}
\newcommand{\bgamma}{\boldsymbol{\Gamma}}
\newcommand{\bupsilon}{\boldsymbol{\Upsilon}}
\newcommand{\ty}{\boldsymbol{y}}
\newcommand{\tx}{\boldsymbol{x}}

\title{Regulation-incorporated Gene Expression Network-based Heterogeneity Analysis}
\author{Rong Li$^1$, Qingzhao Zhang$^2$, Shuangge Ma$^{1*}$ \\
1: Department of Biostatistics, Yale School of Public Health, New Haven, Connecticut, USA. \\
2: Department of Statistics and Data Science, School of Economics, Wang Yanan Institute for Studies in Economics, and Fujian Key Lab of Statistics, Xiamen University, Xiamen, China.}
\date{}

\begin{document}

\maketitle

\begin{quotation}
\noindent {\it Abstract:}
Gene expression-based heterogeneity analysis has been extensively conducted. In recent studies, it has been shown that network-based analysis, which takes a system perspective and accommodates the interconnections among genes, can be more informative than that based on simpler statistics. Gene expressions are highly regulated. Incorporating regulations in analysis can better delineate the ``sources’’ of gene expression effects. Although conditional network analysis can somewhat serve this purpose, it does render enough attention to the regulation relationships. In this article, significantly advancing from the existing heterogeneity analyses based only on gene expression networks, conditional gene expression network analyses, and regression-based heterogeneity analyses, we propose heterogeneity analysis based on gene expression networks (after accounting for or ``removing’’ regulation effects) as well as regulations of gene expressions. A high-dimensional penalized fusion approach is proposed, which can determine the number of sample groups and parameter values in a single step. An effective computational algorithm is proposed. It is rigorously proved that the proposed approach enjoys the estimation, selection, and grouping consistency properties. Extensive simulations demonstrate its practical superiority over closely related alternatives. In the analysis of two breast cancer datasets, the proposed approach identifies heterogeneity and gene network structures different from the alternatives and with sound biological implications.

\vspace{9pt}
\par
\end{quotation}\par

\section{Introduction}

Many complex diseases are intrinsically heterogeneous, with samples having the same disease diagnosis behaving differently. In early studies, heterogeneity analysis is often based on low-dimensional clinical and demographic measurements. With the development of high-throughput profiling, omics measurements, which may more informatively capture disease biology, have been increasingly used in heterogeneity analysis \citep{Lee2021Towards}. Among the various omics measurements, gene expressions have drawn special attention because of important biological implications, broad availability of data, and promising empirical results. Through a series of studies \citep{Church2019Investigating, Pio2022Integrating}, gene expression-based heterogeneity analysis has demonstrated significant successes. It can be supervised and unsupervised, and the two types of analysis serve different purposes. In this study, we conduct unsupervised heterogeneity analysis, under which different sample groups have different gene expression properties.

Some gene expression-based heterogeneity analyses, especially some early ones \citep{Leek2007Capturing, Church2019Investigating}, are based on simple data characteristics such as mean and variance. In multiple recent studies, it has been shown that gene expression network (graph)-based analysis can take a system perspective and lead to more informative heterogeneity structures \citep{Tang2018Prognostic, Pio2022Integrating}. Here it is noted that network-based heterogeneity analysis can also accommodate information on mean and variance. The existing network-based heterogeneity analysis studies are mostly based on the Gaussian Graphical Model (GGM) technique, and there have been two main families of approaches. The first family is based on the finite mixture model technique \citep{Hao2018Simultaneous}, and a common challenge is how to determine the number of sample groups. The second family is based on the fusion technique \citep{Radchenko2017Convex}, which may provide a more “straightforward” way of determining the number of groups. 

\begin{figure}[!htp]
    \centering
    \includegraphics[width=0.45\textwidth, angle=0]{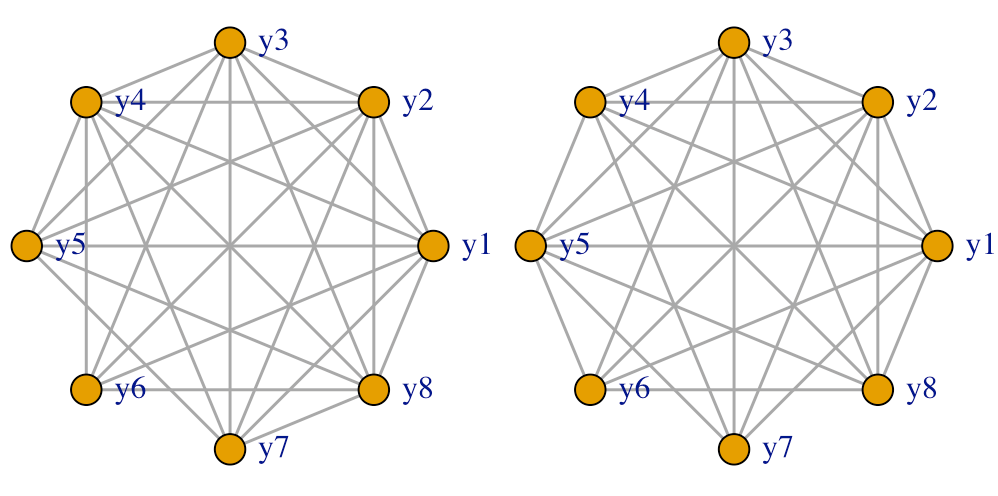}
    \includegraphics[width=0.45\textwidth, angle=0]{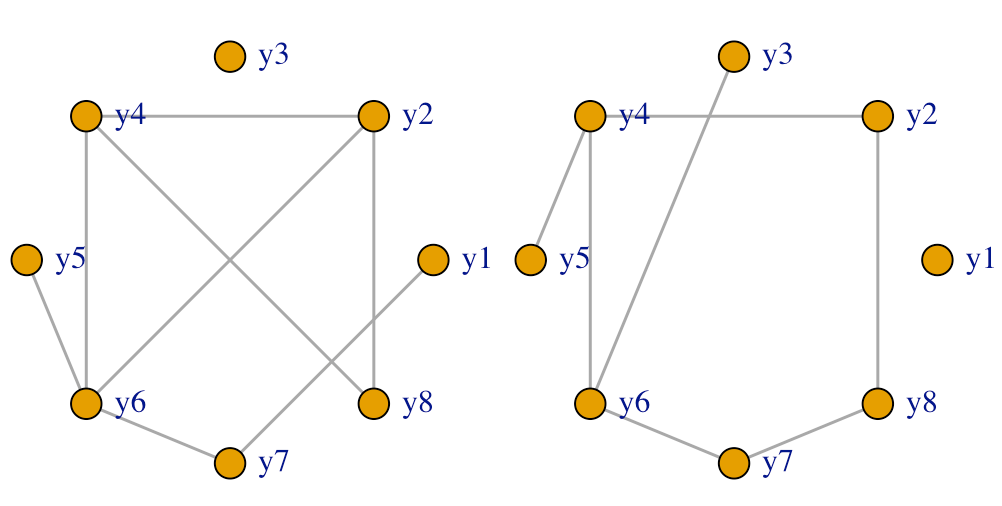}
    \caption{Schematic example: gene expression networks  for two sample groups before (left two) and after (right two) accounting for regulators.}
    \label{fig:toyexample}
\end{figure}

Gene expressions are highly regulated by multiple types of regulators (methylation, microRNAs, etc.). Published studies have suggested that the interconnections among gene expressions, as reflected in networks, can be attributed to regulators as well as ``net connections’’ \citep{Kagohara2018Epigenetic}.  As schematically presented in Figure \ref{fig:toyexample}, gene expression networks without accounting for regulators (left two plots) can be denser and hence less lucid than those accounting for regulatory effects (right two plots). With the growing popularity of multiomics studies (that collect data on gene expressions and their regulators), multiple strategies/approaches have been developed for the collective analysis of gene expression and regulator data. Examples include pooling multiple types of data and jointly modeling \citep{Lee2021Towards}, analyzing regulation relationships for example using regression \citep{Seal2020Estimating}, and others. In the context of network analysis, conditional approaches, for example conditional GGM, have been developed for studying gene expression interconnections with account for regulators \citep{Yin2011Sparse,Sohn2012Joint,Cai2013Covariate,Wang2015Joint}. In conditional analysis, especially under the context of heterogeneity analysis, regulation relationships often serve as a ``middle step’’ and do not play an important role \citep{Huang2018Joint, Lartigue2021Mixture}. 

In this article, our goal is to further advance gene expression network-based heterogeneity analysis by developing a new approach that can more effectively accommodate regulator data. This has been made possible by the increasing availability of multiomics data and motivated by the successes of existing gene expression network-based heterogeneity analyses as well as their limitation in accounting for regulation relationships. This study has a solid ground. Specifically, it belongs to the family of network-based heterogeneity analysis techniques and can enjoy similar merits as \cite{Danaher2014Joint} and \cite{Hao2018Simultaneous}. It is built on the GGM technique – note that, following \cite{Cai2013Covariate}, the normality assumption can be relaxed to make the proposed analysis more broadly applicable. Similar to \cite{Ren2022Gaussian}, it is built on the sparse penalized fusion technique \citep{Ma2017Concave}, can ``automatically’’ determine the number of sample groups, and has advantages over the finite mixture modeling approaches. 

On the other hand, this study also advances from the existing ones in multiple important aspects. First, it considers gene expression network interconnections after accounting for regulator effects. As schematically shown in published studies \citep{Wytock2013Sparse} and the right two plots of Figure \ref{fig:toyexample}, such interconnections can be sparser and easier to interpret. Additionally, they may reflect more essential gene relationships \citep{Sohn2012Joint}.
Second, significantly different from most conditional analyses, we more explicitly model the gene expression-regulator relationships and, more importantly, include such relationships in defining the heterogeneity structure. 
This has been motivated by the findings that such regulations have important implications for exploring the ``hidden'' sources of variants in complex diseases, as dysregulations can be directly associated with disease risk, progression, prognosis, etc. Published literature has also stressed that identifying heterogeneous genetic regulatory mechanisms is essential in the precision medicine era \citep{Kagohara2018Epigenetic}. 
Third, with both gene expression networks and gene expression-regulator relationships, and along with heterogeneity, the proposed analysis can face more technical challenges. Computational and theoretical developments, although may share some similar spirit with the existing studies, can be more complicated and demand careful investigations. 
Last but not least, as demonstrated in our data analysis, this study may deliver a practically useful new approach and findings for deciphering heterogeneity of complex diseases. 
It is noted that, although developed in the context of gene expressions and their regulators, the proposed analysis can have much broader applications. For example, for some other types of omics data (e.g., proteins), heterogeneity analysis and network-based analysis have also been conducted, and there are also upstream measurements with regulatory relationships. Another example may be human disease network analysis, where demographic variables, environmental factors, lifestyle, and others can be viewed as ``regulators’’.

The rest of this article is organized as follows. In Section 2, we introduce the proposed method and present its rationale, computation, and theoretical properties. Simulation study is conducted in Section 3 to gauge performance and compare with alternatives. Data analysis is presented in Section 4. Concluding remarks are presented in Section 5. Additional computational, theoretical, and numerical results are provided in Appendix. 

\section{Methods}

Suppose that the observations $(\ty_i, \tx_i), i=1,\dots,n$ are independent, where $\ty_i=(y_{i1}, \cdots, y_{ip})$ and $\tx_i=(1, x_{i1}, \cdots, x_{i,q+1})$. In our analysis, $\ty_i$'s are gene expressions, and $\tx_i$'s are regulators. As noted in conditional analysis and other analyses \citep{Tabor2002Candidate}, the collection of regulators does not need to be complete, in the sense that $\tx_i$ does not need to include all or a specific set of regulators. Let $\boldsymbol{X}=(\tx_1^T, \dots, \tx_n^T)$ denote the deterministic design matrix. Assume that the $n$ subjects belong to $K_0$ groups, where the group memberships and value of $K_0$ are unknown. For the $l$-th group, consider the regulation model:
\[
\ty = \bgamma_l \tx + \boldsymbol{\epsilon},
\]
where $\boldsymbol{\Gamma}_l$ is the $p \times (q+1)$ coefficient matrix, and $\boldsymbol{\epsilon} \in \mathbb{R}^p$ with zero mean and covariance matrix $\boldsymbol{\Sigma}_l$. Let $\boldsymbol{\Theta}_l = \boldsymbol{\Sigma}_l^{-1}$ be the $l$-th precision matrix and  $\boldsymbol{\Omega}_l=\mbox{vec}(\boldsymbol{\Gamma}_l, \boldsymbol{\Theta}_l)=(\gamma_{11,l},\dots,\gamma_{1(q+1),l},\dots,\gamma_{p(q+1),l}, \\
\theta_{11,l}, \dots, \theta_{1p,l}, \dots, \theta_{pp,l})^T$ be its vectorized representation. 
Conditional on $\tx$, assume that $\ty$ follows a multivariate normal distribution $ N(\boldsymbol{\Gamma}_l\tx, \boldsymbol{\Theta}_l^{-1})$, that is,
\[
f_l(\ty; \tx, \boldsymbol{\Omega}_l) = (2\pi)^{-p/2}|\boldsymbol{\Theta}_l|^{1/2}\exp\left\{ -\frac{1}{2}(\ty-\boldsymbol{\Gamma}_l\tx)^T\boldsymbol{\Theta}_l (\ty-\boldsymbol{\Gamma}_l\tx) \right\}.
\]

Although it is difficult to know $K_0$, it is often easy to specify an ``upper bound’’ $K > K_0$. To be cautious, $K$ can be taken as a relatively large number. With $K$ groups, we denote $\boldsymbol{\Omega}=(\boldsymbol{\Omega}_1, \cdots, \boldsymbol{\Omega}_K)^T$ and consider the mixture distribution: 
\[
f(\ty;\tx, \boldsymbol{\Omega})= \sum_{l=1}^K \pi_l f_l(\ty; \tx, \boldsymbol{\Omega}_l),
\]
where $\pi_l$ is the mixture probability of the $l$-th group. Denote $\boldsymbol{\pi}=(\pi_1, \dots, \pi_K)^T$, which is also unknown.

For parameter estimation and determination of the heterogeneity structure, we propose the penalized objective function: 
\begin{equation}\label{eq:obj}
\mathcal{L}(\boldsymbol{\Omega}, \boldsymbol{\pi}|\boldsymbol{Y}, \boldsymbol{X})=\frac{1}{n}\sum_{i=1}^n \log \left( \sum_{l=1}^K \pi_l f_l(\ty_i; \tx_i, \boldsymbol{\Omega}_l) \right) - \mathcal{P}(\boldsymbol{\Omega}).
\end{equation}
Here, the penalty is proposed as:
\begin{equation}\label{eq:penalty}
\begin{split}
 \mathcal{P}(\boldsymbol{\Omega})= 
& \sum_{l=1}^K \sum_{j \neq m}p(|\theta_{jm,l}|, \lambda_1) +  \sum_{l=1}^K \sum_{j=1}^p \sum_{m=1}^{q+1} p(|\gamma_{jm,l}|, \lambda_2) \\
&  + \sum_{l<l^\prime} p\left( (\Vert \boldsymbol{\Theta}_l - \boldsymbol{\Theta}_{l^\prime}\Vert_F^2 + \Vert \boldsymbol{\Gamma}_l - \boldsymbol{\Gamma}_{l^\prime}\Vert_F^2 )^{1/2}, \lambda_3 \right).
\end{split}
\end{equation}
$\theta_{jm,l}$ is the $jm$-th entry of the $l$-th precision matrix $\boldsymbol{\Theta}_l$. $\gamma_{jm,l}$ is the $jm$-th entry of the $l$-th coefficient matrix. $\Vert \cdot \Vert_F$ is the Frobenius norm. $p(\cdot,\lambda)$ is a concave penalty with tuning parameter $\lambda>0$. 
Consider: 
\[
(\hat{\boldsymbol{\Omega}}, \hat{\boldsymbol{\pi}}) = \arg\max_{\boldsymbol{\Omega}, \boldsymbol{\pi}}\mathcal{L}(\boldsymbol{\Omega}, \boldsymbol{\pi}|\boldsymbol{Y}, \boldsymbol{X}).
\]
Denote $\hat{\boldsymbol{\Omega}}_1, \dots, \hat{\boldsymbol{\Omega}}_{\hat{K}_0}$ as the unique values of $\hat{\boldsymbol{\Omega}}_1, \dots, \hat{\boldsymbol{\Omega}}_K$. Then it is concluded that there are $\hat{K}_0$ groups, with corresponding parameter values $\hat{\boldsymbol{\Omega}}_1, \dots, \hat{\boldsymbol{\Omega}}_{\hat{K}_0}$. The sparsity patterns of the precision matrix estimates directly correspond to the structures of the networks. Specifically, if and only if the $(j,m)$-th entry of the estimate for $\boldsymbol{\Theta}_l$ is zero, the corresponding two genes are not connected conditional on the other genes after removing the shared effects of the regulators in the $l$-th sample group. The sparsity of the estimate of $\boldsymbol{\Gamma}_l$ describes the sparse regulations of the regulators on the gene expressions in the $l$-th sample group. The estimated mixture probabilities can be obtained accordingly. 

\noindent{\bf Rationale}
The proposed modeling has two components: gene expression network and gene expression-regulator relationship. For the first component, we adopt the GGM technique as in multiple published studies. For the second component, we adopt linear regression. Although nonlinear regulations have been proposed, linear regression can be preferred considering the high dimensionality of gene expressions and regulators. It has also been shown to have satisfactory performance  \citep{Yin2011Sparse, Cai2013Covariate}. We adopt penalization for regularized estimation and selection. In (\ref{eq:penalty}), the first two sparse penalties have been commonly adopted \citep{Rothman2010Sparse, Yin2011Sparse}. With the third penalty term, we start with $K (> K_0)$ sample groups and examine if two groups can be shrunk together. By examining the final estimates, we can directly obtain the estimated number of groups as well as model parameters for all groups. The fusion strategy has been adopted in multiple recent heterogeneity analyses and shown to be advantageous over multiple alternatives. Significantly different from the existing heterogeneity analysis \citep{Ren2022Gaussian}, in (\ref{eq:penalty}), the regression coefficient matrices are considered along with the precision matrices – that is, the heterogeneity structure is jointly defined by the gene expression networks and regulation relationships.

\subsection{Computation}
For optimization, we develop an EM + Altering Direction Method of Multipliers (ADMM) algorithm. The complete data log-likelihood function is: 
\[
\frac{1}{n}\sum_{i=1}^n \sum_{l=1}^K \omega_{il}[\log \pi_k + \log f_l(\ty_i; \tx_i, \boldsymbol{\Omega}_l)],
\]
where $\omega_{il}$ is the latent indicator variable showing the group membership of the $i$th sample in the mixture. The EM algorithm maximizes the objective function composed of the above complete data log-likelihood function and penalty function in (\ref{eq:penalty}) iteratively in the following two steps.

Expectation step: here, we compute the conditional expectation of the complete data log-likelihood function with respect to $\omega_{il}$, given the observed data $(\ty_i, \tx_i)$’s and current estimates from the $(t-1)$-th step.  The conditional expectation is:
\begin{equation}\label{eq:estep}
\mathbb{E}_{\boldsymbol{L} | \ty, \tx, \boldsymbol{\Omega}^{(t-1)}}[\mathcal{L}(\boldsymbol{\Omega}, \boldsymbol{\pi} | \boldsymbol{Y}, \boldsymbol{X})]=\frac{1}{n} \sum_{i=1}^{n} \sum_{l=1}^{K} L_{il}^{(t)}
\left\{\log \pi_l+\log f_{l}\left(\ty_{i} ; \tx_i, \boldsymbol{\Omega}_l \right)\right\}-\mathcal{P}(\boldsymbol{\Omega}),
\end{equation}
where $L_{il}^{(t)}$ is the conditional expectation of $\omega_{il}$, which depends on the estimates from the $(t-1)$-th step and can be computed as:
\begin{equation}\label{eq:up_alpha}
L_{il}^{(t)}=\frac{\pi_l^{(t-1)} f_l\left(\ty_{i}; \tx_i, \boldsymbol{\Omega}_l^{(t-1)}\right)}{\sum_{l=1}^{K} \pi_l^{(t-1)} f_l\left(\ty_{i}; \tx_i,  \boldsymbol{\Omega}_l^{(t-1)}\right)}.
\end{equation}

Maximization step: we maximize (\ref{eq:estep}) with respect to $(\boldsymbol{\Omega}, \boldsymbol{\pi})$. For $\boldsymbol{\pi}$, we have: 
\begin{equation}\label{eq:up_pi}
\pi_l^{(t)}=\frac{1}{n} \sum_{i=1}^{n} L_{il}^{(t)}.
\end{equation}
For $\boldsymbol{\Omega}$, we update $\boldsymbol{\Gamma}=(\boldsymbol{\Gamma}_1,\dots,\boldsymbol{\Gamma}_K)^T$ and $\boldsymbol{\Theta}=(\boldsymbol{\Theta}_1, \dots, \boldsymbol{\Theta}_K)^T$ separately. For $\boldsymbol{\Gamma}_l, ~l=1,\dots,K$, maximizing (\ref{eq:estep}) is equivalent to solving:
\begin{equation}\label{eq:up_gamma}
\begin{split}
\left\{\boldsymbol{\Gamma}^{(t)}\right\}
& =\arg\min_{\boldsymbol{\Gamma}}\left(\frac{1}{2n} \sum_{i=1}^{n} \sum_{l=1}^{K} L_{il}^{(t)}
\left\{\left(\ty_i-\boldsymbol{\Gamma}_l \tx_i\right)^T \boldsymbol{\Theta}_l^{(t-1)}(\ty_i-\boldsymbol{\Gamma}_l \tx_i) \right\} \right. \\
& \qquad \left. +\sum_{l=1}^K\sum_{j=1}^p\sum_{m=1}^{q+1} p(|\gamma_{jm,l}|,\lambda_2) +
\sum_{l<l\prime}p \left( (\Vert \boldsymbol{\Theta}_l - \boldsymbol{\Theta}_{l^\prime} \Vert_F^2 + \Vert \boldsymbol{\Gamma}_l -\boldsymbol{\Gamma}_{l^\prime} \Vert_F^2)^{1/2}, \lambda_3 \right) \right).
\end{split}
\end{equation}
We adopt the local quadratic approximation technique. Details are provided in Appendix. For $\boldsymbol{\Theta}_l, ~l=1,\dots,K$, maximizing \eqref{eq:estep} is equivalent to solving:
\begin{equation}\label{eq:up_theta}
\begin{split}
    \left\{ \boldsymbol{\Theta}^{(t)} \right\}
    & = \arg\min_{\boldsymbol{\Theta}} \left\{  \sum_{l=1}^K n_l^{(t)}\left[-\log \mbox{det}(\boldsymbol{\Theta}_l)+ \tr(\textbf{S}_{\Gamma l}^{(t)}\boldsymbol{\Theta}_l) \right] \right. \\
    & \qquad \left. +  \sum_{l=1}^K\sum_{j \neq m}p(|\theta_{jm,l}|,\lambda_1)
     + \sum_{l<l^\prime}p\left( (\Vert \boldsymbol{\Theta}_l - \boldsymbol{\Theta}_{l^\prime} \Vert_F^2 + \Vert \boldsymbol{\Gamma}_l -\boldsymbol{\Gamma}_{l^\prime} \Vert_F^2)^{1/2}, \lambda_3 \right) \right\},
\end{split}
\end{equation}
where $\textbf{S}_{\Gamma l}^{(t)}=\textbf{C}_{yl}^{(t)}-\textbf{C}^{(t)}_{yx,l}\boldsymbol{\Gamma}_l^{(t)T}-\boldsymbol{\Gamma}_l^{(t)}\textbf{C}^{(t)T}_{yx,l}+\boldsymbol{\Gamma}_l^{(t)}\textbf{C}^{(t)}_{xl}\boldsymbol{\Gamma}_l^{(t)T}$, $n_l^{(t)}=\sum_{i=1}^n L_{il}^{(t)}$, and $\textbf{C}_{yl}$, $\textbf{C}_{xl}$, and $\textbf{C}_{yx,l}$ are weighted covariance matrices:
\[
\textbf{C}^{(t)}_{yl}=\sum_{i=1}^n L_{il}^{(t)}\ty_i\ty_i^T/\sum_{i=1}^n L_{il}^{(t)}, 
\textbf{C}^{(t)}_{yx,l}=\sum_{i=1}^n L_{il}^{(t)}\ty_i\tx_i^T/\sum_{i=1}^n L_{il}^{(t)}, 
\textbf{C}^{(t)}_{xl}=\sum_{i=1}^n L_{il}^{(t)}\tx_i\tx_i^T/\sum_{i=1}^n L_{il}^{(t)}.
\]
This optimization is achieved using the ADMM technique (Appendix). Overall, the algorithm contains iterating \eqref{eq:up_alpha}, \eqref{eq:up_pi}, \eqref{eq:up_gamma} and \eqref{eq:up_theta}. The iteration is concluded when the difference between the estimates from two consecutive steps is smaller than a prefixed constant. Satisfactory convergence is observed in all of our numerical studies. For initial values, we resort to the nonparametric mixture approach \cite{Chauveau2016Nonparametric} and observe satisfactory performance. 

\noindent{\bf Tuning parameter selection}
For selecting the optimal $\lambda_1$, $\lambda_2$ and $\lambda_3$, we conduct a grid search and minimize the Hannan-Quinn information criterion (HQC):
\begin{equation}\label{eq:hqc}
-2\sum_{i=1}^n \log\left[ \sum_{k=1}^{\hat{K}_0} \hat{\pi}_k f_k(\ty_i;\tx_i,\hat{\bgamma}_k,\hat{\btheta}_k)\right] + \sum_{k=1}^{\hat{K}_0} \log(\log(n))df_k,
\end{equation}
where $df_k$ is the total number of nonzero parameters in $\hat{\bgamma}_k$ and $\hat{\btheta}_k, k=1, \dots, \hat{K}_0$.

\subsection{Theoretical properties}

Denote the true parameter values as $\boldsymbol{\Upsilon}^*=(\boldsymbol{\Upsilon}_1^*, \dots, \boldsymbol{\Upsilon}_{K_0}^*)^T$ and $\boldsymbol{\Upsilon}_k^*=\mbox{vec}(\boldsymbol{\Gamma}_k^*, \boldsymbol{\Theta}_k^*)$ for $k=1,\dots,K_0$. 
Define $\mathcal{S}_k=\{(j,m): \theta_{jm,k}^*\neq 0, ~ 1\leq j \neq m \leq p \}$ 
and the sparsity parameter $s=\max\{|\mathcal{S}_k|, ~ k=1,\dots, K_0 \}$.
Similarly, define $\mathcal{D}_k=\{(j,m): \gamma_{jm,k}^*\neq 0, ~ 1\leq j \leq p, ~ 1\leq m \leq q+1 \}$ 
and $d=\max\{|\mathcal{D}_k|, ~ k=1,\dots, K_0 \}$. The following conditions are assumed. 

\begin{enumerate}
\item[(C1)] 
For some positive constants $\beta_1$ and $\beta_2$, $0<\beta_1 < \min_{1\leq k\leq K_0} \lambda_{\min}(\boldsymbol{\Theta}_k^*) < \max_{1\leq k\leq K_0}\lambda_{\max}(\boldsymbol{\Theta}_k^*) <\beta_2$, where $\lambda_{\min}(\btheta_k^*)$ and $\lambda_{\max}(\btheta_k^*)$ are the smallest and largest eigenvalues of $\btheta_k^*$, respectively.

\item[(C2)]
$\Vert \boldsymbol{\Theta}^* \Vert_{\infty}=\max_{k=1,\dots,K_0}\Vert \boldsymbol{\Theta}_k^* \Vert_{\infty}$ and
$\Vert \boldsymbol{\Gamma}^* \Vert_{\infty}=\max_{k=1,\dots,K_0}\Vert \boldsymbol{\Gamma}_k^* \Vert_{\infty}$ are bounded. 

\item[(C3)] 
The design matrix $\textbf{X}=(\textbf{X}_1, \cdots, \textbf{X}_{q+1})$ satisfies $\max_{j} \Vert\textbf{X}_j \Vert_2=O(\sqrt{n}), ~j=1,\dots,q+1$.
For each $k=1, \dots, K_0$, let $\textbf{X}_{\mathcal{D}_k}$ be the sub-matrix of $\textbf{X}$ with the support of coefficient matrix $\mathcal{D}_k$, and $\textbf{X}_{\mathcal{D}_k^C}$ is the corresponding complement. Define $L_k(\ty; \tx, \bupsilon^*)=\pi_k f_k(\ty; \tx, \bupsilon_k^*)/\sum_{k=1}^K \pi_k f_k(\ty; \tx, \bupsilon_k^*)$, $\mathbb{E}(L_k(\ty; \tx, \bupsilon^*))=\int L_k(\ty; \tx, \bupsilon^*)f(\ty|\tx,\bupsilon_k^*)d\ty$ and $\textbf{G}_k=\mbox{diag}(\mathbb{E}(L_k(\ty_i; \tx_i, \bupsilon^*)))$ is a $n \times n$ diagonal matrix with $\mathbb{E}(L_k(\ty_i; \tx_i, \bupsilon^*))$ as its elements. Denote $\Vert \textbf{B} \Vert_{2,\infty}=\max_{\Vert v\Vert_2}\Vert \textbf{B}\textbf{v} \Vert_\infty$. For a positive constant $C_0$ and $\alpha_1 \in [0, 1/2)$,
\[
\lambda_{\min}(\textbf{X}_{\mathcal{D}_k}^T \textbf{G}_k \textbf{X}_{\mathcal{D}_k}/n) \ge C_0,
\qquad
\left\lVert (\textbf{X}_{\mathcal{D}_k^C}^T \textbf{G}_k \textbf{X}_{\mathcal{D}_k})(\textbf{X}_{\mathcal{D}_k}^T \textbf{G}_k \textbf{X}_{\mathcal{D}_k})^{-1} \right\rVert_{2, \infty} \leq O(n^{\alpha_1}).
\]

\item[(C4)] 
Minimal signal condition: 
\[
\begin{split}
& \min\left\{ \{|\gamma_{jm,k}^*|: (j,m) \in \mathcal{D}_k, k=1,\dots, K_0 \}, \{ |\theta_{jm,k}^*|: (j,m) \in \mathcal{S}_k, k=1,\dots, K_0\} \right\} \\
& \quad > (a+0.5)\cdot \max\{\lambda_1, \lambda_2 \}.
\end{split}
\]
Denote $b=\min_{1\leq k \neq k^\prime \leq K_0}\Vert \boldsymbol{\Upsilon}_k^* - \boldsymbol{\Upsilon}_{k^\prime}^* \Vert_2$. Then, $b>(a+0.5) \lambda_3$.
\item[(C5)] 
$\lambda_1 \gg \sqrt{\frac{(d+s+p)(\log p +\log q)}{n}}$, $\lambda_2 \gg \sqrt{\frac{(d+s+p)(\log p +\log q)}{n}}$, 
and $
\lambda_3 \gg \sqrt{\frac{(d+s+p)(\log p + \log q)}{n}}$.

\item[(C6)]
The $K_0$ clusters are sufficiently separable such that, with a small $\gamma >0$, 
\[
L_k(\ty; \tx, \bupsilon^*)\cdot L_j(\ty; \tx, \bupsilon^*) \leq \frac{\gamma}{24(K_0-1)\sqrt{\max\{W, W^\prime, W^{\prime \prime}\}}},
\]
for each pair $\{(j,k),1\leq j \neq k \leq K_0\}$. Here, $W=\max_{1\leq k \leq K_0} W_k$, $W^\prime =\max_{1\leq k \leq K_0} W_k^\prime$, and $W^{\prime\prime} =\max_{1\leq k \leq K_0} W_k^{\prime\prime}$, and for each $k=1,\dots,K_0$, 
\[
W_k = \sup_{t\in [0,1]} \mathbb{E}\left\{ \delta_{\boldsymbol{\Upsilon}_{t,k}}(\ty)^T\delta_{\boldsymbol{\Upsilon}_{t,k}}(\ty) \Vert \boldsymbol{\Theta}_k^*(\ty-\boldsymbol{\Gamma}_k^*\tx)\tx\Vert_F^2 \right\},
\]
\[
W_k^\prime = \sup_{t\in [0,1]} \mathbb{E}\left\{ \delta_{\boldsymbol{\Upsilon}_{t,k}}(\ty)^T\delta_{\boldsymbol{\Upsilon}_{t,k}}(\ty) \Vert \boldsymbol{\Theta}_k^{*-1}\Vert_F^2 \right\},
\]
\[
W_k^{\prime\prime} = \sup_{t\in [0,1]} \mathbb{E}\left\{ \delta_{\boldsymbol{\Upsilon}_{t,k}}(\ty)^T\delta_{\boldsymbol{\Upsilon}_{t,k}}(\ty)\Vert (\ty-\boldsymbol{\Gamma}_k^*\tx)(\ty-\boldsymbol{\Gamma}_k^*\tx)^T\Vert_F^2 \right\},
\]
define $\tilde{\boldsymbol{\Upsilon}}_t=\boldsymbol{\Upsilon}^*+t(\boldsymbol{\Upsilon}-\boldsymbol{\Upsilon}^*)$, $\tilde{\boldsymbol{\Upsilon}}_t=(\tilde{\boldsymbol{\Upsilon}}_{t,1}, \dots, \tilde{\boldsymbol{\Upsilon}}_{t,K_0})$, $\tilde{\boldsymbol{\Upsilon}}_{t,k}=\mvec(\tilde{\boldsymbol{\Gamma}}_{t,k}, \tilde{\boldsymbol{\Theta}}_{t,k})$ with $t \in [0,1]$, and for any $\bupsilon \in \mathcal{B}_{\alpha_0}(\bupsilon^*)=\{\bupsilon: \Vert \bupsilon - \bupsilon^* \Vert_2 \leq \alpha_0 \}$: 
\[
\delta_{\boldsymbol{\Upsilon}_{t,k}}(\ty)=
\begin{pmatrix}
\mbox{vec}\left\{ \tilde{\boldsymbol{\Theta}}_{t,k}(\ty-\tilde{\boldsymbol{\Gamma}}_{t,k}\tx)\tx\right\} \\
\frac{1}{2}\mbox{vec}\left\{ \tilde{\boldsymbol{\Theta}}_{t,k}^{-1} - (\ty-\tilde{\boldsymbol{\Gamma}}_{t,k}\tx)(\ty-\tilde{\boldsymbol{\Gamma}}_{t,k}\tx)^T \right\}
\end{pmatrix}.
\]
\item[(C7)]
$\rho(t)=\lambda^{-1}p(t, \lambda)$ is concave in $t \in [0, +\infty)$ with a continuous derivative $\rho^\prime(t)$ satisfying $\rho(0+)=1$, and $\rho^\prime(0+)$ is independent of $\lambda$. There exists a constant $0<a<\infty$ such that $\rho(t)$ is constant for all $t \ge a \lambda$.
\end{enumerate}
Conditions (C1) and (C2) have been commonly assumed in the GGM literature particularly including those on heterogeneity analysis \citep{Hao2018Simultaneous}. The boundedness condition on the coefficients is also common for high-dimensional regression. Condition (C3) is on the design matrix and controls the correlations between variables as well as the correlations between unimportant and important variables in each sample group. It is similar to Condition 4 in \cite{Fan2011Nonconcave}. Condition (C4) specifies the minimal signals and minimal differences across the sample groups. Condition (C5) specifies the orders of the tuning parameters. Condition (C6) is the sufficiently separable condition and requires that a sample belongs to a group with a probability close to either zero or one. Relevant discussions can be found in \cite{Hao2018Simultaneous}. Condition (C7) has been commonly assumed for penalized estimation/selection and is satisfied by SCAD and MCP.

\noindent \textbf{Theorem 1}: 
 Suppose that Conditions (C1)-(C7) hold. If additionally $(d+s+p)(\log p + \log q)/n=o(1)$, then there exists a local maximizer of (\ref{eq:obj}) such as, with probability tending to 1:
\begin{enumerate}
    \item
    $\hat{K}_0=K_0$.
    \item
   $ \sum_{k=1}^{\hat{K}_0} \left(\Vert \hat{\boldsymbol{\Gamma}}_k - \boldsymbol{\Gamma}_k^* \Vert_F + \Vert \hat{\boldsymbol{\Theta}}_k - \boldsymbol{\Theta}_k^* \Vert_F \right) = O_p\left(\sqrt{\frac{(d+s+p)(\log p + \log q)}{n}}\right)$.
    \item Define $\hat{\mathcal{D}}_k=\{(j,m): \hat{\gamma}_{jm,k} \neq 0 \}$ and  $\hat{\mathcal{S}}_k=\{(j,m): \hat{\theta}_{jm,k} \neq 0\}$. Then $\hat{\mathcal{D}}_k = \mathcal{D}_k$ and $\hat{\mathcal{S}}_k=\mathcal{S}_k$ for $k=1,\dots, \hat{K}_0$.
\end{enumerate}
This theorem shows that the proposed approach has the well-desired consistency properties. Specifically, it can consistently identify the number of sample groups, which has not been established in quite a few existing heterogeneity analysis studies. Additionally, it has estimation and variable selection consistency. Although such results are not ``surprising’’ and somewhat similar to those in the existing literature, we note that the present data and model settings are much more complicated and include the existing ones (for example, GGM-based heterogeneity analysis \citep{Hao2018Simultaneous} and under homogeneity, and high-dimensional regression-based heterogeneity analysis \citep{Sun2022Robust}) as special cases, and the theoretical developments are not trivial. The proof is provided in Appendix.

\section{Simulation}

Simulation is conducted to gauge performance of the proposed approach and compare against relevant alternatives. We set $K_0=3$ and consider dimensions $p=q=50$ and $p=q=100$. For sample sizes, we consider three cases: a balanced case with all groups having sample sizes 200, a balanced case with all groups having sample sizes 500, and an imbalanced case with the three groups having sample sizes 150, 200 and 250. Additionally, the following three settings are considered.

\begin{enumerate}
\item[(S1)] 
$\tx_i$ has the first element being 1, and the other elements follow a normal distribution $N(\textbf{0}, \textbf{I}_q)$. The coefficient matrices $\Gamma_1 \neq \Gamma_2 \neq \Gamma_3$. The positions of the nonzero entries are randomly selected, and each entry has a probability proportional to $1/q$ of being nonzero. The nonzero values are generated from $\mbox{Unif}(-1.5, -1) \bigcup \mbox{Unif}(1, 1.5)$. All sample groups have tridiagonal precision matrices with the diagonal elements equal to 1 and the nonzero off-diagonal elements equal to 0.2, 0.3, and 0.4 for the three sample groups, respectively.

\item[(S2)] The precision matrices are generated by the nearest-neighbor networks. Specifically, each network consists of 10 equally-sized disjoint subnetworks (modules), among which eight are shared by the three sample groups. Additionally, the first group shares one module with the second group and another one with the third group. The second group and the third group also have a unique module of their own. The structure of each module is generated by a nearest-neighbor network. We first generate $p/10$ points randomly on a unit square, calculate all $p/10 \times (p/10-1)/2$ pairwise distances, and select $m=2$ nearest neighbors of each point besides itself. The nonzero off-diagonal elements of the precision matrices are located at which the corresponding two points are among the $m$ nearest neighbors of each other. The nonzero values are generated from $\mbox{Unif}(-0.4, -0.1) \bigcup \mbox{Unif}(0.1, 0.4)$. The diagonal elements are all set to 1. The other settings are the same as S1.

\item[(S3)] $\tx_i$’s have categorical distributions. Specifically,  $x_{ij}$ is generated randomly from $\{ 0,1,2\}$ with equal probabilities. The other settings are the same as S1.
\end{enumerate}
The sample size and dimensionality settings are comparable to those in the literature. With the presence of both networks and high-dimensional regressions under heterogeneity, our simulation can be considerably more challenging. It is noted that, although $p$ and $q$ may not seem large, with the precision and regression coefficient matrices for multiple sample groups, the number of unknown parameters is considerably larger than the sample size. Both continuous and categorical regulators are simulated, mimicking, for example, methylation and copy number variation. Two types of network structures are considered, both of which are popular in the literature. When implementing the proposed method, we set $K=6$ -- we have also experimented with a few other values and found similar results. To gauge its performance, we consider the following close alternatives. It is noted that there can be other alternatives. However, the following can be more relevant.

\begin{enumerate}
\item[(a)] 
The strategy is to first conduct clustering and generate sample groups. Then estimation is conducted for each group separately. This can be the most natural choice with the existing tools. Specifically, we use a nonparametric mixture approach \citep{Chauveau2016Nonparametric} for clustering, which outperforms K-means and many other clustering methods. The number of groups is set as $K=3,4,6$, as there is not a simple way for determining its value. For estimation with each group, we apply the conditional Gaussian graphical approach with Lasso penalization (CGLasso) \citep{Yin2011Sparse}. Tuning parameter selection is conducted using BIC as proposed in the literature. 

\item[(b)] This approach is similar to (a), except that the sparse multivariate regression with covariance estimation (MRCE) approach \citep{Rothman2010Sparse} is applied for estimation after clustering. 

\item[(c)]
The mixture of conditional Gaussian graphical model (MCGGM) approach \citep{Lartigue2021Mixture} is applied. With a given number of clusters, it can achieve simultaneous clustering and estimation of the precision matrices as well as the correlation matrices (between gene expressions and regulators). Note that this approach estimates the mutual correlation matrices between $\tx$ and $\ty$, not the regression coefficient matrices $\Gamma$’s.

\item[(d)]
The heterogeneous Gaussian graphical model via penalized fusion (HeteroGGM) approach \citep{Ren2022Gaussian} is applied. It can simultaneously achieve clustering and precision matrix estimation. The number of groups is automatically determined in a way similar to the proposed approach. However, it cannot accommodate the regulations of $\tx$ on $\ty$.
\end{enumerate}

To evaluate performance, we consider the following measures. For grouping accuracy, we consider $\hat{K}_0$ and adjusted Rand index (RI), which measures the similarity between the estimated and true grouping structures. For estimation accuracy, we consider root mean square error (RMSE). Specifically, for the precision matrices,
\[
\mbox{RMSE}(\btheta) = \left\{
\begin{array}{ll}
\frac{1}{K_0}\sum_{k=1}^{K_0} \Vert \hat{\btheta}_k - \btheta_k^*\Vert_F & \hat{K}_0=K_0, \\
\frac{1}{\hat{K}_0} \sum_{l=1}^{\hat{K}_0} \sum_{k=1}^{K_0}\Vert\hat{\btheta}_l -\btheta_k^* \Vert_F \cdot I \\
\left(k=\arg\min_{k^\prime}\{\Vert \hat{\btheta}_l - \btheta_{k^\prime}^*\Vert_F^2+ \Vert \hat{\bgamma}_l - \bgamma_{k^\prime}^*\Vert_F^2 \} \right) &\hat{K}_0 \neq K_0.
\end{array}
\right.
\]
For variable selection accuracy, we consider true/false positive rates (TPR/FPR):
\[
\mbox{TPR}(\btheta) = \left\{
\begin{array}{ll}
\frac{1}{K_0}\sum_{k=1}^{K_0} \frac{\sum_{j<m}I(\theta_{jm,k}^*\neq 0, \hat{\theta}_{jm,k}\neq 0)}{\sum_{j<m}I(\theta_{jm,k}^* \neq 0)} & \hat{K}_0=K_0, \\
\frac{1}{\hat{K}_0} \sum_{l=1}^{\hat{K}_0} \sum_{k=1}^{K_0}\frac{\sum_{j<m}I(\theta_{jm,k}^*\neq 0, \hat{\theta}_{jm,l}\neq 0)}{\sum_{j<m}I(\theta_{jm,k}^* \neq 0)} \cdot I \\
\left(k=\arg\min_{k^\prime}\{\Vert \hat{\btheta}_l - \btheta_{k^\prime}^*\Vert_F^2+ \Vert \hat{\bgamma}_l - \bgamma_{k^\prime}^*\Vert_F^2 \} \right) &\hat{K}_0 \neq K_0,
\end{array}
\right.
\]
\[
\mbox{FPR}(\btheta) = \left\{
\begin{array}{ll}
\frac{1}{K_0}\sum_{k=1}^{K_0} \frac{\sum_{j<m}I(\theta_{jm,k}^*= 0, \hat{\theta}_{jm,k}\neq 0)}{\sum_{j<m}I(\theta_{jm,k}^* = 0)} & \hat{K}_0=K_0, \\
\frac{1}{\hat{K}_0} \sum_{l=1}^{\hat{K}_0} \sum_{k=1}^{K_0}\frac{\sum_{j<m}I(\theta_{jm,k}^*= 0, \hat{\theta}_{jm,l}\neq 0)}{\sum_{j<m}I(\theta_{jm,k}^* = 0)} \cdot I \\
\left(k=\arg\min_{k^\prime}\{\Vert \hat{\btheta}_l - \btheta_{k^\prime}^*\Vert_F^2+ \Vert \hat{\bgamma}_l - \bgamma_{k^\prime}^*\Vert_F^2 \} \right) &\hat{K}_0 \neq K_0.
\end{array}
\right.
\]
The above measures are defined accordingly for the coefficient matrices. 

\begin{figure}[H]
    \centering
    \includegraphics[width=0.8\textwidth, angle=0]{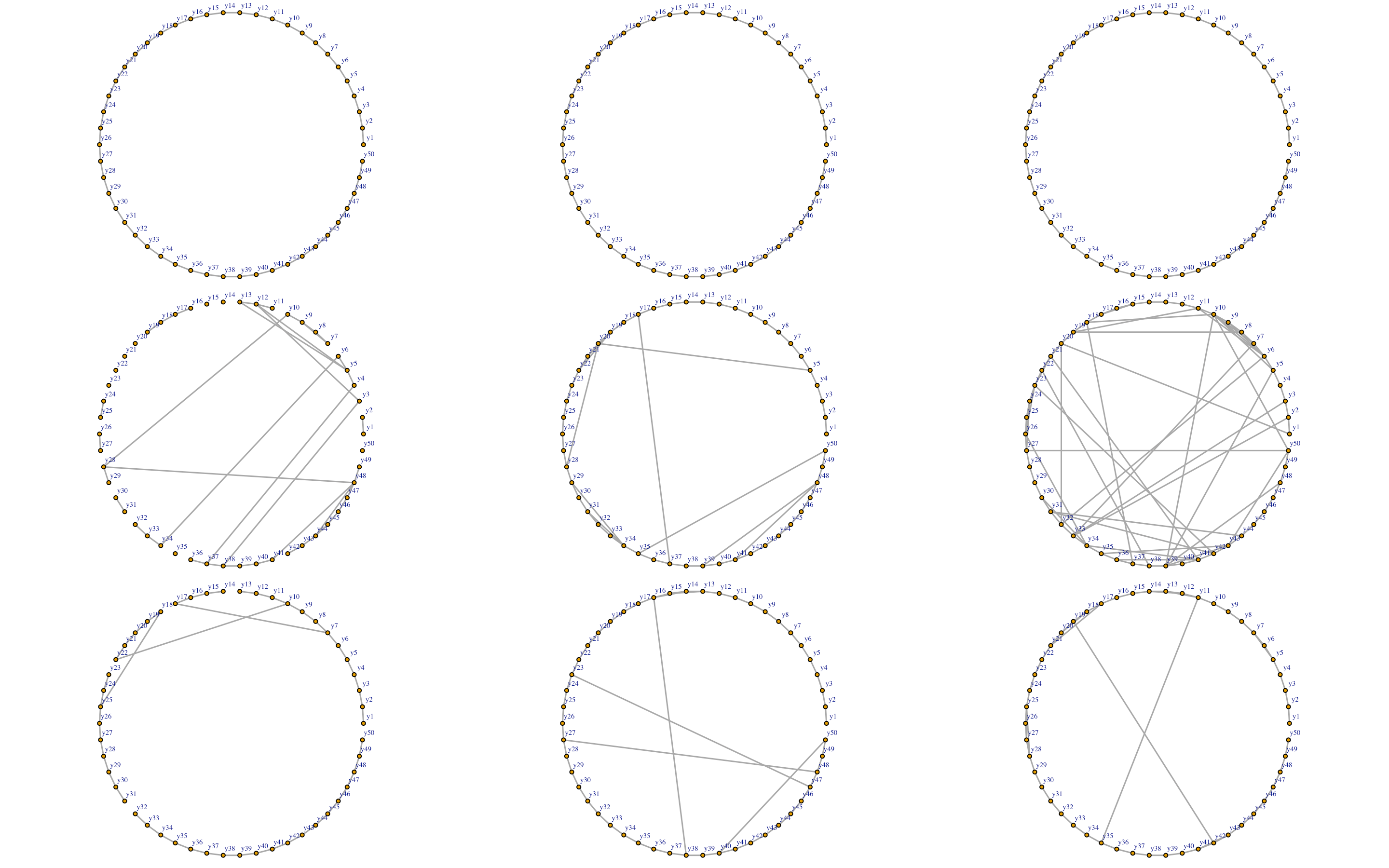}
    \caption{Simulation: true and estimated networks for three sample groups under S1. Top: true networks; Middle: estimation for one replicate with sizes (200, 200, 200); Bottom: estimation for one replicate with sizes (500, 500, 500). 
}\label{fig:graph_single_s1}
\end{figure}
\begin{figure}
    \centering
    \includegraphics[width=0.8\textwidth, angle=0]{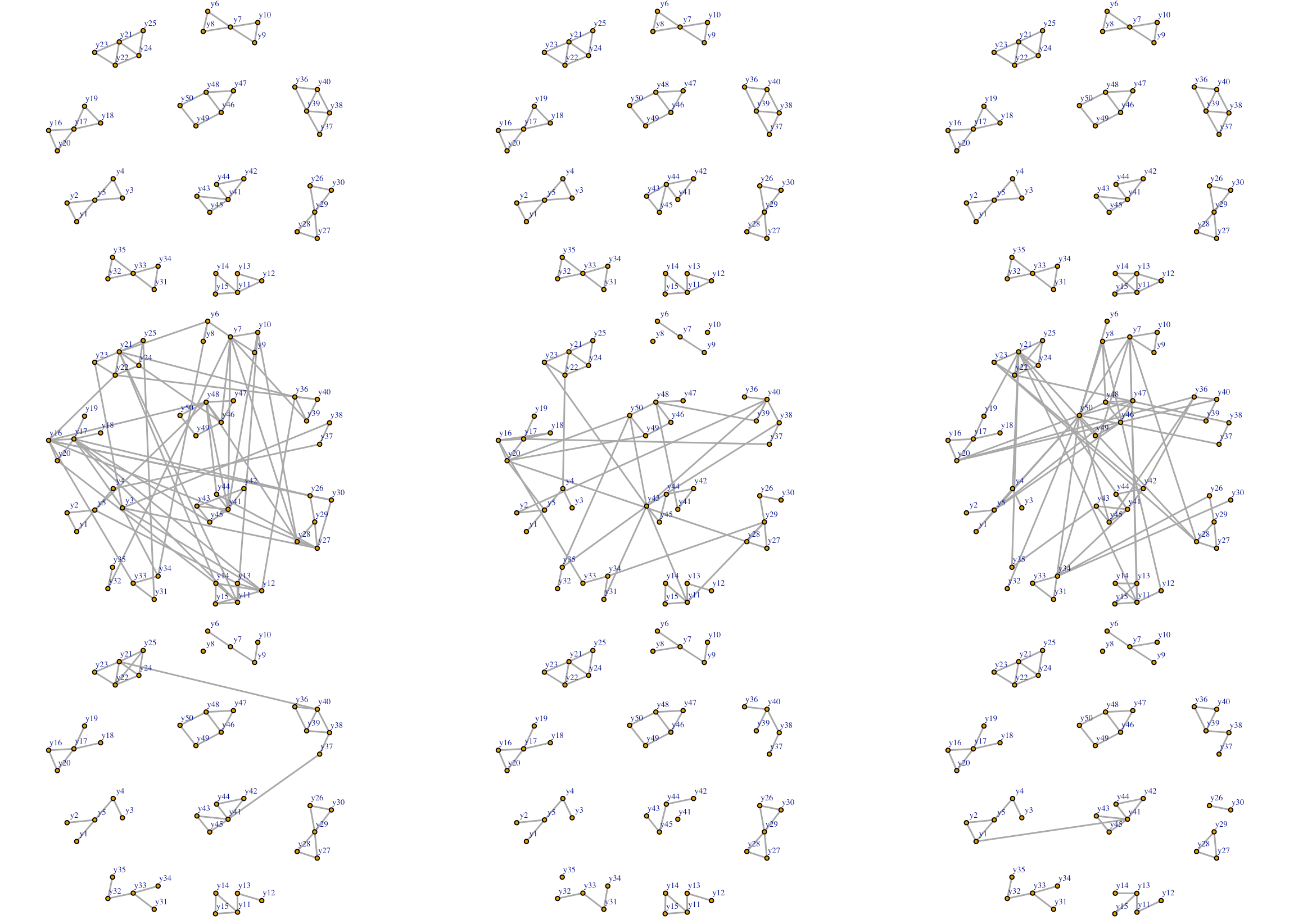}
    \caption{ Simulation: true and estimated networks for three sample groups under S2. Top: true networks; Middle: estimation for one replicate with sizes (200, 200, 200); Bottom: estimation for one replicate with sizes (500, 500, 500).
}\label{fig:graph_single_s2}
\end{figure}
\begin{figure}[!htp]
    \centering
    \includegraphics[width=0.95\textwidth, angle=0]{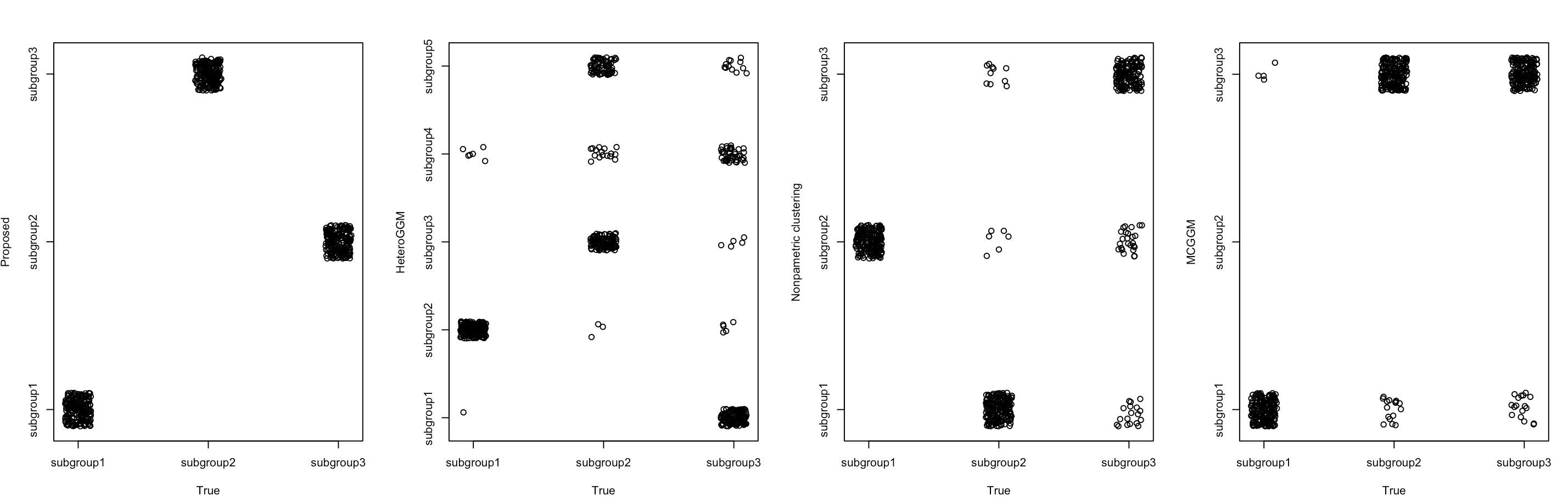}
    \caption{Analysis of one simulation replicate generated under S1 with group sizes (200, 200, 200). 
From left to right: Proposed method, HeteroGGM, nonparametric clustering for CGLasso and MRCE, and MCGGM.}\label{fig:cluster}
\end{figure}

To get some intuition into performance of the proposed and alternative approaches, in Figure \ref{fig:cluster}, for one simulation replicate, we compare grouping performance of different approaches. It is clear that, for this specific replicate, the proposed approach has higher grouping accuracy. Further, in Figures \ref{fig:graph_single_s1} and \ref{fig:graph_single_s2}, we consider one simulation replicate under S1 and S2, respectively. Additionally, we consider two sample size settings. It is observed that the proposed approach generates significantly different network estimations for different sample groups. Under S1 with a relatively simpler structure, performance is already satisfactory under the smaller sample size setting. Under S2, we observe a significant improvement in identification accuracy when sample size increases. 

\clearpage

\begin{table}[!htp]
\caption{Simulation results under S1 with $p=q=50$. In each cell, mean (sd).} \label{tab:s1}
\centering
\renewcommand\arraystretch{0.4}  
\resizebox{\linewidth}{!}{
\begin{tabular}{c c c c c c c c} 
\toprule
$n$ & Method  & &RMSE &TPR &FPR &RI &$\hat{K}_0$ \\
\midrule
\multirow{18}{*}{(200,200,200)} &\multirow{2}{*}{Proposed} & $\btheta$ &1.740(0.613) &0.943(0.047) &0.057(0.014) &\multirow{2}{*}{0.994(0.027)} &\multirow{2}{*}{3.05(0.22)} \\
& &$\bgamma$ &1.531(1.206) &0.961(0.045) &0.008(0.015) & & \\
&\multirow{2}{*}{HeteroGGM} & $\btheta$ &4.146(0.175) &0.980(0.011) &0.894(0.009) &\multirow{2}{*}{0.569(0.250)} &\multirow{2}{*}{4.85(1.27)} \\
& &$\bgamma$ &- &- &- & & \\
&\multirow{2}{*}{CGLasso($K=6$)} & $\btheta$ & 3.632(0.114) &0.861(0.017) &0.266(0.011) &\multirow{2}{*}{0.249(0.096)} &\multirow{2}{*}{6(0)} \\
& &$\bgamma$ &6.083(0.344) &0.962(0.021) &0.108(0.006) & & \\
&\multirow{2}{*}{CGLasso($K=4$)} & $\btheta$ &3.581(0.254) &0.915(0.012)	&0.209(0.027) &\multirow{2}{*}{0.443(0.137)} &\multirow{2}{*}{4(0)} \\
& &$\bgamma$ &4.666(0.855) &0.989(0.013) &0.074(0.011) & & \\
&\multirow{2}{*}{CGLasso($K=3$)} & $\btheta$ &3.638(0.493) &0.938(0.018)	&0.173(0.048) &\multirow{2}{*}{0.502(0.176)} &\multirow{2}{*}{3(0)} \\
& &$\bgamma$ &4.491(1.591) &0.989(0.011) &0.056(0.017) & & \\
&\multirow{2}{*}{MRCE($K=6$)} & $\btheta$
&3.870(0.122) &0.905(0.021)	&0.406(0.024) &\multirow{2}{*}{0.249(0.096)} &\multirow{2}{*}{6(0)} \\
& &$\bgamma$ &6.876(0.347) &0.803(0.054) &0.114(0.016) & & \\
&\multirow{2}{*}{MRCE($K=4$)} & $\btheta$ &3.412(0.373) &0.951(0.021)	&0.292(0.034) &\multirow{2}{*}{0.443(0.137)} &\multirow{2}{*}{4(0)} \\
& &$\bgamma$ &4.677(1.452) &0.907(0.034) &0.156(0.024) & & \\
&\multirow{2}{*}{MRCE($K=3$)} & $\btheta$ &3.197(0.871) &0.971(0.014)	&0.278(0.027) &\multirow{2}{*}{0.502(0.176)} &\multirow{2}{*}{3(0)} \\
& &$\bgamma$ &4.166(2.313) &0.987(0.015) &0.118(0.011) & & \\
&\multirow{2}{*}{MCGGM($K=3$)} & $\btheta$ &3.952(0.614)	&0.794(0.202)	&0.118(0.094) &\multirow{2}{*}{0.395(0.192))} &\multirow{2}{*}{3(0)} \\
& &$\bgamma$ &- &- &- & & \\
\hline

\multirow{18}{*}{(150,200,250)} &\multirow{2}{*}{Proposed} & $\btheta$ &1.457(0.130) &0.950(0.015)	&0.058(0.006) &\multirow{2}{*}{1.000(0.000)} &\multirow{2}{*}{3.00(0.00)} \\
& &$\bgamma$ &0.937(0.512) &0.990(0.021)	&0.003(0.002) & & \\
&\multirow{2}{*}{HeteroGGM} & $\btheta$ &4.152(0.142) &0.986(0.010)	&0.899(0.013) &\multirow{2}{*}{0.547(0.244)} &\multirow{2}{*}{4.50(1.50)} \\
& &$\bgamma$ &- &- &- & & \\
&\multirow{2}{*}{CGLasso($K=6$)} & $\btheta$ &3.681(0.158) &0.889(0.020)	&0.263(0.018) &\multirow{2}{*}{0.212(0.072)} &\multirow{2}{*}{6(0)} \\
& &$\bgamma$ &5.919(0.699) &0.967(0.025)	&0.105(0.010) & & \\
&\multirow{2}{*}{CGLasso($K=4$)} & $\btheta$ &3.458(0.282) &0.926(0.018)	&0.200(0.027) &\multirow{2}{*}{0.417(0.119)} &\multirow{2}{*}{4(0)} \\
& &$\bgamma$ &4.464(0.813) &0.991(0.019)	&0.068(0.011) & & \\
&\multirow{2}{*}{CGLasso($K=3$)} & $\btheta$ &3.645(0.533) &0.948(0.016)	&0.176(0.050) &\multirow{2}{*}{0.431(0.232)} &\multirow{2}{*}{3(0)} \\
& &$\bgamma$ &4.619(1.766) &0.987(0.021)	&0.057(0.019) & & \\
&\multirow{2}{*}{MRCE($K=6$)} & $\btheta$
&3.929(0.623) &0.905(0.028)	&0.343(0.040) &\multirow{2}{*}{0.212(0.072)} &\multirow{2}{*}{6(0)} \\
& &$\bgamma$ &6.535(0.709)	&0.839(0.085)	&0.146(0.022) & & \\
&\multirow{2}{*}{MRCE($K=4$)} & $\btheta$ &3.196(0.324)	&0.957(0.011)	&0.311(0.045) &\multirow{2}{*}{0.417(0.119)} &\multirow{2}{*}{4(0)} \\
& &$\bgamma$ &4.485(0.870)	&0.977(0.042)	&0.160(0.018) & & \\
&\multirow{2}{*}{MRCE($K=3$)} & $\btheta$ &3.346(0.650)	&0.973(0.012)	&0.284(0.046) &\multirow{2}{*}{0.431(0.232)} &\multirow{2}{*}{3(0)} \\
& &$\bgamma$ &4.557(1.894)	&0.973(0.055)	&0.145(0.019) & & \\
&\multirow{2}{*}{MCGGM($K=3$)} & $\btheta$ &4.355(1.579)	&0.768(0.195)	&0.120(0.097) &\multirow{2}{*}{0.394(0.166)} &\multirow{2}{*}{3(0)}\\
& &$\bgamma$ &- &- &- & & \\
 \hline

 \multirow{18}{*}{(500,500,500)} &\multirow{2}{*}{Proposed} & $\btheta$ &0.754(0.035)	&0.999(0.002)	&0.034(0.004) &\multirow{2}{*}{1.000(0.000)} &\multirow{2}{*}{3.00(0.00)} \\
& &$\bgamma$ &0.327(0.023)	&1.000(0.000)	&0.001(0.000) & & \\
&\multirow{2}{*}{HeteroGGM} & $\btheta$ &4.049(0.105)	&0.991(0.004)	&0.879(0.006) &\multirow{2}{*}{0.660(0.087)} &\multirow{2}{*}{5.80(0.70)} \\
& &$\bgamma$ &- &- &- & & \\
&\multirow{2}{*}{CGLasso($K=6$)} & $\btheta$ &3.090(0.165)	&0.934(0.024)	&0.110(0.030) &\multirow{2}{*}{0.470(0.049)} &\multirow{2}{*}{6(0)} \\
& &$\bgamma$ &3.389(0.355)	&0.998(0.004)	&0.056(0.020) & & \\
&\multirow{2}{*}{CGLasso($K=4$)} & $\btheta$ &2.924(0.145)	&0.963(0.021)	&0.089(0.034) &\multirow{2}{*}{0.677(0.025)} &\multirow{2}{*}{4(0)} \\
& &$\bgamma$ &2.707(0.258)	&0.999(0.002)	&0.042(0.009) & & \\
&\multirow{2}{*}{CGLasso($K=3$)} & $\btheta$ &2.618(0.166)	&0.968(0.021)	&0.046(0.018) &\multirow{2}{*}{0.809(0.020)} &\multirow{2}{*}{3(0)} \\
& &$\bgamma$ &1.813(0.067)	&1.000(0.000)	&0.024(0.001) & & \\
&\multirow{2}{*}{MRCE($K=6$)} & $\btheta$
&2.671(0.128)	&0.985(0.009)	&0.266(0.026) &\multirow{2}{*}{0.470(0.049)} &\multirow{2}{*}{6(0)} \\
& &$\bgamma$ &3.235(0.353)	&0.995(0.009)	&0.131(0.012) & & \\
&\multirow{2}{*}{MRCE($K=4$)} & $\btheta$ &2.481(0.161)	&0.997(0.003)	&0.221(0.027) &\multirow{2}{*}{0.677(0.025)} &\multirow{2}{*}{4(0)} \\
& &$\bgamma$ &2.519(0.287)	&0.999(0.003)	&0.096(0.008) & & \\
&\multirow{2}{*}{MRCE($K=3$)} & $\btheta$ &2.220(0.113)	&0.995(0.011)	&0.163(0.012) &\multirow{2}{*}{0.809(0.020)} &\multirow{2}{*}{3(0)} \\
& &$\bgamma$ &1.621(0.092)	&1.000(0.000)	&0.058(0.003) & & \\
&\multirow{2}{*}{MCGGM($K=3$)} & $\btheta$ &4.256(0.479)	&0.797(0.231)	&0.071(0.048) &\multirow{2}{*}{0.405(0.190)} &\multirow{2}{*}{3(0)} \\
& &$\bgamma$ &- &- &- & & \\
\bottomrule	
\end{tabular}}
\end{table}

More definitive results are based on 100 replicates for each setting. The summary results for setting S1 and $p=q=50$ are provided in Table \ref{tab:s1}. The results for the other settings are presented in Tables \ref{tab:s1_p100}-\ref{tab:s3_p100} in Appendix. The proposed approach is observed to have competitive performance across the whole spectrum of simulation. As a representative example, we consider Table \ref{tab:s1}, the setting with group sizes (150, 200, 250). The proposed approach is able to accurately identify the number of sample groups, while HeteroGGM, without accounting for the regulations, over-estimates with a mean of 4.5. HeteroGGM has a satisfactory TPR value for the precision matrices, however, much inferior RMSE and FPR values. The other alternatives all have much inferior estimation performance with much larger RMSEs. They can have acceptable identification performance, especially when the number of sample groups is correctly specified – this can be very difficult in practice. In general, their identification accuracy is worse than the proposed. The alternatives fail to accurately identify the grouping structures. For this specific setting, the proposed approach has an RI value of 1, HeteroGGM has an average RI of 0.547, and the other alternatives all have RI values below 0.5.

\section{Data analysis}

Breast cancer has one of the highest incidence rates, and extensive profiling studies have been conducted on breast cancer. Gene expression data has been collected and analyzed in quite a few studies, among which some are multiomics \citep{Tang2018Prognostic, Lin2020Classifying}. We analyze data collected in the Molecular Taxonomy of Breast Cancer International Consortium (METABRIC) study \citep{Pereira2016Somatic} and refer to the published literature \citep{Curtis2012Genomic} for information on sample and data collection and processing. 

Gene expression and copy number alteration (CNA) measurements are available for 1,758 samples. In principle, the proposed analysis can be conducted with all gene expression and CNA measurements. Considering the limited sample size and large number of unknown parameters, we conduct a ``candidate gene’’ analysis \citep{Tabor2002Candidate} and focus on genes in the KEGG hsa05224 pathway. This pathway is named as ``breast cancer’’ and contains well-known breast cancer related genes such as ESR, MYC, WTN, EGFR, KRAS, HRAS, NRAS, MAPK, and NOTCH. It has been examined in quite a few published studies \citep{Dai2016Cancer}, although it is noted that the perspectives taken in the published studies are significantly different from the proposed. A total of 147 genes belong to this pathway. Among them, two are not measured in the METABRIC study. As such, a total of 145 gene expressions and their corresponding CNAs are available for analysis. We refer to \cite{Curtis2012Genomic, Pereira2016Somatic} for the preprocessing of gene expression and CNA measurements.

\begin{figure}[!htp]
    \centering
    \includegraphics[width=0.3\textwidth, angle=0]{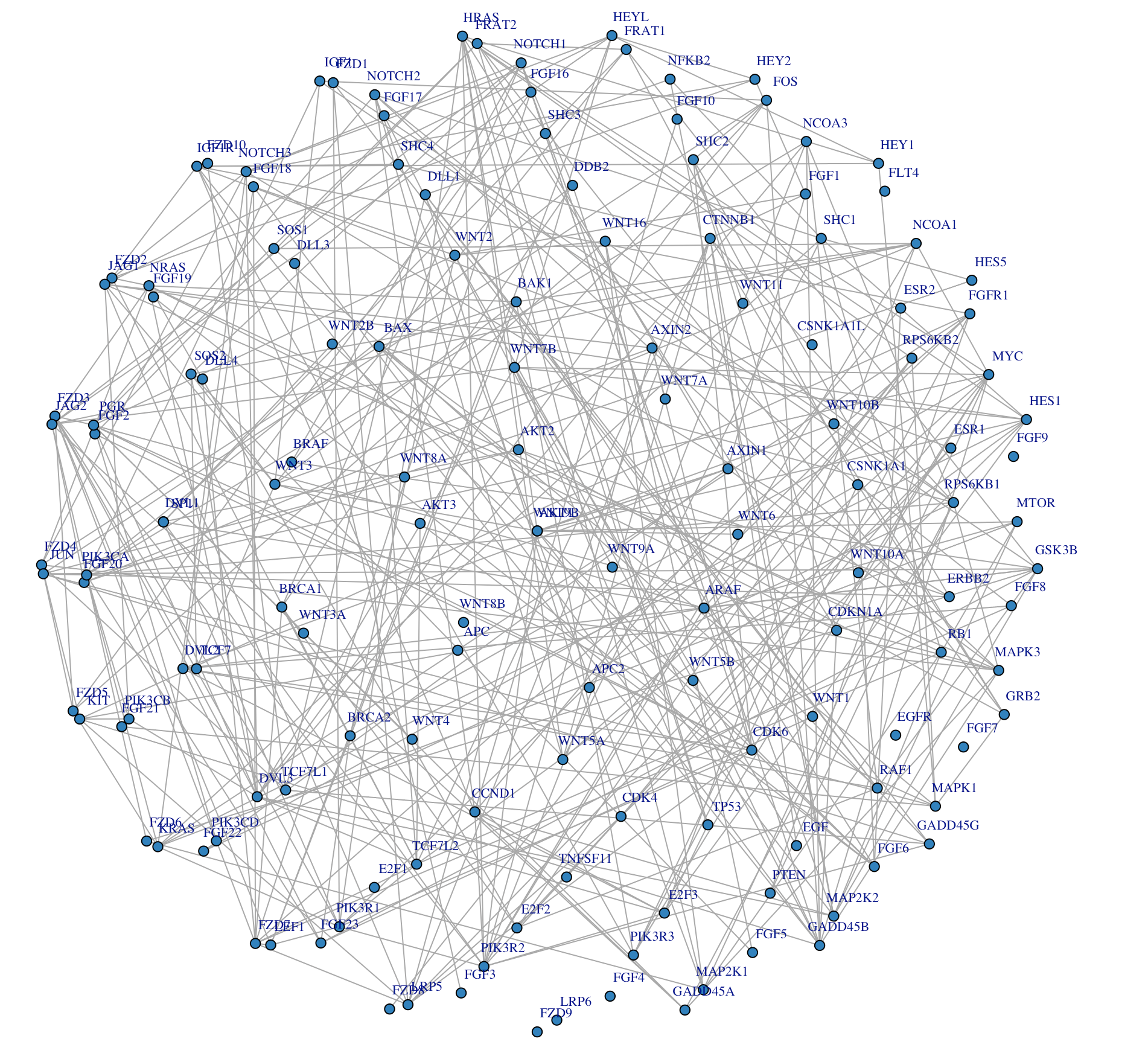}
     \includegraphics[width=0.3\textwidth, angle=0]{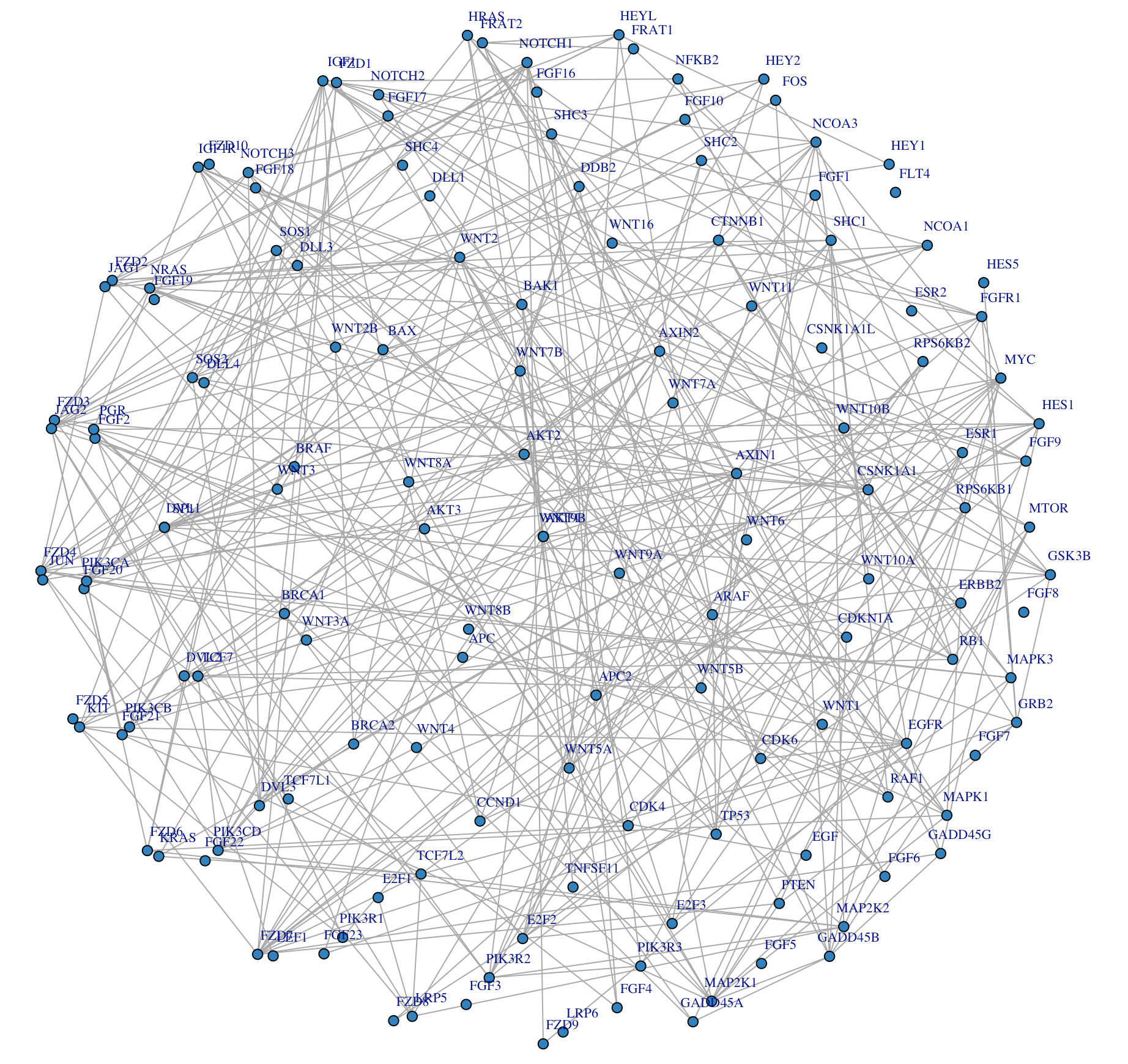}
    \includegraphics[width=0.3\textwidth, angle=0]{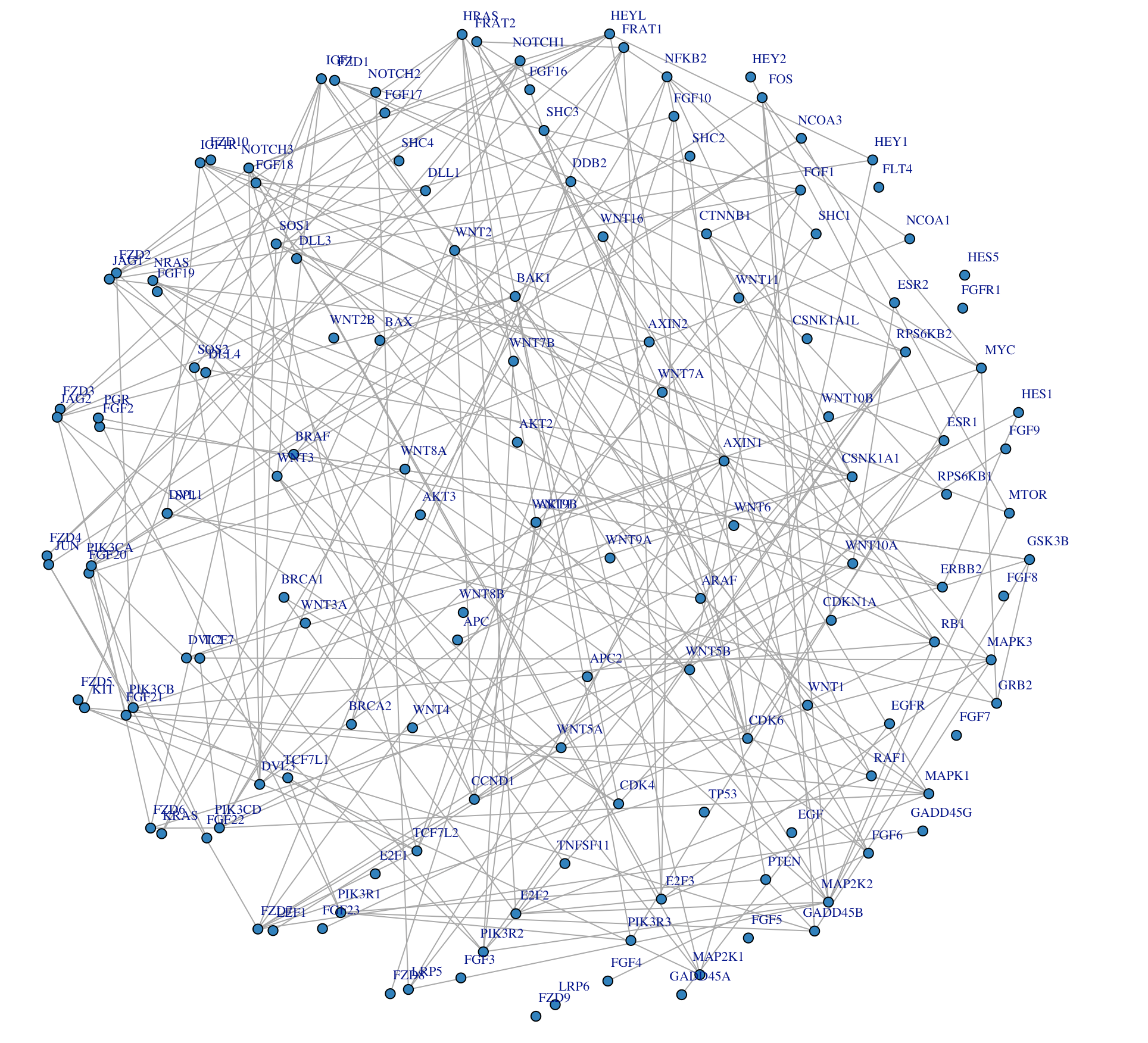} \\
     \includegraphics[width=0.3\textwidth, angle=0]{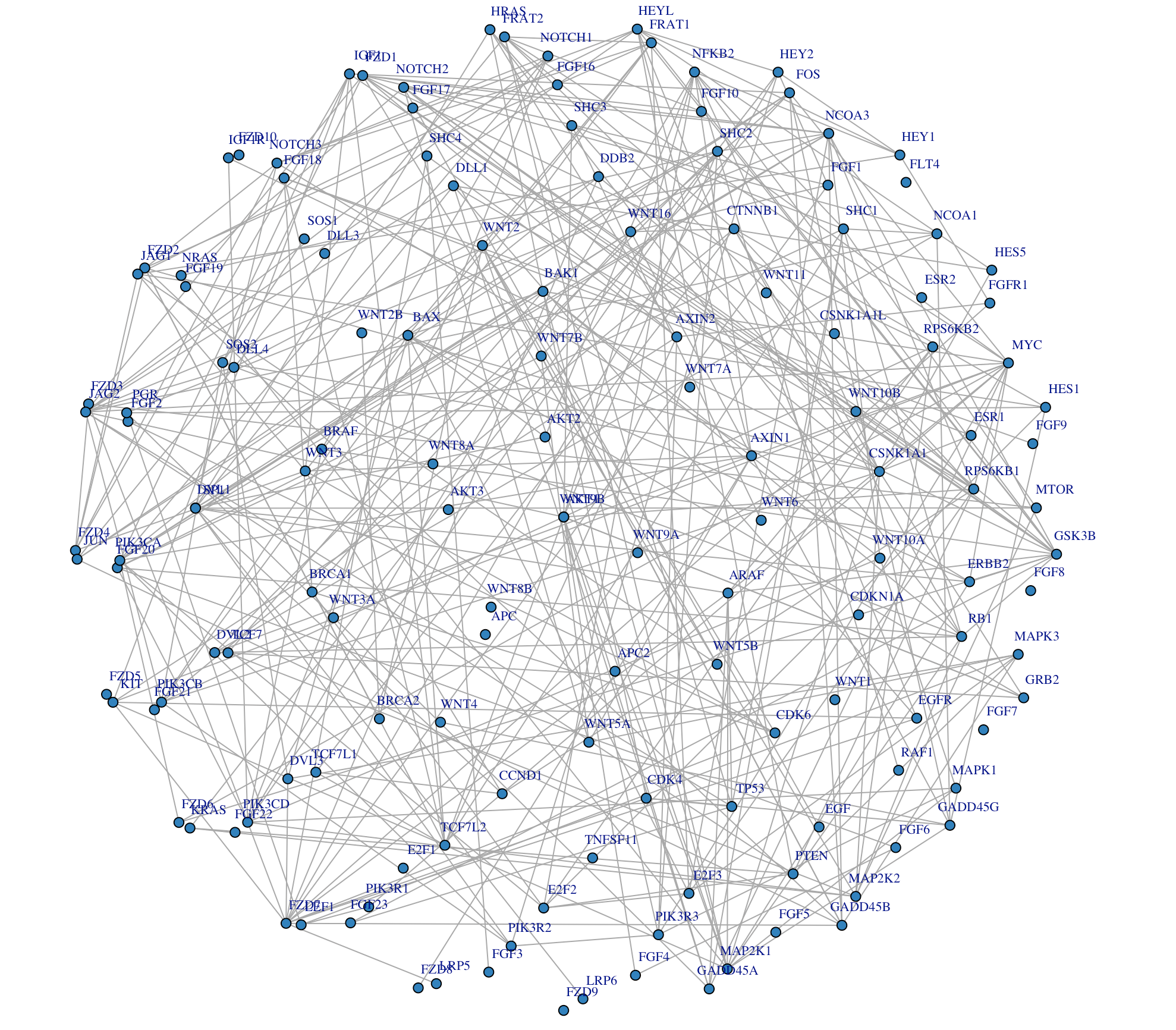}
    \includegraphics[width=0.3\textwidth, angle=0]{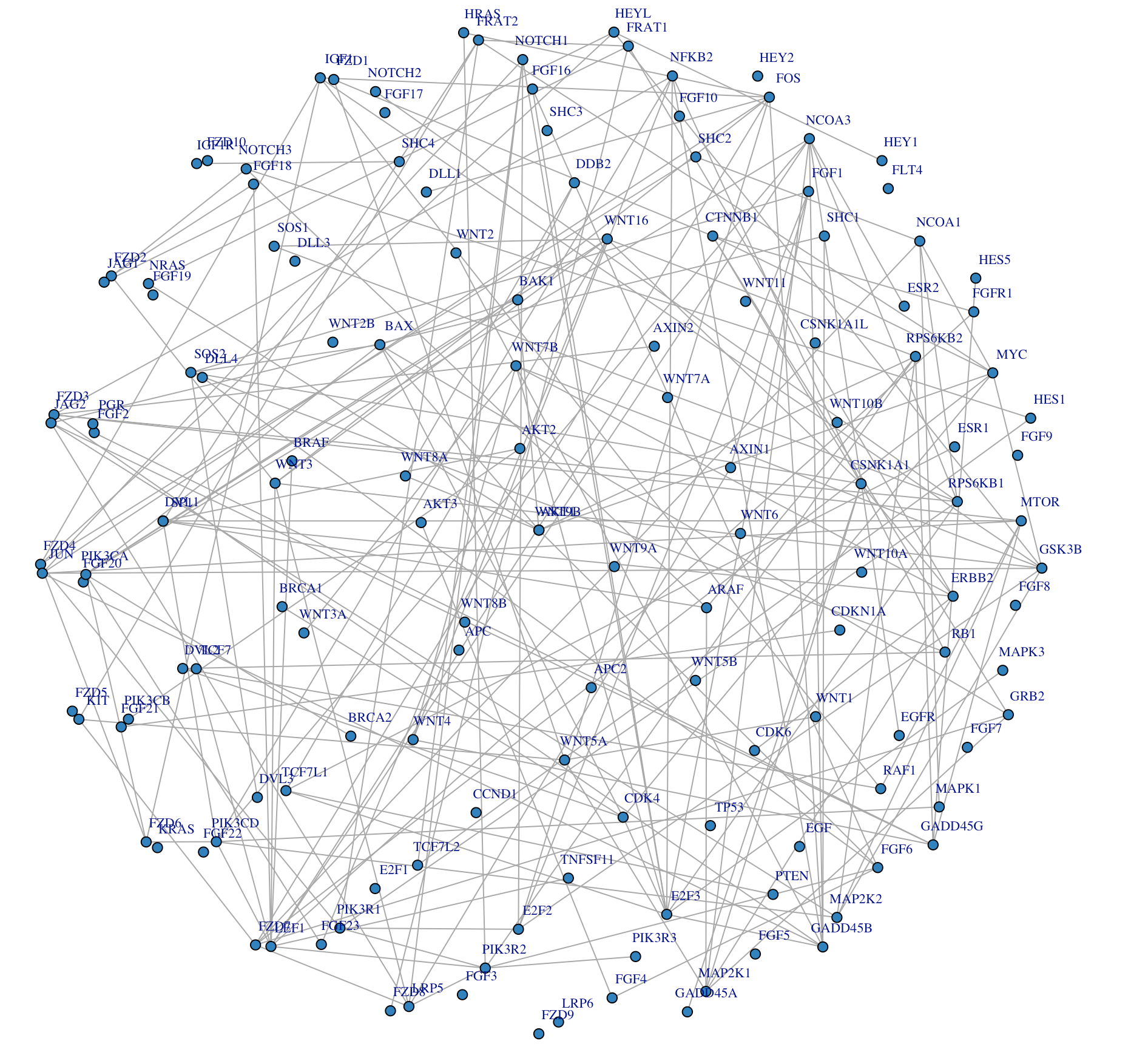}
     \includegraphics[width=0.3\textwidth, angle=0]{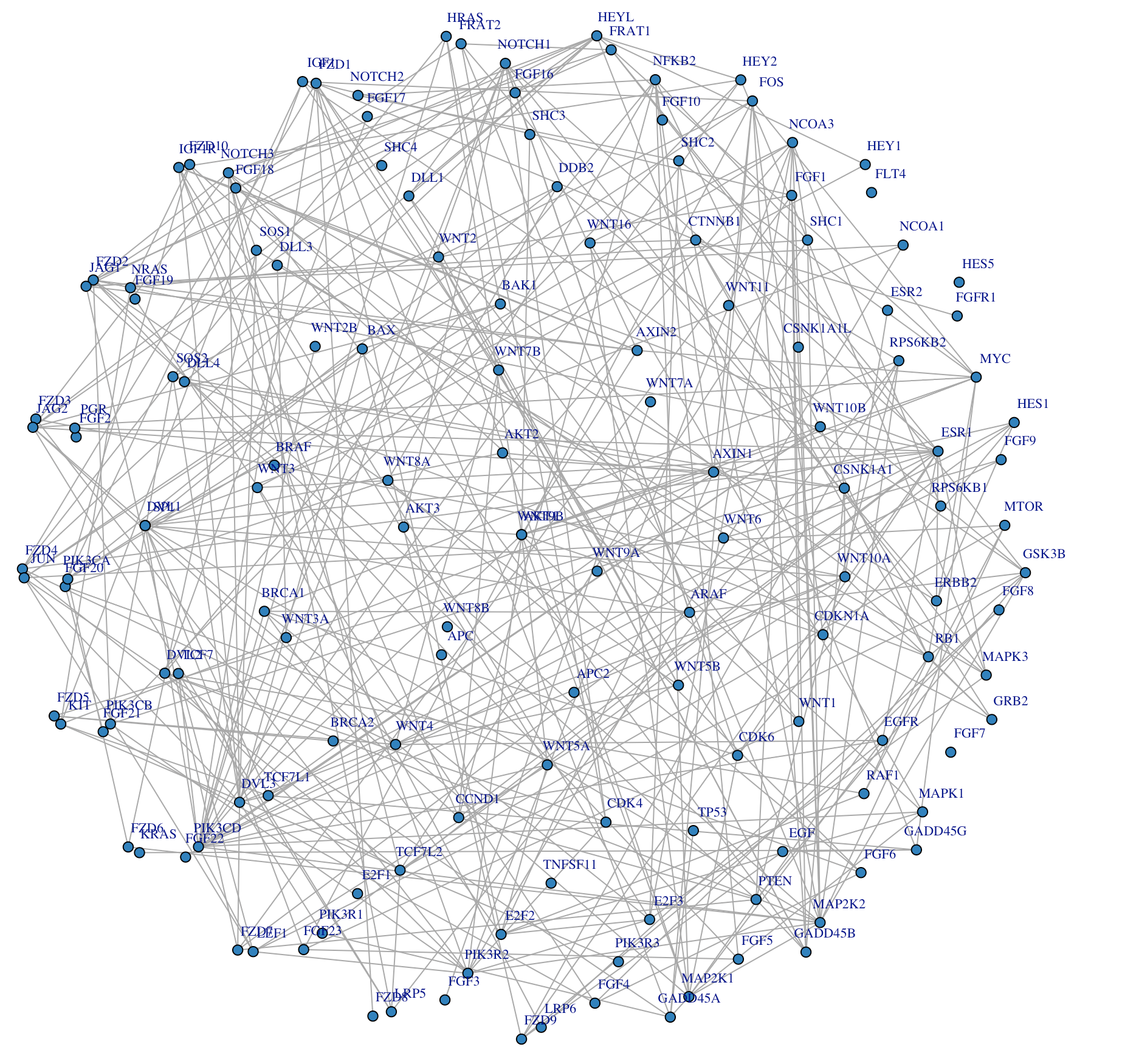}
    \caption{Analysis of METABRIC data: network structures for the six sample groups.}
    \label{fig:metabric_graph}
\end{figure}

\begin{figure}[!htp]
    \centering
    \includegraphics[width=0.85\textwidth, angle=0]{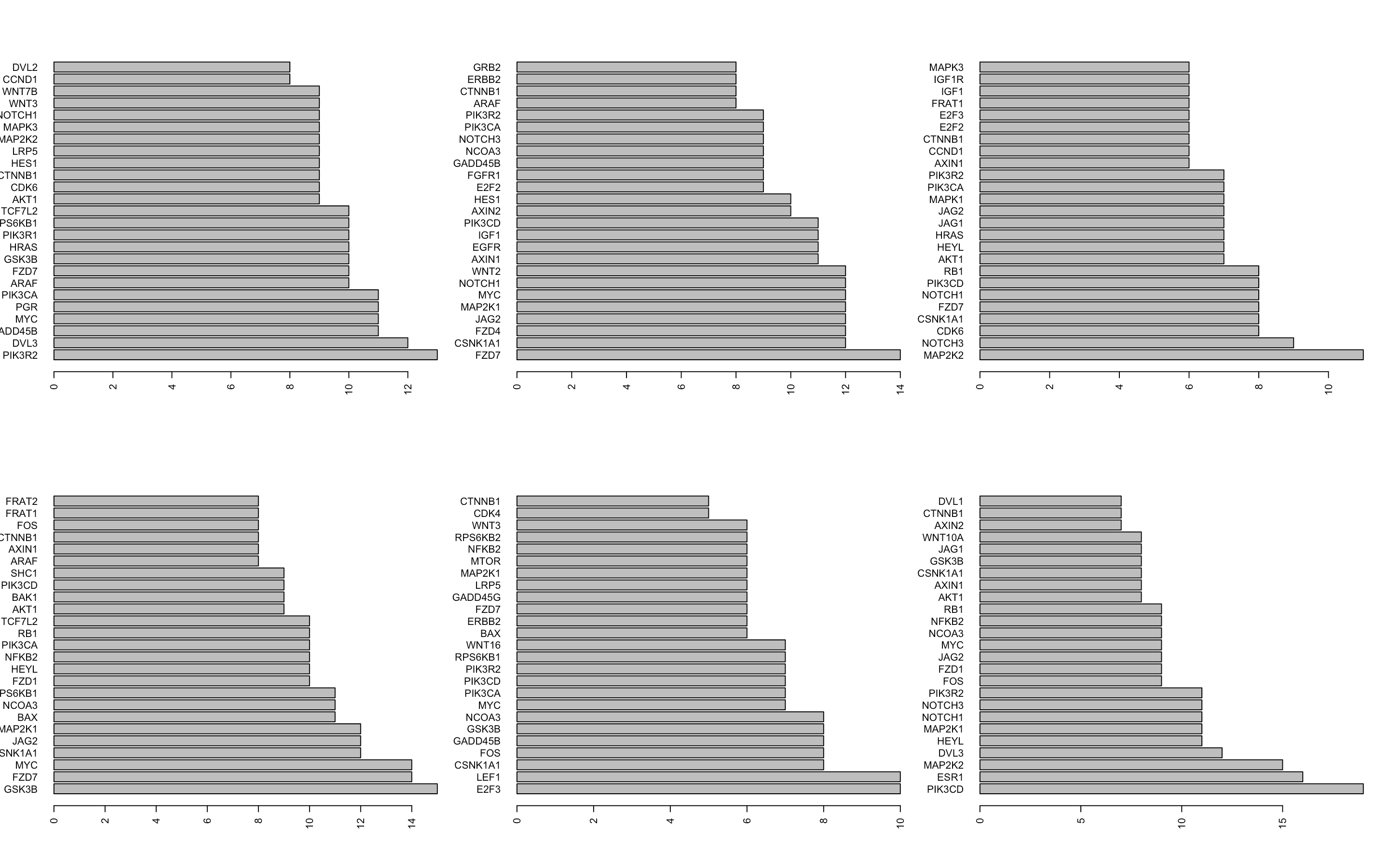}
    \caption{Analysis of METABRIC data: genes with the highest degrees for the six sample groups.}
    \label{fig:metabric_degree}
\end{figure}

\begin{figure}[!htp]
    \centering
    \includegraphics[width=0.7\textwidth, angle=0]{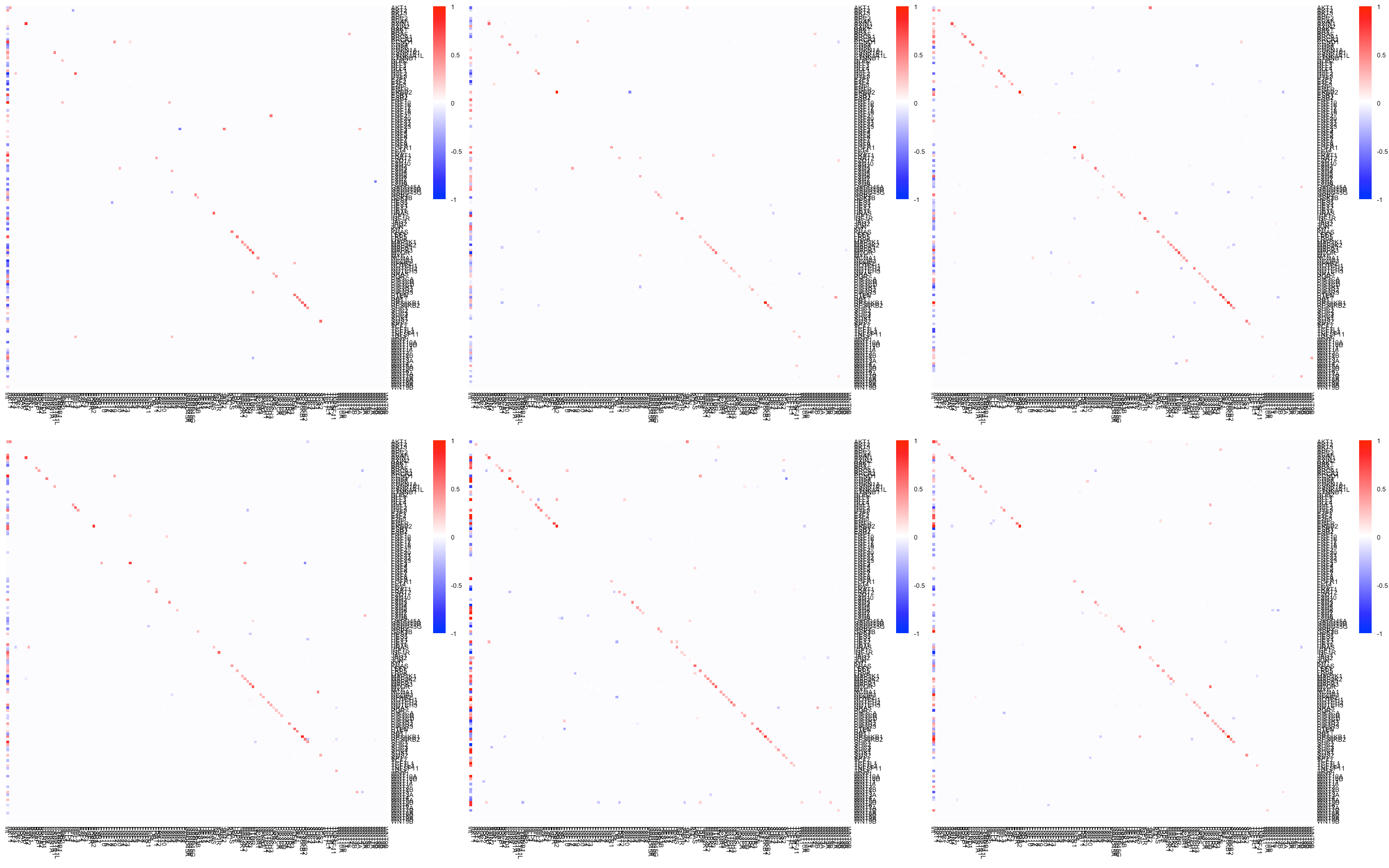}
    \caption{Analysis of METABRIC data: heatmaps of the estimated coefficient matrices for the six sample groups. }
    \label{fig:metabric_heatmap}
\end{figure}

When implementing the proposed approach, we set $K=10$. A total of six sample groups are identified, with sizes 201, 387, 356, 303, 248, and 263. Detailed sample grouping information is available from the authors. For those six groups, the estimated gene expression networks are presented in Figure \ref{fig:metabric_graph}. The six networks have 684, 676, 432, 638, 380, and 652 edges, and Table \ref{tab:metabric_size} suggests that they have small to moderate numbers of overlapping edges. Genes with the highest degrees are presented in Figure \ref{fig:metabric_degree}, and significant differences are observed across the sample groups. For example, gene PIK3CD, which plays a critical role in some solid tumors including breast cancer, is an isolated node in the first network but a key hub node in the other networks, especially the sixth one. Other genes that behave differently in different sample groups include ESR1, DVL3, PGR, RPS6KB1, EGFR, FZD7 and FGFR1. There are also genes that behave similarly in all sample groups, such as CSNK1A1, E2F3 and MYC – they are established breast cancer markers and have high degrees in all the networks. The estimated coefficient matrices are presented in Figure \ref{fig:metabric_heatmap}, where we observe notable differences. In addition, it is observed that the cis regulations are usually the strongest, which is as expected. The regulation relationships are sparse, and there are a few trans regulations. 

\begin{table}[!htp]
    \centering
    \caption{Analysis of  METABRIC data: numbers of edges and overlapping edges for the six sample groups.}\label{tab:metabric_size}
    \renewcommand\arraystretch{0.5}  
    \begin{tabular}{c c c c c c c}
    \toprule
    &Group 1 &Group 2 &Group 3 &Group 4 &Group 5 &Group 6 \\
    \midrule
    Group 1 &684 &134 &118 &140 &68 &110 \\
    Group 2 & &676 &166 &178 &94 &170 \\
    Group 3 & & &432 &136 &72 &148 \\
    Group 4 & & & &638 &96 &152 \\
    Group 5 & & & & &380 &94 \\
    Group 6 & & & & & &652 \\
    \bottomrule
    \end{tabular}
\end{table}

\begin{figure}[!htp]
    \centering
    \includegraphics[width=0.45\textwidth, angle=0]{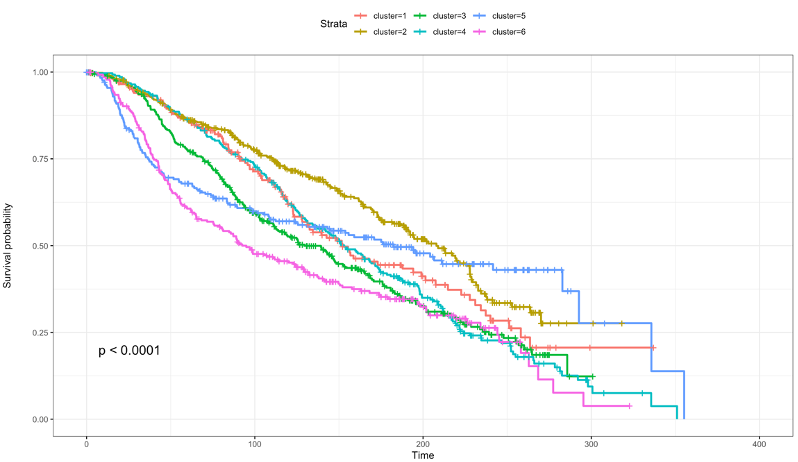}
     \includegraphics[width=0.45\textwidth, angle=0]{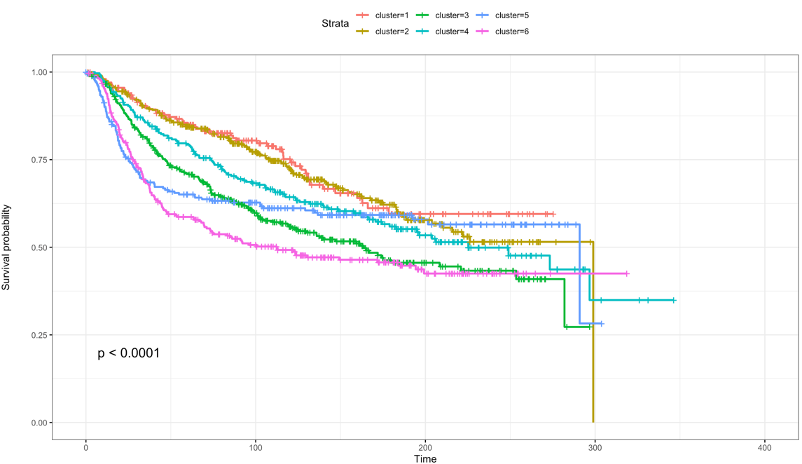}
    \caption{Analysis of METABRIC data: comparison of survival. Left: overall survival; Right: relapse free survival.}
    \label{fig:metabric_surv}
\end{figure}

The proposed analysis is unsupervised. Review of relevant literature suggests that there is a lack of way for determining whether the identified sample groups and their differences (in gene expression network and regulation relationship) are clinically sensible. Here, to provide ``indirect support’’, we compare key clinical features across the identified groups. In Table \ref{tab:metabric_anova}, we report the analysis of variance results for tumor size, mutation count, and tumor burden, all of which have significant clinical implications. In Figure \ref{fig:metabric_surv}, we further compare overall survival and relapse free survival. Significant differences are observed, suggesting that the six sample groups have notable clinical differences. Breast cancer can be classified as luminal A, luminal B, HER2-enriched, basal-like, and Claudin-low \citep{Prat2015Clinical}. In Table \ref{tab:metabric_ri}, we compare the identified six groups against these five subtypes. The Rand index between these two types of grouping is 0.736, suggesting certain consistency. For example, the basal-like ones are mostly in Group 5 identified by the proposed approach, and the HER2-enriched ones are mostly in Group 6. On the other hand, it is recognized that these two groupings also have notable differences. For example, the Claudin-low ones are almost equally presented in Group 2 and Group 5. 

\begin{table}[!htp]
    \centering
    \caption{Analysis of METABRIC data: analysis of variance for selected clinical variables. 
}\label{tab:metabric_anova}
\renewcommand\arraystretch{0.5}  
    \begin{tabular}{c c}
    \toprule
    Clinical variable & p-value \\
    \midrule
    Tumor size & $<0.001$ \\
    Mutation count & $<0.001$ \\
    TMB nonsynonymous & $<0.001$ \\ 
    \bottomrule
    \end{tabular}
\end{table} 

\begin{table}[!htp]
    \centering
    \caption{Analysis of METABRIC data: comparison with subtype. }\label{tab:metabric_ri}
    \renewcommand\arraystretch{0.5}
    \begin{tabular}{c c c c c c c c}
    \toprule
    &Group 1 &Group 2 &Group 3 &Group 4 &Group 5 &Group 6 &Sum \\
    \midrule
    Basal &0 &4 &4 &2 &157 &32 &199 \\
    Claudin-low &2 &80 &4 &0 &79 &34 &199 \\
    Her2 &2 &11 &32 &20 &9 &146 &220 \\
    LumA &114 &250 &80 &219 &2 &14 &679 \\
    LumB &83 &42 &236 &62 &1 &37 &461 \\
    Sum &201 &387 &356 &303 &248 &263 &1758\\
    \bottomrule
    \end{tabular}
\end{table}

\begin{figure}[!htp]
    \centering
    \includegraphics[width=0.85\textwidth, angle=0]{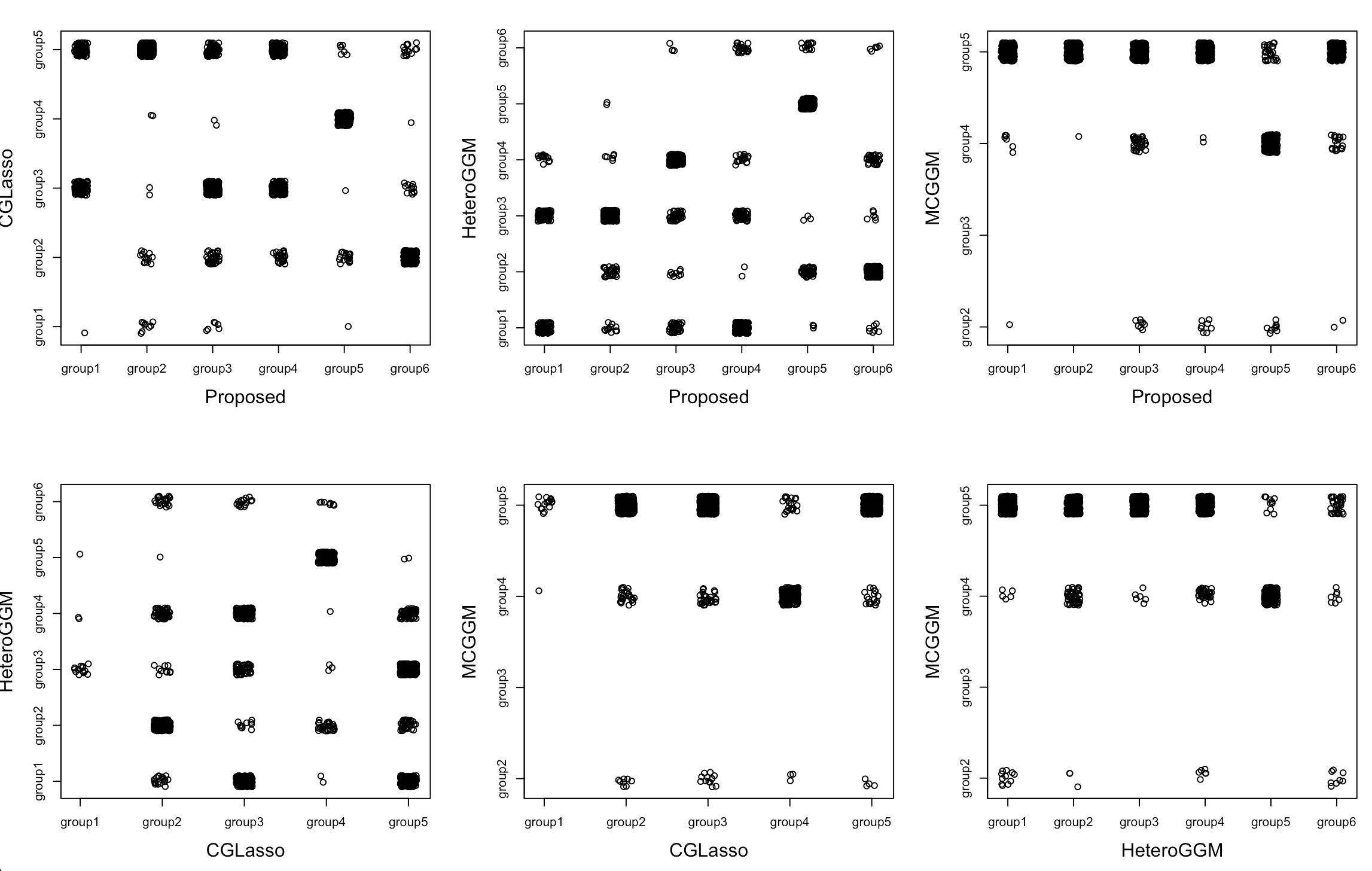} \\
    \caption{Analysis of METABRIC data using different approaches: comparison of grouping results. 
}
\label{fig:metabric_cluster}
\end{figure}

Data is also analyzed using the alternative approaches. With HeteroGGM, the number of sample groups is data-dependently selected to be six. For the other alternatives, we fix the number of groups as five for better comparability. Here it is noted that MCGGM generates two empty groups, leading to three nontrivial ones.
The heterogeneity analysis comparisons are summarized in Table \ref{tab:metabric_rid} and Figure \ref{fig:metabric_cluster}. It is observed that different approaches lead to significantly different groupings. Specifically, CGLasso and HeteroGGM have stronger overlappings with the proposed approach, while MCGGM generates highly imbalanced groups. The Rand index values between the five Claudin subtypes and the alternative approaches are 0.726 (CGLasso), 0.730 (HeteroGGM), and 0.485 (MCGGM). The five networks generated by CGLasso have 590, 132, 96, 142 and 154 edges. Those generated by HeteroGGM have 594, 616, 662, 446, 752 and 3340 edges. And those generated by MCGGM have 238, 70 and 472 edges. It is apparent that the network structures are also significantly different. More detailed results are available from the authors.

\begin{table}[!htp]
 \caption{Analysis of METABRIC data using different approaches. Diagonal: sample group sizes. Off-diagonal: Rand index. }
    \label{tab:metabric_rid}
    \centering
    \renewcommand\arraystretch{0.5}
    \resizebox{\linewidth}{!}{
    \begin{tabular}{c c c c c}
    \toprule
    &Proposed &CGLasso &HeteroGGM &MCGGM \\
    \midrule
    Proposed &201/387/356/303/248/263 &0.777 &0.834 &0.423 \\
    CGLasso & &17/327/531/226/657 &0.771 &0.508 \\
    HeteroGGM & & &380/291/515/338/185/49 &0.452 \\
    MCGGM & & & &0/28/0/275/1455 \\
    \bottomrule
    \end{tabular}}
\end{table}

\section{Discussion}

With gene expression and regulator data, we have developed a new heterogeneity analysis approach that is based on high-dimensional conditional relationships as well as high-dimensional regulations. This analysis/approach includes multiple existing ones as special cases and can be more comprehensive/informative. Theoretical developments not only provide a solid foundation for the proposed approach but also may advance complex high-dimensional statistics – it is noted that there have been limited developments that collectively conduct conditional network and regression analysis, especially in the challenging context of heterogeneity analysis. We have convincingly demonstrated the practical effectiveness of the proposed approach. This study can be potentially extended in multiple ways, for example, to accommodate other network constructions, other modeling for regulations, and other types of data.

\bibliographystyle{apalike}      
\bibliography{ref_arxiv} 

\section{Appendix}

\subsection{Additional Details on Computation}
\setcounter{equation}{0}

We show the detailed calculations to update the estimate of $\boldsymbol{\Theta}^{(t)}$ and $\boldsymbol{\Gamma}^{(t)}$, solving for the optimal solution of equations \eqref{eq:up_gamma} and \eqref{eq:up_theta} in the $t$-th maximization step in the EM algorithm.  

\noindent
\textbf{Update of $\{\boldsymbol{\Gamma}\}$ in the EM algorithm}

Recall that maximizing the conditional expectation with respect to $\boldsymbol{\Gamma}$ in the $t$-th step is equivalent to solving:
\[
\begin{split}
\left\{\boldsymbol{\Gamma}^{(t)}\right\}
& =\arg\min_{\boldsymbol{\Gamma}}\left(\frac{1}{2n} \sum_{i=1}^{n} \sum_{l=1}^{K} L_{il}^{(t)}
\left\{\left(\ty_i-\boldsymbol{\Gamma}_l \tx_i\right)^T \boldsymbol{\Theta}_l^{(t-1)}(\ty_i-\boldsymbol{\Gamma}_l \tx_i) \right\} \right. \\
& \left. +\sum_{l=1}^K\sum_{j=1}^p\sum_{m=1}^{q+1} p(|\gamma_{jm,l}|,\lambda_2) +
\sum_{l<l\prime}p \left( (\Vert \boldsymbol{\Theta}_l^{(t-1)} - \boldsymbol{\Theta}_{l^\prime}^{(t-1)} \Vert_F^2 + \Vert \boldsymbol{\Gamma}_l -\boldsymbol{\Gamma}_{l^\prime} \Vert_F^2)^{1/2}, \lambda_3 \right) \right).
\end{split}
\]
With the local quadratic approximation, it can be rewritten as:
\[
\begin{split}
\left\{\boldsymbol{\Gamma}^{(t)}\right\}
& =\arg\min_{\boldsymbol{\Gamma}}\left(\frac{1}{2n} \sum_{i=1}^{n} \sum_{l=1}^{K} L_{il}^{(t)}
\left\{\left(\ty_i-\boldsymbol{\Gamma}_l \tx_i\right)^T \boldsymbol{\Theta}_l^{(t-1)}(\ty_i-\boldsymbol{\Gamma}_l \tx_i) \right\} \right. \\
& \qquad \left. +\sum_{l=1}^K\sum_{j=1}^p\sum_{m=1}^{q+1} \frac{1}{2}\frac{p^\prime(|\gamma_{jm,l}^{(t-1)}|,\lambda_2)}{|\gamma_{jm,l}^{(t-1)}|}\gamma_{jm,l}^2 +
\sum_{l<l\prime}\frac{1}{2}\frac{p^\prime(\tau_{ll^\prime}^{(t-1,t-1)}, \lambda_3)}{\tau_{ll^\prime}^{(t-1,t-1)}} \Vert \boldsymbol{\Gamma}_l -\boldsymbol{\Gamma}_{l^\prime} \Vert_F^2) \right).
\end{split}
\]
The update of $\boldsymbol{\Gamma}^{(t)}$ is as follows. For $j=1,\dots, p$, $q=1,\dots,q+1$ and $l=1\dots,K$,
\[
\gamma_{jm,l}^{(t)}=\frac{h_{jm,l}^{(t-1)}+n\tilde{v}_{jm,l}^{(t-1)}}{n\hat{v}_{jm,l}^{(t-1)}+n_l\theta_{jj,l}^{(t-1)}C_{xl,mm}^{(t-1)}+np^\prime\left( |\gamma_{jm,l}^{(t-1)}|, \lambda_2\right)/|\gamma_{jm,l}^{(t-1)}| },
\]
where 
\[
\begin{split}
h_{jm,l}^{(t-1)}=
& -n_l^{(t)}\left[\sum_{g\neq j}^p\theta_{gj,l}^{(t-1)}\left(\sum_{h=1}^q C_{xl,mh}^{(t-1)}\gamma_{gh,l}^{(t-1)} \right) + \theta_{jj,l}^{(t-1)} \left( \sum_{h \neq m}^q C_{xl,mh}^{(t-1)}\gamma_{gh,l}^{(t-1)}\right) \right] \\
& \quad + n_l \boldsymbol{e}_m^T\left(\textbf{C}_{yx,l}^{(t-1)T} \boldsymbol{\Theta}_l^{(t-1)}\right)\boldsymbol{e}_j,
\end{split}
\]
\[
\tilde{v}_{jm,l}^{(t-1)}=\sum_{l<l^\prime}\frac{p^\prime(\tau_{ll^\prime}^{(t-1,t-1)}, \lambda_3)}{\tau_{ll^\prime}^{(t-1,t-1)}}\gamma_{jm,l}^{(t-1)},
\]
\[
\hat{v}_{jm,l}^{(t-1)}=\sum_{l<l^\prime}\frac{p^\prime(\tau_{ll^\prime}^{(t-1,t-1)}, \lambda_3)}{\tau_{ll^\prime}^{(t-1,t-1)}},
\]
\[
\tau_{ll^\prime}^{(t-1,t-1)}=\left( \Vert \boldsymbol{\Theta}_l^{(t-1)} - \boldsymbol{\Theta}_{l^\prime}^{(t-1)} \Vert_F^2 + \Vert \boldsymbol{\Gamma}_l^{(t-1)} - \boldsymbol{\Gamma}_{l^\prime}^{(t-1)} \Vert_F^2 \right)^{1/2},
\]
the weighted covariance matrices $\textbf{C}^{(t-1)}_{yl}=\sum_{i=1}^n L_{il}^{(t-1)}\ty_i\ty_i^T/\sum_{i=1}^n L_{il}^{(t-1)}$, $\textbf{C}^{(t-1)}_{yx,l}=\sum_{i=1}^n L_{il}^{(t-1)}\ty_i\tx_i^T/\sum_{i=1}^n L_{il}^{(t-1)}$, $
\textbf{C}^{(t-1)}_{xl}=\sum_{i=1}^n L_{il}^{(t-1)}\tx_i\tx_i^T/\sum_{i=1}^n L_{il}^{(t-1)}$ with $C_{xl,mh}^{(t-1)}$ as its $mh$-th entry, and $n_l^{(t)}=\sum_{i=1}^n L_{il}^{(t)}$. 
$\boldsymbol{e}_m$ denotes a $(q+1)\times 1$ vector with the $m$-th entry being 1 and the others being 0, and $\boldsymbol{e}_j$ denotes a $p\times 1$ vector with the $j$-th entry being 1 and the others being 0. 

\noindent
\textbf{Update of $\{\boldsymbol{\Theta}\}$ in the EM algorithm}

Recall that maximizing the conditional expectation with respect to $\boldsymbol{\Theta}$ in the $t$-th step is equivalent to solving: 
\[
\begin{split}
    \left\{ \boldsymbol{\Theta}^{(t)} \right\}
    & = \arg\min_{\boldsymbol{\Theta}} \left\{  \sum_{l=1}^K n_l^{(t)}\left[-\log \mbox{det}(\boldsymbol{\Theta}_l)+ \tr(\textbf{S}_{\Gamma l}^{(t)}\boldsymbol{\Theta}_l) \right] \right. \\
    & \qquad \left. +  \sum_{l=1}^K\sum_{j \neq m}p(|\theta_{jm,l}|,\lambda_1)
     + \sum_{l<l^\prime}p\left( (\Vert \boldsymbol{\Theta}_l - \boldsymbol{\Theta}_{l^\prime} \Vert_F^2 + \Vert \boldsymbol{\Gamma}_l^{(t)} -\boldsymbol{\Gamma}_{l^\prime}^{(t)} \Vert_F^2)^{1/2}, \lambda_3 \right) \right\},
\end{split}
\]
where $\textbf{S}_{\Gamma l}^{(t)}=\textbf{C}_{yl}^{(t)}-\textbf{C}^{(t)}_{yx,l}\boldsymbol{\Gamma}_l^{(t)T}-\boldsymbol{\Gamma}_l^{(t)}\textbf{C}^{(t)T}_{yx,l}+\boldsymbol{\Gamma}_l^{(t)}\textbf{C}^{(t)}_{xl}\boldsymbol{\Gamma}_l^{(t)T}$, and $\textbf{C}_{yl}^{(t)}$, $\textbf{C}_{xl}^{(t)}$, and $\textbf{C}_{yx,l}^{(t)}$ are the weighted covariance matrices based on $L_{il}^{(t)}$.
This can be solved using the ADMM algorithm \citep{Danaher2014Joint, Wang2020Efficient}, which can be reformulated as: 
\[
\begin{split}
\arg\min_{\btheta, \boldsymbol{\Xi}}
& \left\{ \sum_{l=1}^K n_l^{(t)}\left[-\log \mbox{det}(\boldsymbol{\Theta}_l)+ \tr(\textbf{S}_{\Gamma l}^{(t)}\boldsymbol{\Theta}_l) \right] + \sum_{l=1}^K\sum_{j \neq m}p(|\xi_{jm,l}|,\lambda_1) \right. \\
& \left. + \sum_{l<l^\prime}p\left( (\Vert \boldsymbol{\Xi}_l - \boldsymbol{\Xi}_{l^\prime} \Vert_F^2 + \Vert \boldsymbol{\Gamma}_l^{(t)} -\boldsymbol{\Gamma}_{l^\prime}^{(t)} \Vert_F^2)^{1/2}, \lambda_3 \right)\right\}, \mbox{s.t.} \boldsymbol{\Xi}_l=\btheta_l, l=1,\dots,K,
\end{split}
\]
where $\{\boldsymbol{\Xi}\}=(\boldsymbol{\Xi}_1, \cdots, \boldsymbol{\Xi}_K)$ and $\boldsymbol{\Xi}_k=(\xi_{jm,k})_{1\leq j,m\leq p}$. The scale augmented Lagrangian form for this problem is:
\[
\begin{split}
& \arg\min_{\btheta, \boldsymbol{\Xi}, \boldsymbol{\Psi}}
\left\{ \sum_{l=1}^K n_l^{(t)}\left[-\log \mbox{det}(\boldsymbol{\Theta}_l)+ \tr(\textbf{S}_{\Gamma l}^{(t)}\boldsymbol{\Theta}_l) \right] + \sum_{l=1}^K\sum_{j \neq m}p(|\xi_{jm,l}|,\lambda_1) \right. \\
& \left. + \sum_{l<l^\prime}p\left( (\Vert \boldsymbol{\Xi}_l - \boldsymbol{\Xi}_{l^\prime} \Vert_F^2 + \Vert \boldsymbol{\Gamma}_l^{(t)} -\boldsymbol{\Gamma}_{l^\prime}^{(t)} \Vert_F^2)^{1/2}, \lambda_3 \right) + \frac{\kappa}{2}\sum_{l=1}^K\Vert \btheta_l - \boldsymbol{\Xi}_l + \boldsymbol{\Psi}_l \Vert_F^2 - \frac{\kappa}{2}\sum_{l=1}^K \Vert \boldsymbol{\Psi}_l \Vert_F^2
\right\},
\end{split}
\]
where $\boldsymbol{\Psi}=(\boldsymbol{\Psi}_1, \dots, \boldsymbol{\Psi}_K)$ are dual variables. $\kappa$ is the penalty parameter, and we set $\kappa=1$. We update $\btheta$, $\boldsymbol{\Xi}$ and $\boldsymbol{\Psi}$ iteratively with initial values $\btheta^{(0)}_l=\textbf{I}$, $\boldsymbol{\Xi}_l^{(0)}=\boldsymbol{\Psi}_l^{(0)}=\textbf{0}$.

First, update $\btheta^{(m)}_l$ for $l=1,\dots,K$ by solving: 
\[
\arg\min_{\btheta} \left( \sum_{l=1}^K n_l^{(t)}\left[-\log \mbox{det}(\boldsymbol{\Theta}_l)+ \tr(\textbf{S}_{\Gamma l}^{(t)}\boldsymbol{\Theta}_l) \right] + 
\frac{\kappa}{2}\sum_{l=1}^K\Vert \btheta_l - \boldsymbol{\Xi}_l^{(m-1)} + \boldsymbol{\Psi}_l^{(m-1)} \Vert_F^2 \right).
\]
The closed-form solution is given by:
\[
\btheta_l^{(m)}=\boldsymbol{U}\tilde{\boldsymbol{D}}\boldsymbol{U}^T,
\]
where $\boldsymbol{U}\boldsymbol{D}\boldsymbol{U}^T$ is the eigen-decomposition of $\textbf{S}_{\Gamma l}^{(t)} - \kappa\boldsymbol{\Xi}_l^{(m-1)}/n_l^{(t)} + \kappa\boldsymbol{\Psi}_l^{(m-1)}/n_l^{(t)}$, $\tilde{\boldsymbol{D}}$ is a diagonal matrix with the $j$-th diagonal element $n_l^{(t)}\left[-D_{jj}+(D_{jj}^2 + 4\kappa/n_l^{(t)})^{1/2} \right]/2\kappa$, and $D_{jj}$ is the $j$-th diagonal element of $\boldsymbol{D}$.
Note that $\btheta_l^{(m)}$’s are not necessarily symmetric. They can be symmetrized as:
\[
\theta_{jm,l}^{(m)} = \theta_{jm,l}^{(m)}I(|\theta_{jm,l}^{(m)}|\leq \theta_{mj,l}^{(m)}) +  \theta_{mj,l}^{(m)}I(|\theta_{jm,l}^{(m)}| >  \theta_{mj,l}^{(m)}).
\]

Second, update $\boldsymbol{\Xi}^{(m)}_l$ for $l=1,\dots,K$ by solving:
\begin{eqnarray*}
&&\arg\min_{\boldsymbol{\Xi}} \left( 
\frac{\kappa}{2}\sum_{l=1}^K\Vert \boldsymbol{\Xi}_l - \boldsymbol{Z}_l \Vert_F^2
+ \sum_{l=1}^K\sum_{j \neq m}p(|\xi_{jm,l}|,\lambda_1)
\right.\\
&&~~~~~~~~~~~~\left. + \sum_{l<l^\prime}p\left( (\Vert \boldsymbol{\Xi}_l - \boldsymbol{\Xi}_{l^\prime} \Vert_F^2 + \Vert \boldsymbol{\Gamma}_l^{(t)} -\boldsymbol{\Gamma}_{l^\prime}^{(t)} \Vert_F^2)^{1/2}, \lambda_3 \right) \right),
\end{eqnarray*}
where $\boldsymbol{Z}_l=\btheta_l^{(m)}-\boldsymbol{\Psi}_l^{(m-1)}$. This can be done using the sparse alternating minimization algorithm (S-AMA), which has reformulation:
\[
\arg\min_{\boldsymbol{\Xi}}\left( 
\frac{\kappa}{2} \sum_{j=1}^{p^2}\Vert \boldsymbol{\xi}_{(j)} -\boldsymbol{z}_{(j)} \Vert_2^2 
+ \sum_{j=1}^{p^2}\sum_{l=1}^K p(|\xi_{jl}|, \lambda_1)\cdot I(j \in \mathcal{O}) 
+\sum_{r \in \mathcal{E}}p((\eta_r^{(t)}+\Vert \boldsymbol{v}_r\Vert_2^2)^{1/2}, \lambda_3)
\right), \]
\[
\mbox{s.t.}~ \mvec\boldsymbol{\Xi}_l - \mvec\boldsymbol{\Xi}_{l^\prime} - \boldsymbol{v}_r=0,
\]
where $\boldsymbol{\xi}_{(j)}, \boldsymbol{z}_{(j)} \in \mathbb{R}^K$ are the $j$-th columns of $(\mvec\boldsymbol{\Xi}_1, \cdots, \mvec\boldsymbol{\Xi}_K)^T$ and $(\mvec\boldsymbol{Z}_1, \cdots, \mvec\boldsymbol{Z}_K)^T$, respectively, $j=1,\dots,p^2$, and $\xi_{jl}$ is the $l$-th element of $\boldsymbol{\xi}_{(j)}$. $\mathcal{O}=\{j:j\neq d(p+1), d=0,1,\dots,p-1\}$ is the index set of the off-diagonal elements of the precision matrices. $\mathcal{E}=\{(l,l^\prime):1\leq l,l^\prime \leq K\}$, and $\eta_r^{(t)}=\Vert \boldsymbol{\Gamma}_l^{(t)} -\boldsymbol{\Gamma}_{l^\prime}^{(t)} \Vert_F^2$. The augmented Largrangian form for this problem is: 
\[
\begin{split}
\arg\min_{\boldsymbol{\Xi}, \boldsymbol{V}, \boldsymbol{\Delta}}
& \frac{\kappa}{2}\sum_{j=1}^{p^2}\Vert \boldsymbol{\xi}_{(j)} -\boldsymbol{z}_{(j)} \Vert_2^2 + 
\sum_{j=1}^{p^2}\sum_{l=1}^K p(|\xi_{jl}|, \lambda_1)\cdot I(j \in \mathcal{O}) 
+\sum_{r \in \mathcal{E}}p((\eta_r^{(t)}+\Vert \boldsymbol{v}_r\Vert_2^2)^{1/2}, \lambda_3) \\
& \sum_{r \in \mathcal{E}}\langle \boldsymbol{\delta}_r, \boldsymbol{v}_r-\mvec\boldsymbol{\Xi}_k + \mvec\boldsymbol{\Xi}_{k^\prime}\rangle
+ \frac{\kappa^\prime}{2}\sum_{r\in\mathcal{E}}\Vert \boldsymbol{v}_r-\mvec\boldsymbol{\Xi}_k + \mvec\boldsymbol{\Xi}_{k^\prime} \Vert_2^2,
\end{split}
\]
where $\boldsymbol{V}=(\boldsymbol{v}_1, \dots, \boldsymbol{v}_{|\mathcal{E}|})$, and the dual variables $\boldsymbol{\Delta}=(\boldsymbol{\delta}_1,\dots,\boldsymbol{\delta}_{|\mathcal{E}|})$. $\kappa^\prime$ is a penalty parameter, and we set $\kappa^\prime=0$ when solving for $\boldsymbol{\Xi}$.
S-AMA solves one block of variables at a time and updates iteratively. 

Updating $\boldsymbol{\Xi}$ requires solving $p^2$ individual regularization problems:
\[
\min_{\boldsymbol{\xi}_{(j)}} \frac{\kappa}{2}\Vert \boldsymbol{\xi}_{(j)} -\boldsymbol{u}_{(j)}\Vert_2^2
+\sum_{l=1}^K p(|\xi_{jl}|, \lambda_1)\cdot I(j \in \mathcal{O}), 
\]
where $\boldsymbol{u}_{(j)}=\boldsymbol{z}_{(j)}+\sum_{r\in\mathcal{E}}\delta_{rj}(\boldsymbol{e}_l-\boldsymbol{e}_{l^\prime})$, $\delta_{rj}$ is the $j$-th element of $\boldsymbol{\delta}_r$, and $\boldsymbol{e}_k$ is a $K$-dimensional vector with the $k$-th element being 1 and the other elements being 0. This has a closed-form solution with the MCP penalty:
\[
\hat{\xi}_{jl}=I(j \in \mathcal{O}) \cdot 
\begin{cases}
\frac{S(u_{lj}, \lambda_2/\kappa)}{1-1/(a\kappa)} &\mbox{if} |u_{lj}|\leq a\lambda_2 \\
u_{lj}, &\mbox{if} |u_{lj}|> a\lambda_2
\end{cases}
+ I(j \notin \mathcal{O})\cdot u_{lj},
\]
where $S(t,\lambda)=(|t|-\lambda)_+ \cdot \mbox{sign}(t)$. 

Updating $\boldsymbol{V}$ requires solving the separate regularization problems:
\[
\hat{\boldsymbol{v}}_r = \arg\min_{\boldsymbol{v}_r} \frac{\kappa^\prime}{2}\Vert\boldsymbol{v}_r - \boldsymbol{w}_r \Vert_2^2 + p((\eta_r^{(t)}+\Vert \boldsymbol{v}_r\Vert_2^2)^{1/2}, \lambda_3),
\]
where $\boldsymbol{w}_r=\mvec\boldsymbol{\Xi}_l -\mvec\boldsymbol{\Xi}_{l^\prime}-\boldsymbol{\delta}_r/\kappa^\prime$. This solution has a closed-form with the group MCP penalty:
\[
\hat{\boldsymbol{v}}_r =\frac{\boldsymbol{w}_r}{(\eta_r^{(t)}+\Vert \boldsymbol{w}_r\Vert_2^2)^{1/2}} \cdot
\begin{cases}
   \frac{S((\eta_r^{(t)}+\Vert \boldsymbol{w}_r\Vert_2^2)^{1/2}, \lambda_3/\kappa^\prime)}{1-1/(a\kappa^\prime)} &\mbox{if}  (\eta_r^{(t)}+\Vert \boldsymbol{w}_r\Vert_2^2)^{1/2} \leq a\lambda_3 \\
   (\eta_r^{(t)}+\Vert \boldsymbol{w}_r\Vert_2^2)^{1/2} &\mbox{if} (\eta_r^{(t)}+\Vert \boldsymbol{w}_r\Vert_2^2)^{1/2} > a\lambda_3.
\end{cases}
\]
Update $\boldsymbol{\Delta}$ by $\boldsymbol{\delta}_r \leftarrow \boldsymbol{\delta}_r + \kappa^\prime(\boldsymbol{v}_r -\mvec\boldsymbol{\Xi}_l+\mvec\boldsymbol{\Xi}_{l\prime})\cdot I(\Vert\boldsymbol{v}_r \Vert_2 >0), ~r\in \mathcal{E}$.


Lastly, update $\boldsymbol{\Psi}_l^{(m)}$ for $l=1,\dots, K$ by $\boldsymbol{\Psi}_l^{(m)}=\boldsymbol{\Psi}_l^{(m-1)}+\boldsymbol{\Theta}_l^{(m)}-\boldsymbol{\Xi}_l^{(m)}$.

We iteratively update 
$\boldsymbol{\Theta}$, $\boldsymbol{\Xi}$ and $\boldsymbol{\Psi}$ until $\frac{\Vert \boldsymbol{\Theta}_l^{(m)} - \boldsymbol{\Theta}_l^{(m-1)}\Vert_F}{\Vert\boldsymbol{\Theta}_l^{(m-1)} \Vert_F} < 10^{-2}$ and then conclude convergence.


\subsection{Proofs}
\setcounter{equation}{0}

\begin{proof}

Denote the index set of the diagonal components in the $K_0$ precision matrices as $\mathcal{G}=\bigcup_{k=1}^{K_0}\mathcal{G}_k$, $\mathcal{G}_k=\{k[p(q+1)+1], k[p(q+1)+p+2],\dots, k[p(q+1)+p^2]\}$, $k=1,\dots, K_0$, and the index set of the nonzero components of $\boldsymbol{\Upsilon}^*$ as $\mathcal{M}=\bigcup_{k=1}^{K_0} \mathcal{M}_k$ and $\mathcal{M}_k=\mathcal{M}_{\mathcal{D}_k} \cup \mathcal{M}_{\mathcal{S}_k} \cup \mathcal{G}_k$, where:
\[
\mathcal{M}_{\mathcal{D}_k}=\{(k-1)[p(q+1)+p^2]+(j-1)(q+1)+m: (j,m) \in \mathcal{D}_k\},
\]
\[
\mathcal{M}_{\mathcal{S}_k}=\{(k-1)[p(q+1)+p^2] + p(q+1)+(j-1)p + m: (j,m) \in \mathcal{S}_k \}.
\]
Denote $\boldsymbol{\Omega}^*=(\boldsymbol{\Omega}_1^*, \dots, \boldsymbol{\Omega}_K^*)$ with some true group annotation $\mathcal{T}_k^*=\{l: \boldsymbol{\Omega}_l^*=\boldsymbol{\Upsilon}_k^*, 1\leq l \leq K \}, ~ k=1,\dots,K_0$.
Define $|\mathcal{T}_{\min}|=\min_{1\leq k \leq K_0}|\mathcal{T}_k^*|$ and $|\mathcal{T}_{\max}|=\max_{1\leq k\leq K_0}|\mathcal{T}_k^*|$, where $|\mathcal{T}_k^*|$ is the cardinality of $\mathcal{T}_k^*$. Let
\[
\Lambda_{\mathcal{T}^*}=\left\{ \boldsymbol{\Omega} \in \mathbb{R}^{K(p(q+1)+p^2)}: \boldsymbol{\Omega}_l = \boldsymbol{\Omega}_{l^\prime}, \forall l, l^\prime \in \mathcal{T}_k^*, 1\leq k \leq K_0 \right\}.
\]
First, we define the oracle estimator for $\boldsymbol{\Omega}$, $\hat{\boldsymbol{\Omega}}^o \in \mathbb{R}^{K(p(q+1)+p^2)}$, for which the true grouping structure $\{\mathcal{T}_1^*, \dots, \mathcal{T}_{K_0}^*\}$ is known:
\begin{equation}\label{eq:orac_pl}
\begin{split}
\hat{\boldsymbol{\Omega}}^o=
& \arg\max_{\boldsymbol{\Omega}\in \Lambda_{\mathcal{T}^*}} \left\{ \frac{1}{n}\sum_{i=1}^n \log \left( \sum_{l=1}^K \pi_l f_l(\textbf{y}_i| \tx_i, \boldsymbol{\Gamma}_l, \boldsymbol{\Theta}_l) \right) \right.\\
& \left. - \sum_{l=1}^K \sum_{j \neq m}p(|\theta_{jm,l}|, \lambda_1) 
    - \sum_{l=1}^K \sum_{j=1}^p \sum_{m=1}^{q+1} p(|\gamma_{jm,l}|, \lambda_2) \right\},
\end{split}
\end{equation}
and the corresponding distinct values as $\hat{\boldsymbol{\Upsilon}}^o=(\hat{\boldsymbol{\Upsilon}}^o_1, \dots, \hat{\boldsymbol{\Upsilon}}^o_{K_0})$.
To prove the theorem, it is sufficient to establish the following Result 1 and Result 2.

\textbf{Result 1}: Under the assumed conditions, the oracle estimator satisfies: 
\begin{equation}
 \Vert \hat{\boldsymbol{\Upsilon}}^o -\boldsymbol{\Upsilon}^* \Vert_2 = O\left( K_0^2\sqrt{\frac{(d+s+p)(\log p + \log q)}{n}} \right),
\end{equation}
and $\hat{\mathcal{S}}_k^o =\mathcal{S}_k$ as well as $\hat{\mathcal{D}}_k^o =\mathcal{D}_k$ for $k=1,\dots,K_0$.  

\textit{Proof of Result 1}:
We first define $\bupsilon_0=(\bupsilon_{01}, \cdots, \bupsilon_{0,K_0})$, where $\bupsilon_{0k}=\mvec(\bgamma_{\mathcal{D}_k}, \btheta_k)$ with $\bgamma_{\mathcal{D}_k}=\{\gamma_{jm,k}: (j,m) \in \mathcal{D}_k \}$ contains the corresponding coefficients of the true active ones. Then there exists a local maximizer $\hat{\bupsilon}_0$ of objective function:
\begin{equation}\label{eq:pl_k0}
Q_n(\bupsilon_0)= \frac{1}{n}\sum_{i=1}^n\log\left( \sum_{k=1}^{K_0} \pi_k f_k(\ty_i| \tx_i, \bupsilon_{0k}) \right) - \mathcal{P}_1(\bupsilon_0),
\end{equation}
where $\mathcal{P}_1(\boldsymbol{\Upsilon}_0)=\sum_{k=1}^{K_0}\sum_{j\neq m}|\mathcal{T}_k^*|\rho(|\theta_{jm,k}|,\lambda_1)$ and $\rho(t, \lambda)=\lambda^{-1}p(t, \lambda)$.
Denote $\boldsymbol{\omega}=(\omega_{ik})_{n\times K_0}$, where $\omega_{ik}=I(\ty_i \in \mathcal{A}_k)$ is the latent indicator variable designating the component membership of the $i$-th observation in the mixture, and $\mathcal{A}_k$ is the $k$-th group. If $\omega_{ik}$ is available,  objective function (\ref{eq:pl_k0}) can be written as:
\begin{equation}\label{eq:em_orac}
Q_n(\bupsilon_0|\boldsymbol{\omega}, \ty, \tx) =
\frac{1}{n}\sum_{i=1}^n \sum_{k=1}^{K_0} \omega_{ik} \left[
\log \pi_k + \log f_k(\ty_i| \tx_i, \boldsymbol{\Upsilon}_{0k})  \right] -\mathcal{P}_1(\bupsilon_0). 
\end{equation}
$\omega_{ik}$ is a latent Bernoulli variable with expectation $\mathbb{E}(\omega_{ik}| \ty, \tx, \bupsilon)=P(\omega_{ik}=1|\ty, \tx, \bupsilon)$, which is denoted as $L_k(\ty_i; \tx_i, \bupsilon)=\frac{\pi_kf_k(\ty_i| \tx_i,\boldsymbol{\Upsilon})}{\sum_{k=1}^{K_0} \pi_k f_k(\ty_i| \tx_i, \boldsymbol{\Upsilon})}$, and its population version is $\mathbb{E}(L_k(\ty; \tx, \bupsilon))=\int L_k(\ty; \tx, \bupsilon) f(\ty|\tx, \bupsilon)d\ty$. 
In the $t$-th step of the EM algorithm, it is needed to maximize the conditional expectation of (\ref{eq:em_orac}), which is denoted as $\tilde{H}_n(\bupsilon_0|\bupsilon_0^{(t-1)})$: 
\begin{equation}\label{eq:em_H}
\tilde{H}_n(\bupsilon_0|\bupsilon_0^{(t-1)}) = \mathbb{E}_{\boldsymbol{\omega}, \ty, \tx|\bupsilon_0^{(t-1)}}[Q_n(\bupsilon_0|\boldsymbol{\omega}, \ty, \tx)] =
H_n(\bupsilon_0|\bupsilon_0^{(t-1)}) - \mathcal{P}_1(\bupsilon_0),
\end{equation}
where 
\[
H_n(\boldsymbol{\Upsilon}_0|\boldsymbol{\Upsilon}_0^{(t-1)}) =
    \frac{1}{n}\sum_{i=1}^n \sum_{k=1}^{K_0} L_k(\ty_i;\tx_i,\boldsymbol{\Upsilon}_0^{(t-1)}) \left[
\log \pi_k + \log f_k(\ty_i| \tx_i, \boldsymbol{\Upsilon}_0)  \right]. 
\]
Here $L_k(\ty_i;\tx_i,\boldsymbol{\Upsilon}_0^{(t-1)})=\frac{\pi_k^{(t-1)} f_k(\ty_i|\tx_i,\boldsymbol{\Upsilon}_0^{(t-1)})}{\sum_{k=1}^{K_0} \pi_k^{(t-1)} f_k(\ty_i| \tx_i, \boldsymbol{\Upsilon}_0^{(t-1)})}$, which depends on $\pi_k^{(t-1)}$ and $\bupsilon_0^{(t-1)}$.  $\bupsilon_0^{(t-1)}$ is the estimate from the $(t-1)$-th iteration of the EM algorithm, and $\bupsilon_0^{(t-1)}= \mvec(\bgamma_{\mathcal{D}}^{(t-1)}, \btheta^{(t-1)})=\arg\max_{\bupsilon_0}\tilde{H}_n(\bupsilon_0|\bupsilon_0^{(t-2)})$.

\textit{STEP 1}:
Define $\bupsilon_0^*=(\bupsilon_{01}^*, \cdots, \bupsilon_{0,K_0}^*)$, where $\bupsilon_{0k}^*=\mvec(\bgamma^*_{\mathcal{D}_k}, \btheta^*_k)$ with $\bgamma^*_{\mathcal{D}_k}=\{\gamma^*_{jm,k}: (j,m) \in \mathcal{D}_k \}$ is the true value of the nonzero parameters in the coefficient matrices.
We need to show that if $\boldsymbol{\Upsilon}_0^{(t-1)} \in \mathcal{B}_{\alpha}(\boldsymbol{\Upsilon}_0^*)$, then $\Vert \bupsilon_0^{(t)} - \bupsilon^*_0 \Vert_2 \leq \chi$ with probability tending to 1, where $\chi=\frac{4\epsilon}{\varrho} + \iota\Vert\boldsymbol{\Upsilon}_0^{(t-1)}-\boldsymbol{\Upsilon}_0^*  \Vert_2$, $1/6 \leq \iota < 1$. We set $\alpha=O((d+s+p)\sqrt{(\log p +\log q)/n})$ and $\epsilon=O(\sqrt{(d+s+p)(\log p + \log q)/n})$.

It suffices to show that, for
\begin{equation}
    q(\textbf{v})= \tilde{H}_n(\boldsymbol{\Upsilon}^*_0 + \textbf{v}|\boldsymbol{\Upsilon}_0^{(t-1)}) - \tilde{H}_n(\boldsymbol{\Upsilon}^*_0|\boldsymbol{\Upsilon}_0^{(t-1)}),
\end{equation}
$P(\mbox{sup}_{\textbf{v} \in C(\chi)} q(\textbf{v}) < 0) \rightarrow 1$, where $C(\chi)=\{\textbf{v}: \Vert \textbf{v} \Vert_2=\chi\}$. Note that for a sufficiently large $n$, $\chi \leq 2\alpha$.

Next, we first show that for any $\boldsymbol{\Upsilon}_0$, with probability at least $1-\delta$, for all $\boldsymbol{\Upsilon}_0^\prime=\mvec(\bgamma_{\mathcal{D}}^\prime, \btheta^\prime) \in \{\boldsymbol{\Upsilon}_0^\prime: \Vert \boldsymbol{\Upsilon}_0^\prime-\boldsymbol{\Upsilon}^*_0 \Vert_2 \leq \alpha_0 \}$,
\begin{equation}\label{eq:h1}
 H_n(\boldsymbol{\Upsilon}_0^\prime|\boldsymbol{\Upsilon}_0) - H_n(\boldsymbol{\Upsilon}_0^*|\boldsymbol{\Upsilon}_0) \leq
    \left \langle \nabla H_n(\boldsymbol{\Upsilon}_0^*|\boldsymbol{\Upsilon}_0), \boldsymbol{\Upsilon}_0^\prime-\boldsymbol{\Upsilon}_0^* \right \rangle - \frac{\varrho}{2}\Vert \boldsymbol{\Upsilon}_0^\prime-\boldsymbol{\Upsilon}_0^* \Vert_2^2,
\end{equation}
with a sufficiently large $n$, where $\varrho=c\cdot \min(\beta_1C_0, (\beta_2+\alpha_0)^{-2}/4)$ and the gradient $\nabla H_n(\boldsymbol{\Upsilon}_0^*|\boldsymbol{\Upsilon}_0)$ is taken with respect to the first variable in $H_n(\cdot|\cdot)$.

For $k=1,\dots, K_0$, we define:
$$\nabla_{\bupsilon_{0k}^\prime} H_n(\boldsymbol{\Upsilon}^\prime_{0k}|\boldsymbol{\Upsilon}_0)=\left( [\mbox{vec}(\nabla _{\boldsymbol{\Gamma}_{\mathcal{D}_k}^\prime} H_n(\boldsymbol{\Upsilon}^\prime_{0k}|\boldsymbol{\Upsilon}_0))]^T, [\mbox{vec}(\nabla _{\boldsymbol{\Theta}_k^\prime} H_n(\boldsymbol{\Upsilon}^\prime_{0k}|\boldsymbol{\Upsilon}_0))]^T\right)^T,$$
$\nabla _{\boldsymbol{\Gamma}_{\mathcal{D}_k}^\prime} H_n(\boldsymbol{\Upsilon}^\prime_{0k}|\boldsymbol{\Upsilon}_0)
=\frac{1}{n}\sum_{i=1}^n \left[ L_k(\textbf{y}_i; \tx_i, \bupsilon_0)\boldsymbol{\Theta}_k^\prime(\textbf{y}_i-\boldsymbol{\Gamma}_{\mathcal{D}_k}^\prime \textbf{x}_{i\mathcal{D}_k})\textbf{x}_{i\mathcal{D}_k}^T \right]$, and \\
$ \nabla _{\boldsymbol{\Theta}_k^\prime} H_n(\boldsymbol{\Upsilon}^\prime_{0k}|\boldsymbol{\Upsilon}_0)
=\frac{1}{2n}\sum_{i=1}^n \left[ L_k(\textbf{y}_i; \tx_i, \bupsilon_0)\boldsymbol{\Theta}_k^{\prime-1} \right]
-\frac{1}{2n}\sum_{i=1}^n \left[ L_k(\textbf{y}_i; \tx_i, \bupsilon_0) (\textbf{y}_i-\boldsymbol{\Gamma}^\prime_{\mathcal{D}_k}\textbf{x}_{i\mathcal{D}_k})(\textbf{y}_i-\boldsymbol{\Gamma}^\prime_{\mathcal{D}_k}\textbf{x}_{i\mathcal{D}_k})^T\right]$, 
where $\tx_{i\mathcal{D}_k}=\{x_{ij}, j\in \mathcal{D}_k\}$. Note that:
\[
\begin{split}
& H_n(\boldsymbol{\Upsilon}_{0k}^\prime|\boldsymbol{\Upsilon}_0) - H_n(\boldsymbol{\Upsilon}_{0k}^*|\boldsymbol{\Upsilon}_0)
 = \frac{1}{n}\sum_{i=1}^n\left\{
L_k(\textbf{y}_i; \tx_i, \bupsilon_0)\left[\frac{1}{2}\log\det(\boldsymbol{\Theta}_k^\prime)-\frac{1}{2}\log\det(\boldsymbol{\Theta}_k^*) \right. \right.  \\
&\quad  \left. \left. \quad +\frac{1}{2}(\textbf{y}_i-\boldsymbol{\Gamma}_{\mathcal{D}_k}^*\textbf{x}_{i\mathcal{D}_k})^T\boldsymbol{\Theta}_k^*(\textbf{y}_i-\boldsymbol{\Gamma}_{\mathcal{D}_k}^*\textbf{x}_{i\mathcal{D}_k}) - \frac{1}{2}(\textbf{y}_i-\boldsymbol{\Gamma}_{\mathcal{D}_k}^\prime\textbf{x}_{i\mathcal{D}_k})^T\boldsymbol{\Theta}_k^\prime(\textbf{y}_i-\boldsymbol{\Gamma}_{\mathcal{D}_k}^\prime\textbf{x}_{i\mathcal{D}_k}) \right]\right\}.
\end{split}
\]
If we define $h(\boldsymbol{\Gamma}, \boldsymbol{\Theta})=\frac{1}{2}(\textbf{y}_i-\boldsymbol{\Gamma}\textbf{x}_{i\mathcal{D}_k})^T\boldsymbol{\Theta}(\textbf{y}_i-\boldsymbol{\Gamma}\textbf{x}_{i\mathcal{D}_k})$, then we can show that: 
\[
 H_n(\boldsymbol{\Upsilon}_{0k}^\prime|\boldsymbol{\Upsilon}_0) - H_n(\boldsymbol{\Upsilon}_{0k}^*|\boldsymbol{\Upsilon}_0) - 
    \left \langle \nabla_{\boldsymbol{\Upsilon}_{0k}^\prime} H_n(\boldsymbol{\Upsilon}_{0k}^*|\boldsymbol{\Upsilon}_0), \boldsymbol{\Upsilon}_{0k}^\prime-\boldsymbol{\Upsilon}_{0k}^* \right \rangle = H_1 + H_2,
\]
with
\[
H_1 = \frac{1}{n}\sum_{i=1}^n\left\{ L_k(\ty_i;\tx_i,\bupsilon_0) \left[ h(\bgamma_{\mathcal{D}_k}^*, \btheta_k^*)- h(\bgamma_{\mathcal{D}_k}^\prime, \btheta_k^*)\right]\right\} - [ \mvec(\bgamma_{\mathcal{D}_k}^\prime - \bgamma_{\mathcal{D}_k}^*)]^T\nabla_{\bgamma_{\mathcal{D}_k}^*}H_n(\bupsilon_{0k}^*|\bupsilon_0),
\]
\begin{eqnarray*}
&&H_2 = \frac{1}{n}\sum_{i=1}^n \left\{L_k(\ty_i;\tx_i, \bupsilon_0)\left[ \frac{1}{2}\log\det(\btheta_k^\prime) - \frac{1}{2}\log \det(\btheta_k^*) + h(\bgamma_{\mathcal{D}_k}^\prime, \btheta_k^*) - h(\bgamma_{\mathcal{D}_k}^\prime, \btheta_k^\prime)\right] \right\} \\
&&~~~~~~~~~~~~- 
[ \mvec(\btheta_k^\prime - \btheta_k^*)]^T\nabla_{\btheta_k^*}H_n(\bupsilon_{0k}^*|\bupsilon_0).
\end{eqnarray*}

As for $H_1$, if we define $g_1(\bgamma_{\mathcal{D}_k})=-\frac{1}{n}\sum_{i=1}^nL_k(\ty_i; \tx_i, \bupsilon_0)h(\bgamma_{\mathcal{D}_k}, \btheta_k^*)$ and note that
\[
\begin{split}
H_1 
& = g_1(\bgamma_{\mathcal{D}_k}^\prime)-g_1(\bgamma_{\mathcal{D}_k}^*)-\left \langle\mvec(\nabla g_1(\bgamma_{\mathcal{D}_k}^*)), \mvec(\bgamma_{\mathcal{D}_k}^\prime-\bgamma_{\mathcal{D}_k}^*) \right \rangle \\
& = \frac{1}{2}[\mvec(\bgamma_{\mathcal{D}_k}^\prime-\bgamma_{\mathcal{D}_k}^*)]^T\nabla^2g_1(\textbf{Z})[\mvec(\bgamma_{\mathcal{D}_k}^\prime-\bgamma_{\mathcal{D}_k}^*)]
\end{split}
\]
according to Taylor’s expansion, with $\textbf{Z}=t\bgamma_{\mathcal{D}_k}^\prime+(1-t)\bgamma_{\mathcal{D}_k}^*$ and $t\in [0,1]$, then:
\[
-\nabla^2g_1(\textbf{Z})= \frac{1}{n} \sum_{i=1}^n L_k(\ty_i; \tx_i, \bupsilon_0)\btheta_k^* \otimes (\tx_{i\mathcal{D}_k} \tx_{i\mathcal{D}_k}^T),
\]
\begin{equation}\label{eq:exp_samp}
\begin{split}
& \lambda_{\min}(-\nabla^2g_1(\textbf{Z})) 
\geq \lambda_{\min}(\btheta_k^*) \lambda_{\min} \left( \frac{1}{n} \sum_{i=1}^n L_k(\ty_i; \tx_i, \bupsilon_0) \tx_{i\mathcal{D}_k} \tx_{i\mathcal{D}_k}^T \right) \\
& \quad \geq \lambda_{\min}(\btheta_k^*) \left[ \lambda_{\min}(\frac{1}{n}\textbf{X}_{\mathcal{D}_k}^T \textbf{G}_k \textbf{X}_{\mathcal{D}_k}) - \left\lVert \frac{1}{n} \sum_{i=1}^n L_k(\ty_i; \tx_i, \bupsilon_0) \tx_{i\mathcal{D}_k} \tx_{i\mathcal{D}_k}^T - \frac{1}{n}\textbf{X}_{\mathcal{D}_k}^T \textbf{G}_k \textbf{X}_{\mathcal{D}_k} \right\rVert_F \right],
\end{split}
\end{equation}
where $\textbf{G}=\mbox{diag}(\mathbb{E}(L_k(\ty_i; \tx_i, \bupsilon_0))$ is a diagonal matrix. 
Note that $L_{ik}(\ty_i; \tx_i, \bupsilon_0) \tx_{i\mathcal{D}_k} \tx_{i\mathcal{D}_k}^T$ is bounded by some matrix $\textbf{A}$, and define $M^2=\Vert \textbf{A} \Vert^2$, where $\Vert \cdot\Vert$ is the spectral norm. According to matrix Hoeffding’s inequality, we have:
\begin{equation}\label{eq:matrixhoe}
P\left\{ \lambda_{\max}\left( \frac{1}{n}\sum_{i=1}^n L_k(\ty_i; \tx_i, \bupsilon_0) \tx_{i\mathcal{D}_k} \tx_{i\mathcal{D}_k}^T - \frac{1}{n}\textbf{X}_{\mathcal{D}_k}^T \textbf{G}_k \textbf{X}_{\mathcal{D}_k} \right) \ge t \right\} \ge 1-d_k\exp(-nt^2/8M^2),
\end{equation}
where $d_k=|\mathcal{D}_k|$. If we let $t=\sqrt{\frac{8M^2}{n}\log\frac{d_k K_0}{\delta}}$, then 
$$\left\lVert \frac{1}{n} \sum_{i=1}^n L_k(\ty_i; \tx_i, \bupsilon_0) \tx_{i\mathcal{D}_k} \tx_{i\mathcal{D}_k}^T - \frac{1}{n}\textbf{X}_{\mathcal{D}_k}^T \textbf{G}_k \textbf{X}_{\mathcal{D}_k} \right\rVert_F \leq \sqrt{\frac{8d_kM^2}{n}\log\frac{d_k K_0}{\delta}}$$ 
with probability at least $1-\delta/K_0$, 
and $\sqrt{\frac{8d_kM^2}{n}\log\frac{d_k K_0}{\delta}}=o(1)$. 
Therefore, by Condition (C1) and (C3), we have $H_1 \leq -\beta_1C_0 \Vert \mvec(\bgamma_{\mathcal{D}_k}^\prime - \bgamma_{\mathcal{D}_k}^*) \Vert_2^2/2$ with probability at least $1-\delta/K_0$.

As for $H_2$, if we define $g_2(\btheta_k)=\frac{1}{n}\sum_{i=1}^n \left[ L_k(\ty_i; \tx_i,\bupsilon_0) \left\{\frac{1}{2}\log\det(\btheta_k)-h(\bgamma_{\mathcal{D}_k}^\prime, \btheta_k) \right\}\right]$, then 
\[
H_2=g_2(\btheta_k^\prime)-g_2(\btheta_k^*)-\left \langle\mvec(\nabla g_2(\btheta_k^*)), \mvec(\btheta_k^\prime-\btheta_k^*) \right \rangle = 
\frac{1}{2}[\mvec(\btheta_k^\prime-\btheta_k^*)]^T\nabla^2g_2(\textbf{Z})[\mvec(\btheta_k^\prime-\btheta_k^*)],
\]
according to Taylor’s expansion, where $\textbf{Z}=t\btheta_k^\prime+(1-t)\btheta_k^*$ with $t\in [0,1]$. In other words, if we define $\boldsymbol{\Delta}^\prime=\btheta_k^\prime-\btheta_k^*$, then $\textbf{Z}=\btheta_k^* + t\boldsymbol{\Delta}^\prime$.
According to the proof of Lemma 9 in \citet{Hao2018Simultaneous}, under Condition (C1),
\[
\begin{split}
& \lambda_{\min}(-\nabla^2g_2(\textbf{Z})) = \frac{1}{2}L_k(\ty_i;\tx_i,\bupsilon_0)\Vert \btheta_k^* + t\boldsymbol{\Delta}^\prime \Vert_2^{-2} 
\ge \frac{1}{2}L_k(\ty_i;\tx_i,\bupsilon_0)[\Vert \btheta_k^* \Vert_2 + \Vert t\boldsymbol{\Delta}^\prime \Vert_2]^{-2} \\
& \quad \ge \frac{1}{2}L_k(\ty_i;\tx_i,\bupsilon_0)(\beta_2 + \alpha_0)^{-2}.
\end{split}
\]
Therefore, it can be obtained that:
\[
H_2 \leq -\frac{1}{2n}\sum_{i=1}^n L_k(\ty_i; \tx_i, \bupsilon_0) \frac{(\beta_2 + \alpha_0)^{-2}}{2}\Vert\mvec(\btheta_k^\prime-\btheta_k^*)\Vert_2^2.
\]
Since $0\leq L_k(\ty_i; \tx_i, \bupsilon_0)\leq 1$, according to Hoeffding's inequality, 
\[
P\left(  \left\lvert \frac{1}{n}\sum_{i=1}^n \left[ L_k(\ty_i; \tx_i, \bupsilon_0)-\mathbb{E}(L_k(\ty; \tx, \bupsilon_0)) \right] \right\rvert \leq t \right) \ge 1-2\exp(-2nt^2).
\]
If we let $t=\sqrt{\frac{1}{2n}\log\frac{2K_0}{\delta}}$, then
\begin{equation}\label{eq:L}
\left \lvert \frac{1}{n}\sum_{i=1}^n[L_k(\ty_i;\tx_i,\bupsilon_0)-\mathbb{E}(L_k(\ty; \tx, \bupsilon_0))] \right \rvert \leq \sqrt{\frac{1}{2n}\log\frac{2K_0}{\delta}},
\end{equation}
with probability at least $1-\delta/K_0$. Since $\sqrt{\frac{1}{2n}\log\frac{2K_0}{\delta}}=o(1)$ 
and $0 \leq \mathbb{E}(L_k(\ty; \tx, \bupsilon_0))\leq 1$, there exists some constant $c$ such that $-\frac{1}{n}\sum_{i=1}^n L_k(\ty_i; \tx_i, \bupsilon_0) \leq -c$ when $n$ is large enough. Then, we have $H_2 \leq -\frac{1}{4}c(\beta_2+\alpha_0)^{-2}\Vert\mvec(\btheta_k^\prime-\btheta_k^*)\Vert_2^2$ with probability at least $1-\delta/K_0$.

Combining with the upper bound of $H_1$, we have:
\[
H_1 + H_2 \leq -c\cdot\frac{\min(\beta_1 C_0, 0.5(\beta_2 + \alpha_0)^{-2})}{2} \Vert \boldsymbol{\Upsilon}_{0k}^\prime - \boldsymbol{\Upsilon}_{0k}^* \Vert_2^2,
\]
with probability at least $1-\delta/K_0$. It further implies that (\ref{eq:h1}) holds with probability at least $1-\delta$, if we define $\varrho=c\cdot \min(\beta_1C_0, 0.5(\beta_2+\alpha_0)^{-2})$.

Therefore, it can be obtained that:
\begin{equation}\label{eq:q}
\begin{split}
    q(\textbf{v}) 
    & = \tilde{H}_n(\boldsymbol{\Upsilon}^*_0 + \textbf{v}|\boldsymbol{\Upsilon}_0^{(t-1)}) - \tilde{H}_n(\boldsymbol{\Upsilon}^*_0|\boldsymbol{\Upsilon}_0^{(t-1)})  \\
    & = H_n(\boldsymbol{\Upsilon}^*_0 + \textbf{v}|\boldsymbol{\Upsilon}_0^{(t-1)}) - H_n(\boldsymbol{\Upsilon}^*_0|\boldsymbol{\Upsilon}_0^{(t-1)})
    -\lambda_1[\mathcal{P}_1(\bupsilon_0^*+\textbf{v})-\mathcal{P}_1(\bupsilon_0^*)] \\
    & \leq \left \langle \nabla H_n(\boldsymbol{\Upsilon}_0^*|\boldsymbol{\Upsilon}_0^{(t-1)}), \textbf{v} \right \rangle - \lambda_1\left[ \mathcal{P}_1(\boldsymbol{\Upsilon}_0^* + \textbf{v}) - \mathcal{P}_1(\boldsymbol{\Upsilon}_0^*)\right] \\
    & = \left \langle \nabla H_n(\boldsymbol{\Upsilon}_0^*|\boldsymbol{\Upsilon}_0^{(t-1)}) - \nabla H(\boldsymbol{\Upsilon}_0^*|\boldsymbol{\Upsilon}_0^{(t-1)}) + \nabla H(\boldsymbol{\Upsilon}_0^*|\boldsymbol{\Upsilon}_0^{(t-1)}) - \nabla H(\boldsymbol{\Upsilon}_0^*|\boldsymbol{\Upsilon}_0^*), \textbf{v} \right \rangle \\
    & \quad - \lambda_1\left[ \mathcal{P}_1(\boldsymbol{\Upsilon}_0^* + \textbf{v}) - \mathcal{P}_1(\boldsymbol{\Upsilon}_0^*)\right]
    - \frac{\varrho}{2} \Vert \textbf{v} \Vert_2^2 \\
    & = q_1 + q_2 + q_3,
\end{split}   
\end{equation}
where
\[
q_1  = \left \langle \nabla_{\boldsymbol{\Upsilon}^*_{\mathcal{S}^C}} H_n(\boldsymbol{\Upsilon}_0^*|\boldsymbol{\Upsilon}_0^{(t-1)}) - \nabla_{\boldsymbol{\Upsilon}^*_{\mathcal{S}^C}} H(\boldsymbol{\Upsilon}_0^*|\boldsymbol{\Upsilon}_0^{(t-1)}), \textbf{v}_{\mathcal{S}^C}\right \rangle 
- \lambda_1\left[ \mathcal{P}_1(\boldsymbol{\Upsilon}^*_{\mathcal{S}^C} + \textbf{v}_{\mathcal{S}^C}) - \mathcal{P}_1(\boldsymbol{\Upsilon}^*_{\mathcal{S}^C})\right],
\]
\[
q_2 = \left \langle \nabla_{\boldsymbol{\Upsilon}^*_{\mathcal{M}}} H_n(\boldsymbol{\Upsilon}_0^*|\boldsymbol{\Upsilon}_0^{(t-1)}) - \nabla_{\boldsymbol{\Upsilon}^*_{\mathcal{M}}} H(\boldsymbol{\Upsilon}_0^*|\boldsymbol{\Upsilon}_0^{(t-1)}), \textbf{v}_{\mathcal{M}}\right \rangle
- \lambda_1\left[ \mathcal{P}_1(\boldsymbol{\Upsilon}^*_{\mathcal{M}} + \textbf{v}_{\mathcal{M}}) - \mathcal{P}_1(\boldsymbol{\Upsilon}^*_{\mathcal{M}})\right],
\]
\[
q_3 = 
\left \langle \nabla H(\boldsymbol{\Upsilon}^*_0|\boldsymbol{\Upsilon}_0^{(t-1)}) - \nabla H(\boldsymbol{\Upsilon}^*_0|\boldsymbol{\Upsilon}_0^*), \textbf{v} \right \rangle - \frac{\varrho}{2}\Vert \textbf{v} \Vert_2^2.
\]

With respect to $q_1$, if $|v_{kj}| \leq a\lambda_1$, then $\lambda_1 |\mathcal{T}_k^*|\rho(|v_{kj}|, \lambda_1) \ge \lambda_1 C_{\lambda_1}|\mathcal{T}_{\min}||v_{kj}|$ for a constant $C_{\lambda_1}>0$. And if $|v_{kj}| > a\lambda_1$, then $\lambda_1 |\mathcal{T}_k^*|\rho(|v_{kj}|, \lambda_1) \ge \frac{1}{2}a\lambda_1^2|\mathcal{T}_{\min}|$. Thus, 
\[
\begin{split}
q_1
& \leq \sum_{k=1}^{K_0}\sum_{j \in \mathcal{S}_k^C} \left( \Vert \nabla H_n(\boldsymbol{\Upsilon}_0^*|\boldsymbol{\Upsilon}_0^{(t-1)}) - \nabla H(\boldsymbol{\Upsilon}_0^*|\boldsymbol{\Upsilon}_0^{(t-1)})\Vert_\infty - \lambda_1C_{\lambda_1}|\mathcal{T}_{\min}| \right)|v_{kj}|\cdot I(|v_{kj}| \leq a\lambda_1) \\
& + \sum_{k=1}^{K_0}\sum_{j \in \mathcal{S}_k^C} \left( \Vert \nabla H_n(\boldsymbol{\Upsilon}_0^*|\boldsymbol{\Upsilon}_0^{(t-1)}) - \nabla H(\boldsymbol{\Upsilon}_0^*|\boldsymbol{\Upsilon}_0^{(t-1)})\Vert_\infty \Vert \textbf{v} \Vert_\infty - \frac{1}{2}a \lambda_1^2 |\mathcal{T}_{\min}| \right) \cdot I(|v_{kj}| > a\lambda_1). 
\end{split}
\]
It is easy to show that under Conditions (C2) and (C3), $\Vert \nabla H_n(\boldsymbol{\Upsilon}_0^*|\boldsymbol{\Upsilon}_0^{(t-1)}) - \nabla H(\boldsymbol{\Upsilon}_0^*|\boldsymbol{\Upsilon}_0^{(t-1)}) \Vert_\infty = O_p(\sqrt{(\log p + \log q)/n})$.
Note that $\Vert \textbf{v}\Vert_\infty \leq \frac{4\epsilon}{\varrho}+\iota\alpha$, $\alpha=O((d+s+p)\sqrt{(\log p +\log q)/n})$, and $\lambda_1 \gg O(\sqrt{(d+s+p)(\log p +\log q)/n})$. 
Therefore, $q_1<0$.

Now consider $q_2$. Denote $u_\mathcal{M}(\boldsymbol{\Upsilon}_0^*|\boldsymbol{\Upsilon}_0^{(t-1)})=\nabla_{
\boldsymbol{\Upsilon}^*_\mathcal{M}}H_n(\boldsymbol{\Upsilon}_0^*|\boldsymbol{\Upsilon}_0^{(t-1)}) - \nabla_{
\boldsymbol{\Upsilon}^*_\mathcal{M}}H(\boldsymbol{\Upsilon}_0^*|\boldsymbol{\Upsilon}_0^{(t-1)})$.
\begin{equation}\label{eq:u_m}
    \left \langle u_\mathcal{M}(\boldsymbol{\Upsilon}_0^*|\boldsymbol{\Upsilon}_0^{(t-1)}), \textbf{v}_{\mathcal{M}} \right \rangle
    \leq \Vert u_{\mathcal{M}}(\boldsymbol{\Upsilon}_0^*|\boldsymbol{\Upsilon}_0^{(t-1)}) \Vert_\infty \Vert \textbf{v}_{\mathcal{M}} \Vert_2
   \leq \epsilon_1\sqrt{K_0(d+s+p)} \Vert \textbf{v}\Vert_2,
\end{equation}
where $\epsilon_1=C\sqrt{(\log p + \log q)/n}$. 
In addition, 
\begin{equation}\label{eq:pena1}
    -\lambda_1\left[ \mathcal{P}_1(\boldsymbol{\Upsilon}^*_{\mathcal{M}} + \textbf{v}_{\mathcal{M}}) - \mathcal{P}_1(\boldsymbol{\Upsilon}^*_{\mathcal{M}})\right] \leq \lambda_1C^\prime_{p1}|\nabla\mathcal{P}_1(\boldsymbol{\Upsilon}(\mathcal{M})^*)^T\textbf{v}_{\mathcal{M}}|=0,
\end{equation}
for a positive constant $C^\prime_{p1}$ according to the minimal signal condition of the true parameters and Condition (C7). 
Combining Equation (\ref{eq:u_m}) and (\ref{eq:pena1}), we have:
\begin{equation}\label{eq:g1_m}
     q_2 \leq \epsilon_1\sqrt{K_0(d+s+p)} \Vert \textbf{v}\Vert_2.
\end{equation}

With respect to $q_3$, according to Lemma 7 in \citet{Ren2022Gaussian},  under Condition (C6), we have: 
\begin{equation}\label{eq:g2}
q_3 \leq -\frac{\varrho}{2} \Vert \textbf{v} \Vert_2^2 + \tau \cdot \Vert \boldsymbol{\Upsilon}_0^{(t-1)} - \boldsymbol{\Upsilon}_0^* \Vert_2 \Vert \textbf{v} \Vert_2, 
\end{equation}
with probability at least $1-(26K_0^2+8K_0+1)\delta$. Combining the upper bounds of $q_1$, $q_2$ and $q_3$, an upper bound of $q(\textbf{v})$ can be obtained as:
\begin{equation}
    q(\textbf{v}) < -\frac{\varrho}{2}\Vert \textbf{v} \Vert_2^2 + \left( \epsilon_1\sqrt{K_0(d+s+p)} + \tau \cdot \Vert \boldsymbol{\Upsilon}_0^{(t-1)} - \boldsymbol{\Upsilon}_0^* \Vert_2 \right) \Vert \textbf{v} \Vert_2,
\end{equation}
with probability at least $1-(26K_0^2+8K_0+1)/(p+q)$ and $\delta=1/(p+q)$. 

Note that $\epsilon=CK_0^2\sqrt{\frac{(d+s+p)(\log p + \log q)}{n}}$. Thus for a properly chosen positive constant $C$, an upper bound of $q(\textbf{v})$ can be obtained as:
\[
q(\textbf{v})< -\frac{\varrho}{2}\Vert \textbf{v} \Vert_2^2 + \left(\epsilon + \tau \cdot \Vert \boldsymbol{\Upsilon}_0^{(t-1)} - \boldsymbol{\Upsilon}_0^* \Vert_2 \right) \Vert \textbf{v} \Vert_2,
\]
with probability at least $1-(26K_0^2+8K_0+1)/(p+q)$.
It can be obtained that, when $\Vert \textbf{v}\Vert_2 > \frac{2\epsilon}{\varrho} + \frac{2\tau}{\varrho}\Vert \boldsymbol{\Upsilon}_0^{(t-1)} -\boldsymbol{\Upsilon}_0^* \Vert_2$,  $q(\textbf{v}) <0$. Note that $\Vert \textbf{v} \Vert_2=\chi=\frac{4\epsilon}{\varrho}+\iota\Vert \boldsymbol{\Upsilon}_0^{(t-1)} - \boldsymbol{\Upsilon}_0^* \Vert_2$, $1/6 \leq \iota < 1$, and $\tau \leq \varrho/12$.
Therefore, there is a local maximizer $\boldsymbol{\Upsilon}_0^{(t)}$ that follows $\boldsymbol{\Upsilon}_0^{(t-1)}$ and satisfies: if $\boldsymbol{\Upsilon}_0^{(t-1)} \in \mathcal{B}_\alpha(\boldsymbol{\Upsilon}_0^*)$, then $\Vert \boldsymbol{\Upsilon}_0^{(t)}-\boldsymbol{\Upsilon}_0^* \Vert_2 \leq \chi$ with probability at least $1-(26K_0^2+8K_0+1)/(p+q)$.

\textit{STEP 2}: We can show that, if $\boldsymbol{\Upsilon}_0^{(0)} \in \mathcal{B}_\alpha(\boldsymbol{\Upsilon}_0^*)$, for any $t\ge 1$,
\begin{equation}
\begin{split}
    \Vert \boldsymbol{\Upsilon}_0^{(t)}-\boldsymbol{\Upsilon}_0^*\Vert_2 
    & \leq
    \frac{4\epsilon}{\varrho}(\iota^0+\iota^2+\cdots+\iota^{t-1}) + \iota^t\Vert \boldsymbol{\Upsilon}_0^{(0)}-\boldsymbol{\Upsilon}_0^* \Vert_2 \\
    & = \frac{1-\iota^t}{1-\iota}\frac{4\epsilon}{\varrho} + \iota^t\Vert \boldsymbol{\Upsilon}_0^{(0)}-\boldsymbol{\Upsilon}_0^* \Vert_2 \\
    & \leq \frac{8\epsilon}{\varrho} + \iota^t  \Vert \boldsymbol{\Upsilon}_0^{(0)}-\boldsymbol{\Upsilon}_0^*\Vert_2,
\end{split}
\end{equation}
with probability at least $1-t(26K_0^2+8K_0+1)/(p+q)$, where $1/6 \leq \iota < 1$. When $t$ is sufficiently large, i.e., $t \ge T=\log_{1/\iota}\left(\frac{\varrho\Vert \boldsymbol{\Upsilon}_0^{(0)}-\boldsymbol{\Upsilon}_0^*\Vert_2}{8\epsilon} \right)$, $\iota^t  \Vert \boldsymbol{\Upsilon}_0^{(0)}-\boldsymbol{\Upsilon}_0^*\Vert_2$ is dominated by $8\epsilon/\varrho$. So the error of $\hat{\boldsymbol{\Upsilon}}_0$ can be bounded as: 
\begin{equation}\label{eq:mse}
    \Vert \hat{\boldsymbol{\Upsilon}}_0 - \boldsymbol{\Upsilon}_0^*\Vert_2=O\left(\sqrt{\frac{(d+s+p)(\log p + \log q)}{n}})\right),
\end{equation}
with probability at least $1-T(26K_0^2+8K_0+1)/(p+q)$,
which goes to 1 as $p, q$ and $n$ diverge.

\textit{STEP 3}: We show that $\hat{\bupsilon}^o=(\hat{\bupsilon}_0, \hat{\bgamma}_{\mathcal{D}^C})=(\hat{\bupsilon}_0, \textbf{0})$ is a local maximizer of objective function (\ref{eq:orac_pl}), which can also be rewritten as:
\begin{equation}\label{eq:obj_or_re}
\tilde{Q}_n(\bupsilon)= \frac{1}{n}\sum_{i=1}^n\log\left( \sum_{k=1}^{K_0} \pi_k f_k(\ty_i\vert\tx_i, \bupsilon_k) \right) - \mathcal{P}_1(\bupsilon) -\mathcal{P}_2(\bupsilon) = Q_n(\bupsilon) - \mathcal{P}_2(\bupsilon),
\end{equation}
where $\mathcal{P}_1(\boldsymbol{\Upsilon})=\sum_{k=1}^{K_0}\sum_{j\neq m}|\mathcal{T}_k^*|\rho(|\theta_{jm,k}|,\lambda_1)$,  $\mathcal{P}_2(\boldsymbol{\Upsilon})=\sum_{k=1}^{K_0}\sum_{j=1}^p\sum_{m=1}^{q+1} |\mathcal{T}_k^*| \rho(|\gamma_{jm,k}|,\lambda_2)$, and $\rho(t, \lambda)=\lambda^{-1}p(t, \lambda)$.
This indicates that $\Vert \hat{\bupsilon}^o - \bupsilon^* \Vert_2 = O\left(\sqrt{\frac{(d+s+p)(\log p+\log q)}{n}}\right)$ in Result 1 holds due to the conclusion drawn in STEP 2. 
Following Theorem 1 in \citet{Fan2011Nonconcave}, it suffices to show that $(\lambda_2)^{-1}\Vert \textbf{z}\Vert_\infty \leq \rho^\prime(0+)$, where $\textbf{z}=\nabla_{\bgamma_{\mathcal{D}^C}}Q_n(\hat{\bupsilon}^o)$. 
For each $k=1,\dots, K_0$, note that:
\begin{eqnarray}\label{eq:tyexp}
&&  \begin{pmatrix}
     \nabla_{\bupsilon_0}Q_n(\hat{\bupsilon}_k^o) \\
     \nabla_{\bgamma_{\mathcal{D}_k^C}}Q_n(\hat{\bupsilon}_k^o)
  \end{pmatrix}   =
  \begin{pmatrix}
       \nabla_{\bupsilon_0}Q_n(\bupsilon_k^*) \\
     \nabla_{\bgamma_{\mathcal{D}_k^C}}Q_n(\bupsilon_k^*)
  \end{pmatrix} + 
  \begin{pmatrix}
      \nabla^2_{\bupsilon_0 \bupsilon_0}Q_n(\bupsilon_k^*) &  \nabla^2_{\bupsilon_0 \bgamma_{\mathcal{D}_k^C}}Q_n(\bupsilon_k^*)\\
     \nabla^2_{\bgamma_{\mathcal{D}_k^C} \bupsilon_0}Q_n(\bupsilon_k^*) &
      \nabla^2_{\bgamma_{\mathcal{D}_k^C} \bgamma_{\mathcal{D}_k^C}}Q_n(\bupsilon_k^*)
  \end{pmatrix}
  \nonumber \\
&&
 ~~~~~~~~~~~~~~~~~~~~~~~~~~~~~~
  \begin{pmatrix}
     \hat{\bupsilon}_{0k}- \bupsilon_{0k}^* \\
     \textbf{0}
  \end{pmatrix} 
   + 
  \begin{pmatrix}
      R(\Delta)_0 \\
      R(\Delta)_{\mathcal{D}_k^C} 
  \end{pmatrix} = 
  \begin{pmatrix}
      \textbf{0} \\
      \textbf{z}_k
  \end{pmatrix},
\end{eqnarray}
where $R(\Delta)=\left( R(\Delta)_0,  R(\Delta)_{\mathcal{D}_k^C}\right)$ is the residual of the first order Taylor’s expansion of the gradient. From Lemma 3 in \citet{Wytock2013Sparse}, $\Vert R(\Delta)\Vert_\infty$ is bounded by $\Vert \Delta \Vert_\infty = \Vert \hat{\bupsilon}_k^o -\bupsilon_k^* \Vert_\infty^2$. According to (\ref{eq:tyexp}), 
\[
\begin{split}
\textbf{z}_k 
& = \nabla_{\bgamma_{\mathcal{D}_k^C}}Q_n(\bupsilon_k^*) + \nabla^2_{\bgamma_{\mathcal{D}_k^C} \bupsilon_0}Q_n(\bupsilon_k^*)(\hat{\bupsilon}_{0k} - \bupsilon_{0k}^*) + R(\Delta)_{\mathcal{D}_k^C} \\  
& = \frac{1}{n}\sum_{i=1}^n L_k(\ty_i;\tx_i,\bupsilon^*) \btheta_k^* (\ty_i-\bgamma_k^*\tx_i) \tx_{i\mathcal{D}_k^C}^T
+ \\
& \quad \frac{1}{n}\sum_{i=1}^n L_k(\ty_i;\tx_i,\bupsilon^*) \btheta_k^* \otimes \tx_{i\mathcal{D}_k^C}\tx_{i\mathcal{D}_k}^T (\hat{\bupsilon}_{0k} - \bupsilon_{0k}^*) + R(\Delta)_{\mathcal{D}_k^C} \\
& =\boldsymbol{\xi}_1 +\boldsymbol{\xi}_2 + R(\Delta)_{\mathcal{D}_k^C},
\end{split}
\]
\[
\begin{split}
& \Vert \boldsymbol{\xi}_1 \Vert_\infty 
 \leq \frac{1}{n}\sum_{i=1}^n\mathbb{E}(L_k(\ty_i;\tx_i, \bupsilon^*)) \left\lVert \frac{1}{n}\sum_{i=1}^n \btheta_k^* (\ty_i-\bgamma_k^*\tx_i)\tx_{i\mathcal{D}_k^C}^T \right\rVert_\infty \\
& ~~~~~ + \frac{1}{n}\sum_{i=1}^n[L_k(\ty_i;\tx_i,\bupsilon^*)-\mathbb{E}(L_k(\ty_i;\tx_i, \bupsilon^*))] \left\lVert \btheta_k^* (\ty_i-\bgamma_k^*\tx_i)\tx_{i\mathcal{D}_k^C}^T \right\rVert_\infty  \\
& \leq \frac{1}{n}\sum_{i=1}^n \Vert \btheta^*_k \Vert_\infty \left\lVert \boldsymbol{\epsilon}_i\tx_{i\mathcal{D}_k^C}  \right\rVert_\infty
+ \frac{1}{n}\sum_{i=1}^n[L_k(\ty_i;\tx_i,\bupsilon^*)-\mathbb{E}(L_k(\ty_i;\tx_i, \bupsilon^*))] \left\lVert \btheta_k^* \tx_{i\mathcal{D}_k^C}(\ty_i-\bgamma_k^*\tx_i) \right\rVert_\infty. \\
\end{split}
\]
In the first term, 
$\boldsymbol{\epsilon}_i=(\epsilon_{i1}, \dots, \epsilon_{ip})$ and $\epsilon_{ij}=(\ty_i-\bgamma_k^*\tx_i)_j, ~j=1,\dots, p$. According to the Hoeffding's bound,
\[
\begin{split}
P\left(\sum_{i=1}^n \epsilon_{ij}x_{im} \ge \sqrt{(\log p + \log q)n} \right)
& = 1-P\left(\sum_{i=1}^n \epsilon_{ij}x_{im} \leq \sqrt{(\log p +\log q)n} \right) \\
& \ge 1-2\exp\left( -\frac{n(\log p +\log q)}{2\max_{1\leq j \leq q}\Vert\textbf{X}_j \Vert_2^2}\right).
\end{split}
\]
This probability approaches 1, as $\Vert \textbf{X}_j\Vert_2=O(\sqrt{n})$ and $p$, $q$ diverge. 
In the second term, from (\ref{eq:L}), 
\[
P\left( \left\lvert\frac{1}{n}\sum_{i=1}^n[ L_k(\ty_i;\tx_i,\bupsilon_k^*)-\mathbb{E}(L_k(\ty_i;\tx_i, \bupsilon_k^*))] \right\rvert \leq \sqrt{\frac{\log p + \log q}{n}} \right),
\]
with probability approaching 1 when setting $\delta=1/pq$. 
With respect to $\xi_2$, according to (\ref{eq:tyexp}),
\[
\hat{\bupsilon}_{0k} - \bupsilon_{0k}^* = -\left[ \nabla^2_{\bupsilon_0 \bupsilon_0}Q_n(\bupsilon_k^*)\right]^{-1} \left[ \nabla_{\bupsilon_0}Q_n(\bupsilon_k^*) + R(\Delta)_0 \right].
\]
\[
\begin{split}
\boldsymbol{\xi}_2
& = -\frac{1}{n}\sum_{i=1}^n L_k(\ty_i;\tx_i,\bupsilon^*) \btheta_k^* \otimes \tx_{i\mathcal{D}_k^C} \tx_{i\mathcal{D}_k}^T \left[ \nabla^2_{\bupsilon_0 \bupsilon_0}Q_n(\bupsilon_k^*)\right]^{-1} \left[ \nabla_{\bupsilon_0}Q_n(\bupsilon_k^*) + R(\Delta)_0 \right] \\
& = \left[ \frac{1}{n}\sum_{i=1}^n L_k(\ty_i;\tx_i,\bupsilon^*) \btheta_k^* \otimes \tx_{i\mathcal{D}_k^C} \tx_{i\mathcal{D}_k}^T \right] \left[ \frac{1}{n}\sum_{i=1}^n L_k(\ty_i;\tx_i,\bupsilon^*) \btheta_k^* \otimes \tx_{i\mathcal{D}_k} \tx_{i\mathcal{D}_k}^T \right]^{-1} \\
& ~~~~~~~~
\left( \frac{1}{n}\sum_{i=1}^n L_k(\ty_i;\tx_i,\bupsilon^*) \btheta_k^* (\ty_i-\bgamma_k^*\tx_i) \tx_{i\mathcal{D}_k}^T + R(\Delta)_0\right)\\
& =\textbf{I}_p \otimes \left\{\left[ \frac{1}{n}\sum_{i=1}^n L_k(\ty_i;\tx_i,\bupsilon^*)\tx_{i\mathcal{D}_k^C} \tx_{i\mathcal{D}_k}^T \right] \left[ \frac{1}{n}\sum_{i=1}^n L_k(\ty_i;\tx_i,\bupsilon^*)\tx_{i\mathcal{D}_k} \tx_{i\mathcal{D}_k}^T \right]^{-1}\right\}\\
& ~~~~~~~~
\left( \frac{1}{n}\sum_{i=1}^n L_k(\ty_i;\tx_i,\bupsilon^*) \btheta_k^* (\ty_i-\bgamma_k^*\tx_i) \tx_{i\mathcal{D}_k}^T + R(\Delta)_0\right). \\
\end{split}
\]
Similar to the proof in (\ref{eq:exp_samp}) and (\ref{eq:matrixhoe}), with the matrix Hoeffding inequality, we have:
\begin{equation}
\begin{split}
    \Vert \boldsymbol{\xi}_2 \Vert_\infty \leq 
    & \left\lVert (\textbf{X}_{\mathcal{D}_k^C}^T \textbf{G}_k \textbf{X}_{\mathcal{D}_k})(\textbf{X}_{\mathcal{D}_k}^T \textbf{G}_k\textbf{X}_{\mathcal{D}_k})^{-1} \right\rVert_{2,\infty}  \cdot \\
    & \quad \left\lVert \frac{1}{n}\sum_{i=1}^n L_{ik}(\ty_i;\tx_i,\bupsilon^*) \btheta_k^* (\ty_i-\bgamma_k^*\tx_i) \tx_{i\mathcal{D}_k}^T + R(\Delta)_0 \right\rVert_\infty,
\end{split}
\end{equation}
with probability tending to 1. By Condition (C3), we can bound $\Vert \boldsymbol{\xi}_2 \Vert_\infty$ in the same way. Therefore, with probability tending to 1,
\[
\Vert \textbf{z} \Vert_\infty \leq \sqrt{\frac{\log p + \log q}{n}} + n^{\alpha_1}\left( \sqrt{\frac{\log p + \log q}{n}} + \Vert \hat{\bupsilon}_0 - \bupsilon_0^* \Vert_\infty^2 \right)+  \Vert \hat{\bupsilon}_0 - \bupsilon_0^* \Vert_\infty^2 \ll \lambda_2.
\]

\textit{STEP 4}: Here we establish selection consistency for the precision matrices. That is, 
$\hat{\mathcal{S}}_k^o \supseteq \mathcal{S}_k$ and $\mathcal{S}_k \supseteq \hat{\mathcal{S}}_k^o$. \\
First, it is sufficient to show that, for any $(i,j) \in \mathcal{S}_k$ and $k=1,\dots, K_0$, $\hat{\theta}_{ij,k} \neq 0$. 
Note that:
\[
|\hat{\theta}_{ij,k}| \ge |\theta_{ij,k}^*| - |\hat{\theta}_{ij,k}-\theta_{ij,k}^*| \ge |\theta_{ij,k}^*| - \sqrt{\sum_{1\leq i,j\leq p}(\hat{\theta}_{ij,k}-\theta_{ij,k})^2}
\ge |\theta_{ij,k}^*| - \Vert \hat{\boldsymbol{\Upsilon}}^o-\boldsymbol{\Upsilon^*}\Vert_2.
\]
According to the results in STEP 3 and the minimal signal Condition (C4), $|\hat{\theta}_{ij,k}|>0$, which implies that $\hat{\mathcal{S}}_k^o \supseteq \mathcal{S}_k$.

Second, we show that for any $(i,j) \in \mathcal{S}_k^c$ and $k=1,\dots, K_0$, $\hat{\theta}_{ij,k}=0$. 
Consider a local maximizer $\hat{\boldsymbol{\Upsilon}}$ that satisfies (\ref{eq:mse}) and optimizes objective function (\ref{eq:obj_or_re}). The derivative of the objective function with respect to $\theta_{jm,k}$ for $(j,m) \in \mathcal{S}_k^C, k=1,\dots, K_0$ is:
\[
\frac{\partial \tilde{Q}_n(\hat{\boldsymbol{\Upsilon}})}{\partial \theta_{jm,k}} = \mathcal{R}(\hat{\boldsymbol{\Upsilon}}_k) - \lambda_1 \rho^\prime(|\hat{\theta}_{jm,k}|)\mbox{sgn}(\hat{\theta}_{jm,k}),
\]
where $\mathcal{R}(\hat{\boldsymbol{\Upsilon}}_k)=\frac{1}{2n}\sum_{i=1}^n L_k(\ty_i; \tx_i, \hat{\boldsymbol{\Upsilon}}_k)[\hat{\sigma}_{jm,k}-(y_{ij}-(\hat{\bgamma}_k\tx_i)_j)(y_{im}-(\hat{\bgamma}_k\tx_i)_m)]$, and $L_k(\ty_i; \tx_i, \hat{\boldsymbol{\Upsilon}}_k)=\frac{\hat{\pi}_kf_k(\ty_i
|\tx_i, \hat{\boldsymbol{\Upsilon}}_k)}{\sum_{k=1}^{K_0}\hat{\pi}_kf_k(\ty_i| \tx_i, \hat{\boldsymbol{\Upsilon}}_k)}$, $\hat{\sigma}_{jm,k}$ is the $(j,m)$-th element of $\hat{\btheta}_k^{-1}$, and $\mbox{sgn}(\hat{\theta}_{jm,k})$ denotes the sign of $\hat{\theta}_{jm,k}$. 
It can be decomposed that $\mathcal{R}(\hat{\bupsilon}_k) \leq |\hat{\sigma}_{jm,k} -\sigma_{jm,k}^*| + \mathcal{R}^*(\hat{\bupsilon}_k)$,
where $\sigma_{jm,k}^*$ denotes the true value of $\sigma_{jm,k}$. 
Note that $|\hat{\sigma}_{jm,k} -\sigma_{jm,k}^*| \leq \Vert \btheta_k^* -\hat{\btheta}_k \Vert_2 \leq \Vert \hat{\bupsilon} - \bupsilon^*\Vert_2$. 
\[
\mathcal{R}^*(\hat{\bupsilon}_k) = \frac{1}{2n}\sum_{i=1}^n L_k(\ty_i; \tx_i, \hat{\boldsymbol{\Upsilon}}_k)[\sigma^*_{jm,k}-(y_{ij}-(\hat{\bgamma}_k\tx_i)_j)(y_{im}-(\hat{\bgamma}_k\tx_i)_m)],
\]
which can be bounded as $|\mathcal{R}^*(\hat{\bupsilon}_k)| \leq \frac{1}{2}\vert\mathcal{R}_1^* \vert + \frac{1}{2}\vert \mathcal{R}_2^*\vert$, where
\[
\mathcal{R}_1^*=\frac{1}{n}\sum_{i=1}^nL_k(\ty_i; \tx_i, \boldsymbol{\Upsilon}^*_k)[\sigma_{jm,k}^*-(y_{ij}-\bgamma_k^*\tx_i)_j)(y_{im}-(\bgamma_k^*\tx_i)_m)],
\]
\[
\begin{split}
\mathcal{R}_2^*
& =\frac{1}{n}\sum_{i=1}^n[L_k(\ty_i; \tx_i, \hat{\boldsymbol{\Upsilon}}_k)-L_k(\ty_i; \tx_i, \boldsymbol{\Upsilon}^*_k)][(y_{ij}-(\hat{\bgamma}_k\tx_i)_j)(y_{im}-(\hat{\bgamma}_k\tx_i)_m)] \\
&~~~~
+\frac{1}{n}\sum_{i=1}^nL_k(\ty_i; \tx_i, \boldsymbol{\Upsilon}^*_k)[(y_{ij}-(\bgamma^*_k\tx_i)_j)((\bgamma_k^*\tx_i)_j-(\hat{\bgamma}_k\tx_i)_j)]\\
& ~~~~~~~~ + \frac{1}{n}\sum_{i=1}^nL_k(\ty_i; \tx_i, \boldsymbol{\Upsilon}^*_k)[(y_{im}-(\bgamma^*_k\tx_i)_m)((\bgamma_k^*\tx_i)_m-(\hat{\bgamma}_k\tx_i)_m)] \\
& ~~~~~~~~~~~~ + \frac{1}{n}\sum_{i=1}^nL_k(\ty_i; \tx_i, \boldsymbol{\Upsilon}^*_k)[((\bgamma_k^*\tx_i)_j-(\hat{\bgamma}_k\tx_i)_j)((\bgamma_k^*\tx_i)_m-(\hat{\bgamma}_k\tx_i)_m)].
\end{split}
\]
Note that $L_k(\ty_i;\tx_i,\cdot)$ is continuous. Then,
\[
\mathcal{R}_2^*=O\left\{ \sup_{i,k}[L_k(\ty_i; \tx_i, \hat{\boldsymbol{\Upsilon}}_k)-L_k(\ty_i; \tx_i, \boldsymbol{\Upsilon}^*_k)] + \sup_k \Vert \hat{\bgamma}_k - \bgamma_k^* \Vert_\infty \right\} \lesssim \Vert \hat{\bupsilon} - \bupsilon^*\Vert_2,
\]
where $a \lesssim b$ if $a \leq Db$ for some positive constant $D$.
As for $\mathcal{R}_1^*$, we have 
$$ \frac{1}{2n}\sum_{i=1}^n\mathbb{E}[L_k(\ty_i; \tx_i, \boldsymbol{\Upsilon}^*_k)](\btheta_k^{*-1})-\frac{1}{2n}\sum_{i=1}^n\mathbb{E}[L_k(\ty_i; \tx_i, \boldsymbol{\Upsilon}^*_k)(\ty-\bgamma_k^*\tx)(\ty-\bgamma_k^*\tx)^T]=0.$$
Thus, $\mathcal{R}_1^*$ can be rewritten as:
\[
\begin{split}
\mathcal{R}_1^* 
& = \frac{1}{n}\sum_{i=1}^n \left( L_k(\ty_i; \tx_i, \boldsymbol{\Upsilon}^*_k)\sigma_{jm,k}^* - \mathbb{E}[L_k(\ty_i; \tx_i, \boldsymbol{\Upsilon}^*_k)]\sigma_{jm,k}^* \right)\\
&- \frac{1}{n}\sum_{i=1}^n \left(L_k(\ty_i; \tx_i, \boldsymbol{\Upsilon}^*_k)[(y_{ij}-\bgamma_k^*\tx_i)_j)(y_{im}-(\bgamma_k^*\tx_i)_m)] \right.\\
& \quad \left. - \mathbb{E}[L_k(\ty_i; \tx_i, \boldsymbol{\Upsilon}^*_k)(\ty-\bgamma_k^*\tx)_j(\ty-\bgamma_k^*\tx)_m^T] \right),
\end{split}
\]
note that we have $|\mathcal{R}_1^*|=O(\sqrt{\log p/n})$. 
In summary, it can be obtained that: 
\[
|\mathcal{R}(\hat{\boldsymbol{\Upsilon}}_k)| \lesssim \Vert\hat{\boldsymbol{\Upsilon}}_0 - \boldsymbol{\Upsilon}^*\Vert_2 \lesssim \sqrt{\frac{(d+s+p)(\log p + \log q)}{n}}.
\]
By Conditions (C5) and (C7), we have:
\[
\lambda_1\rho^\prime(|\hat{\theta}_{jm,k}|) \gg C_{\lambda_1} \sqrt{\frac{(d+s+p)(\log p + \log q)}{n}},
\]
for $\hat{\theta}_{jm,k}$ in a small neighborhood of 0 and some positive constant $C_{\lambda_1}$. Therefore, if $\hat{\theta}_{jm,k}$ lies in a small neighborhood of 0, the sign of $\frac{\partial Q_n(\hat{\boldsymbol{\Upsilon}})}{\partial \theta_{jm,k}}$ only depends on $\mbox{sgn}(\hat{\theta}_{jm,k})$, with probability tending to 1. Then,  variable selection consistency of the precision matrix estimators can be proved.

\textbf{Result 2}: Assume that $\lambda_3 \gg \sqrt{\frac{(s+p)(\log p + \log q)}{n}}$,  $b=\min_{1\leq k \neq k^\prime \leq K_0}\Vert \boldsymbol{\Upsilon}_k^* - \boldsymbol{\Upsilon}_{k^\prime}^* \Vert_2 >a \lambda_3$, and the conditions in Result 1 hold. Then there exists $\hat{\boldsymbol{\Omega}}$, a local maximizer of $\mathcal{L}(\boldsymbol{\Omega}, \boldsymbol{\pi}|\textbf{Y})$ defined in equation (\ref{eq:obj}) that satisfies: 
\begin{equation}
    P(\hat{\boldsymbol{\Omega}}=\hat{\boldsymbol{\Omega}}^o) \rightarrow 1.
\end{equation}

\textit{Proof of Result 2}: 
Denote $\hat{\boldsymbol{\Omega}}^o$ as the maximizer of (\ref{eq:orac_pl}). 
According to Result 1, we have:
\[
\Vert \hat{\boldsymbol{\Omega}}^o - \boldsymbol{\Omega}^* \Vert_2 =O_p\left( K\Vert \hat{\boldsymbol{\Upsilon}}^o - \boldsymbol{\Upsilon}^* \Vert_2 \right) = O_p(\epsilon_n),
\]
where $\epsilon_n = \sqrt{(d+s+p)(\log p + \log q)/n}$. 
Denote the locations of the non-zero coefficients in $\boldsymbol{\Omega}_l^*$ as $\mathcal{W}_l, l=1, \dots, K$. Consider two neighborhood sets of $\boldsymbol{\Omega}^*$:
\[
\mathcal{C}=\left\{ \boldsymbol{\Omega} \in \mathbb{R}^{K(p(q+1)+p^2)}: \sup_l \Vert\boldsymbol{\Omega}_l -\boldsymbol{\Omega}_l^* \Vert_2 \leq \epsilon_n \right\},
\]
\[
\mathcal{C}_0=\left\{ \boldsymbol{\Omega} \in \mathbb{R}^{K(p(q+1)+p^2)}: \sup_l \Vert\boldsymbol{\Omega}_l -\boldsymbol{\Omega}_l^* \Vert_2 \leq \epsilon_n, \Omega_{lj}=0, j\notin \mathcal{W}_l, l=1, \dots, K \right\}.
\]

By Result 1, there exists an event $E_1$ in which
$\sup_l\Vert \hat{\boldsymbol{\Omega}}_l^o - \boldsymbol{\Omega}_l^* \Vert_2 \leq \epsilon_n, \hat{\Omega}_{lj}^o=0, j \notin \mathcal{W}_l, l=1,\dots, K$, 
and $P(E_1^C) \rightarrow 0$. Thus $\hat{\boldsymbol{\Omega}}^o \in \mathcal{C}_0$. 
Let $G: \Lambda_{\mathcal{T}^*} \rightarrow \mathbb{R}^{K_0(p(q+1)+p^2)}$ be the mapping such that $G(\boldsymbol{\Omega})$ is the $K_0(p(q+1)+p^2)$ vector consisting of the $K_0$ groups with each group having dimension $p(q+1)+p^2$, and its $l$-th vector component equal to the common value of $\boldsymbol{\Omega}_l$ for $l \in \mathcal{T}_k^*$. 
Let $\breve{G}: \mathbb{R}^{K(p(q+1)+p^2)} \rightarrow \mathbb{R}^{K_0(p(q+1)+p^2)}$ be the mapping such that $\breve{G}(\boldsymbol{\Omega})=\{|\mathcal{T}_k^*|^{-1}\sum_{l \in \mathcal{T}_k^*} \boldsymbol{\Omega}_l^T, l=1, \dots, K\}^T$. For any $\boldsymbol{\Omega}$, denote $\breve{\boldsymbol{\Omega}}=G^{-1}\left( \breve{G}(\boldsymbol{\Omega})\right) \in \Lambda_{\mathcal{T}^*}$.
For any $\boldsymbol{\Omega}^{(0)}$, define $\boldsymbol{\Omega}$ with $\Omega_{lj}=\Omega_{lj}^{(0)}$ for $j \in \mathcal{W}_l$ and $\Omega_{lj}=0$ for $j \notin \mathcal{W}_l$, $l=1,\dots, K$. Clearly, if $\boldsymbol{\Omega}^{(0)} \in \mathcal{C}$, then $\boldsymbol{\Omega} \in \mathcal{C}_0$. 

With objective function (\ref{eq:orac_pl}), remove the constraint of oracle group membership $\boldsymbol{\Omega} \in \Lambda_{\mathcal{T}^*}$:
\[
\mathcal{Q}(\boldsymbol{\Omega})= \frac{1}{n}\sum_{i=1}^n \log \left( \sum_{l=1}^K \pi_l f_l(\textbf{y}_i; \tx_i, \boldsymbol{\Gamma}_l, \boldsymbol{\Theta}_l) \right)
- \sum_{l=1}^K \sum_{j \neq m}p(|\theta_{jm,l}|, \lambda_1) 
- \sum_{l=1}^K \sum_{j=1}^p \sum_{m=1}^{q+1} p(|\gamma_{jm,l}|, \lambda_2),
\]
\[
\mathcal{P}_3(\boldsymbol{\Omega})=\sum_{1 \leq l < l^\prime \leq K}p(\Vert\boldsymbol{\Omega}_l -\boldsymbol{\Omega}_{l^\prime} \Vert_2, \lambda_3).
\]
Then, $\mathcal{L}(\boldsymbol{\Omega})=\mathcal{Q}(\boldsymbol{\Omega})-\mathcal{P}_3(\boldsymbol{\Omega})$. 

If we can show the following two results, then we have that $\hat{\boldsymbol{\Omega}}^o$ is a strict local maximizer of objective function (\ref{eq:obj}) with probability converging to 1. With results (i) and (ii), for any $\boldsymbol{\Omega}^{(0)} \in \mathcal{C} \cap \mathcal{C}_n$ and $\boldsymbol{\Omega}^{(0)} \neq \hat{\boldsymbol{\Omega}}^o$ on $\mathcal{C} \cap \mathcal{C}_n$, where $\mathcal{C}_n$ is the neighborhood of $\hat{\boldsymbol{\Omega}}^o$, we have $\mathcal{L}(\boldsymbol{\Omega}^{(0)}) \leq  \mathcal{L}(\hat{\boldsymbol{\Omega}}^o)$. So $\hat{\boldsymbol{\Omega}}^o$ is a strict local maximizer of objective function (\ref{eq:obj}) on event $E_1$ with $P(E_1) \rightarrow 1$ and a sufficiently large $n$.
\begin{enumerate}
    \item[(i)] 
    On event $E_1$, for and $\breve{\boldsymbol{\Omega}} \in \mathcal{C}_0$ and $\breve{\boldsymbol{\Omega}} \neq \hat{\boldsymbol{\Omega}}^o$, $\mathcal{L}(\breve{\boldsymbol{\Omega}}) < \mathcal{L}(\hat{\boldsymbol{\Omega}}^o)$.
    \item[(ii)]
    On event $E_1$, there is a neighborhood of $\hat{\boldsymbol{\Omega}}^o$, denoted as $\mathcal{C}_n$, such that $\mathcal{L}(\boldsymbol{\Omega}^{(0)}) 
    \leq \mathcal{L}(\boldsymbol{\Omega}) \leq \mathcal{L}(\breve{\boldsymbol{\Omega}})$, for any $\boldsymbol{\Omega}^{(0)} \in \mathcal{C} \cap \mathcal{C}_n$ and a sufficiently large $n$.
\end{enumerate}

By $\lambda_3 \gg \sqrt{\frac{(d+s+p)(\log p + \log q)}{n}}$ and $b > a\lambda_3$,
the penalty function $\sum_{1\leq l < l^\prime \leq K}p(\Vert\boldsymbol{\Omega}_l -\boldsymbol{\Omega}_{l^\prime} \Vert_2, \lambda_3)$ is a constant that does not depend on $\boldsymbol{\Omega}$.
We further impose the constraint $\boldsymbol{\Omega} \in \Lambda_{\mathcal{T}^*}$ on $\mathcal{Q}(\boldsymbol{\Omega})$ and $\mathcal{P}_3(\boldsymbol{\Omega})$, denoted as $\mathcal{Q}^T(\boldsymbol{\Omega})$ and $\mathcal{P}^T_3(\boldsymbol{\Omega})$, respectively. Define $\mathcal{L}^T(\boldsymbol{\Omega})=\mathcal{Q}^T(\boldsymbol{\Omega}) - \mathcal{P}^T_3(\boldsymbol{\Omega})$. When $\breve{\boldsymbol{\Omega}} \in \Lambda_{\mathcal{T}^*}$, we have $\mathcal{L}(\breve{\boldsymbol{\Omega}})=\mathcal{L}^T(\breve{\boldsymbol{\Omega}})$.
Since $\mathcal{P}_3(\boldsymbol{\Omega})$ is a constant and $\hat{\boldsymbol{\Omega}}^o$ is the unique maximizer of $\mathcal{Q}^T(\boldsymbol{\Omega})$, for any $\breve{\boldsymbol{\Omega}} \in \mathcal{C}_0$, $\mathcal{Q}^T(\breve{\boldsymbol{\Omega}}) < \mathcal{Q}^T(\hat{\boldsymbol{\Omega}}^o)$. Therefore,    $\mathcal{L}(\breve{\boldsymbol{\Omega}}) < \mathcal{L}(\hat{\boldsymbol{\Omega}}^o)$, and the result in (i) is proved.

Next, we prove result (ii). First, we show that $\mathcal{L}(\boldsymbol{\Omega}^{(0)}) \leq \mathcal{L}(\boldsymbol{\Omega})$.
Given a positive sequence $\phi_n$, consider: 
\[
\mathcal{C}_n=\left\{ \boldsymbol{\Omega} \in \mathbb{R}^{K(p(q+1)+p^2)}: \sup_l \Vert \boldsymbol{\Omega}_l -\hat{\boldsymbol{\Omega}}_l^o \vert_\infty \leq \phi_n \right\}.
\]
For any $\boldsymbol{\Omega}^{(0)} \in \mathcal{C} \cap \mathcal{C}_n$ and the corresponding $\boldsymbol{\Omega} \in \mathcal{C}_0 \cap \mathcal{C}_n$, $\Vert \boldsymbol{\Omega}^{(0)}_l - \boldsymbol{\Omega}^{(0)}_{l^\prime} \Vert_2 \ge \Vert \boldsymbol{\Omega}_l - \boldsymbol{\Omega}_{l^\prime} \Vert_2$. Therefore, $-\mathcal{P}_3(\boldsymbol{\Omega}^{(0)}) \leq - \mathcal{P}_3(\boldsymbol{\Omega})$. 
Moreover, for the non-zero part $\mathcal{W}$, $\boldsymbol{\Omega}^{(0)}=\boldsymbol{\Omega}$, and the difference between $\boldsymbol{\Omega}^{(0)}$ and $\boldsymbol{\Omega}$ lies in the zero entries of $\boldsymbol{\Omega}$ in $\mathcal{W}^C$. And the corresponding values of $\boldsymbol{\Omega}^{(0)}$ are small enough and can be controlled by $\phi_n$. According to the proof of selection consistency, when $\hat{\theta}_{jm,l}$ or $\hat{\gamma}_{jm,l}$ lies in a small neighborhood of 0, $\partial \mathcal{Q}(\hat{\boldsymbol{\Upsilon}})/\partial \theta_{jm,l} < 0$, which indicates that $\mathcal{Q}(\hat{\boldsymbol{\Upsilon}})$ is a decreasing function in terms of $\hat{\theta}_{jm,l}$ and $\hat{\gamma}_{jm,l}$ in the neighborhood of 0. Thus $\mathcal{Q}(\boldsymbol{\Omega}^{(0)}) \leq \mathcal{Q}(\boldsymbol{\Omega})$. Therefore,  $\mathcal{L}(\boldsymbol{\Omega}^{(0)}) \leq \mathcal{L}(\boldsymbol{\Omega})$.

Second, we show that $\mathcal{L}(\boldsymbol{\Omega}) \leq \mathcal{L}(\breve{\boldsymbol{\Omega}})$. For $\boldsymbol{\Omega} \in \mathcal{C}_0 \cap \mathcal{C}_n$ and $\breve{\boldsymbol{\Omega}} \in \mathcal{C}_0$, by Taylor's expansion, 
\[
\mathcal{L}(\boldsymbol{\Omega}) - \mathcal{L}(\breve{\boldsymbol{\Omega}}) = l_1-l_2,
\]
where
\[
l_1 = \sum_{l=1}^K \textbf{D}_l^T(\boldsymbol{\Omega}_l - \breve{\boldsymbol{\Omega}}_l), \textbf{D}_l^T=(\textbf{D}_{1l}^T, \textbf{D}_{2l}^T)^T\mathcal{I}_{\mathcal{W}_l},
\]
\[
\textbf{D}_{1l} = \frac{1}{n}\sum_{i=1}^n L_l(\textbf{y}_i; \tx_i, \tilde{\boldsymbol{\Omega}}_l)\tilde{\boldsymbol{\Theta}}_l(\textbf{y}_i - \tilde{\boldsymbol{\Gamma}}_l \textbf{x}_i)\textbf{x}_i - \lambda_2 \textbf{s}_2(\tilde{\boldsymbol{\Gamma}}_l),
\]
\[
\textbf{D}_{2l} = \frac{1}{n}\sum_{i=1}^n L_l(\textbf{y}_i; \tx_i, \tilde{\boldsymbol{\Omega}}_l)\frac{1}{2}\left[ \mbox{vec}(\tilde{\boldsymbol{\Theta}}_l)^{-1} - \mbox{vec}(\textbf{y}_i - \tilde{\boldsymbol{\Gamma}}_l\textbf{x}_i)(\textbf{y}_i - \tilde{\boldsymbol{\Gamma}}_l\textbf{x}_i)\right] - \lambda_1 \textbf{s}_1(\tilde{\boldsymbol{\Theta}}_l),
\]
\[
l_2 = \sum_{l=1}^K \frac{\partial \mathcal{P}_3(\tilde{\boldsymbol{\Omega}})}{\partial \boldsymbol{\Omega}_l^T}(\boldsymbol{\Omega}_l - \breve{\boldsymbol{\Omega}}_l),
\]
and $\mathcal{I}_{\mathcal{W}_l}$ is a diagonal matrix with the $j$-th diagonal element $I(j \in \mathcal{W}_l), j=1, \dots, p(q+1)+p^2$, and $L_l(\textbf{y}_i; \tx_i, \tilde{\boldsymbol{\Omega}}_l)=\frac{\pi_l f_l(\textbf{y}_i; \tx_i, \tilde{\boldsymbol{\Omega}})}{\sum_{l=1}^K \pi_l f_l(\textbf{y}_i; \tx_i,  \tilde{\boldsymbol{\Omega}})}$. $\textbf{s}_1(\tilde{\boldsymbol{\Theta}}_l) \in \mathbb{R}^{p^2}$ such that the corresponding element is $\rho^\prime(|\tilde{\theta}_{jm,l}|)\mbox{sgn}(\tilde{\theta}_{jm,l})I(j \neq m)$,  $\textbf{s}_2(\tilde{\boldsymbol{\Gamma}}_l) \in \mathbb{R}^{p(q+1)}$ such that the corresponding element is $\rho^\prime(|\tilde{\gamma}_{jm,l}|)\mbox{sgn}(\tilde{\gamma}_{jm,l})I(j \neq m)$, and $\tilde{\boldsymbol{\Omega}}=\varsigma\boldsymbol{\Omega} + (1-\varsigma)\breve{\boldsymbol{\Omega}}$ for some constant $\varsigma \in (0,1)$. 

For $l_2$, we can obtain:
\begin{equation}
    l_2 \ge \sum_{k=1}^{K_0}\sum_{\{l, l^\prime \in \mathcal{T}_k^*, l<l^\prime\}} \lambda_3 \rho^\prime(4\phi_n) \Vert \boldsymbol{\Omega}_l - \boldsymbol{\Omega}_{l^\prime} \Vert_2.
\end{equation}
For $l_1$, we can obtain :
\begin{equation}
    l_1 \leq \sum_{k=1}^{K_0}\sum_{\{l, l^\prime \in \mathcal{T}_k^*, l<l^\prime\}} |\mathcal{T}_{\min}|^{-1} \sup_l \Vert \textbf{D}_l - \textbf{D}_{l^\prime} \Vert_2 \Vert \boldsymbol{\Omega}_l -\boldsymbol{\Omega}_{l^\prime}\Vert_2.
\end{equation}
It can be shown that:
\begin{equation}
    \sup_l \Vert \textbf{D}_l - \textbf{D}_{l^\prime} \Vert_2 \leq \sqrt{d \cdot \sup_l \Vert \textbf{D}_{1l} - \textbf{D}_{1l^\prime} \Vert_\infty^2 + s \cdot \sup_l \Vert \textbf{D}_{2l} - \textbf{D}_{2l^\prime} \Vert_\infty^2 }.
\end{equation}
According to the minimal signal condition and Condition (C7), 
we have $\textbf{s}_1(\tilde{\boldsymbol{\Theta}}_l)\mathcal{I}_{\mathcal{W}_l} = \textbf{0}$ and $\textbf{s}_2(\tilde{\boldsymbol{\Gamma}}_l)\mathcal{I}_{\mathcal{W}_l} = \textbf{0}$. Note that:
\[
\begin{split}
    \Vert \textbf{D}_{1l} - \textbf{D}_{1l^\prime} \Vert_\infty 
    & \leq \sup_{i} \Vert L_l(\textbf{y}_i; \tx_i, \tilde{\boldsymbol{\Omega}}_l)m_1(\textbf{y}_i; \textbf{x}_i, \tilde{\boldsymbol{\Omega}}_l) - L_l(\textbf{y}_i; \tx_i, \tilde{\boldsymbol{\Omega}}_{l^\prime})m_1(\textbf{y}_i; \textbf{x}_i, \tilde{\boldsymbol{\Omega}}_{l^\prime}) \Vert_\infty \\
    & = \sup_i \Vert L_l(\textbf{y}_i; \tx_i, \tilde{\boldsymbol{\Omega}}_l)\left[ m_1(\textbf{y}_i; \textbf{x}_i, \tilde{\boldsymbol{\Omega}}_l)-m_1(\textbf{y}_i; \textbf{x}_i, \tilde{\boldsymbol{\Omega}}_{l^\prime}) \right] \\
    &\quad + \left[L_l(\textbf{y}_i; \tx_i, \tilde{\boldsymbol{\Omega}}_l)- L_l(\textbf{y}_i; \tx_i, \tilde{\boldsymbol{\Omega}}_{l^\prime}) \right]m_1(\textbf{y}_i; \textbf{x}_i, \tilde{\boldsymbol{\Omega}}_l) \Vert_\infty \\
    & \leq \sup_i \{ L_l(\textbf{y}_i; \tx_i, \tilde{\boldsymbol{\Omega}}_l) \} \sup_i\Vert m_1(\textbf{y}_i; \textbf{x}_i, \tilde{\boldsymbol{\Omega}}_l)-m_1(\textbf{y}_i; \textbf{x}_i, \tilde{\boldsymbol{\Omega}}_{l^\prime}) \Vert_\infty \\
    & \quad + \sup_i \{ L_l(\textbf{y}_i; \tx_i, \tilde{\boldsymbol{\Omega}}_l)- L_l(\textbf{y}_i; \tx_i, \tilde{\boldsymbol{\Omega}}_{l^\prime})\} \sup_i \Vert m_1(\textbf{y}_i; \textbf{x}_i, \tilde{\boldsymbol{\Omega}}_{l^\prime})\Vert_\infty,
\end{split}
\]
where $m_1(\textbf{y}_i; \textbf{x}_i, \tilde{\boldsymbol{\Omega}}_l)=\tilde{\boldsymbol{\Theta}}_l(\textbf{y}_i - \tilde{\boldsymbol{\Gamma}}_l \textbf{x}_i)\textbf{x}_i$. 
In addition, $L_l(\textbf{y}_i; \tx_i, \cdot)$ and $m_1(\textbf{y}_i; \textbf{x}_i, \cdot)$ are continuous. Then, there exists a constant $C^\prime >0$,
\[
\sup_i\Vert m_1(\textbf{y}_i; \textbf{x}_i, \tilde{\boldsymbol{\Omega}}_l)-m_1(\textbf{y}_i; \textbf{x}_i, \tilde{\boldsymbol{\Omega}}_{l^\prime}) \Vert_\infty \leq C^\prime\Vert \tilde{\boldsymbol{\Omega}}_l -\tilde{\boldsymbol{\Omega}}_{l^\prime}  \Vert_\infty, 
\]
\[
\sup_i \{ L_l(\textbf{y}_i; \tx_i, \tilde{\boldsymbol{\Omega}}_l)- L_l(\textbf{y}_i; \tx_i, \tilde{\boldsymbol{\Omega}}_{l^\prime})\} \leq C^\prime \Vert \tilde{\boldsymbol{\Omega}}_l -\tilde{\boldsymbol{\Omega}}_{l^\prime}  \Vert_\infty.
\]

Note that, for $l, l^\prime \in \mathcal{T}_k^*$, $\boldsymbol{\Omega}_l, \boldsymbol{\Omega}_{l^\prime} \in \mathcal{C}_n$. Then,
\[
\Vert \tilde{\boldsymbol{\Omega}}_l -\tilde{\boldsymbol{\Omega}}_{l^\prime}  \Vert_\infty = \varsigma \Vert \boldsymbol{\Omega}_l -\boldsymbol{\Omega}_{l^\prime}  \Vert_\infty \leq 2\phi_n.
\]
Thus, we have:
\[
\sup_l \Vert \textbf{D}_{1l} - \textbf{D}_{1l^\prime} \Vert_\infty = O(\phi_n).
\]
Similarly, 
\[
\begin{split}
 \Vert \textbf{D}_{2l} - \textbf{D}_{2l^\prime} \Vert_\infty 
 & \leq \sup_i \{ L_l(\textbf{y}_i; \tx_i, \tilde{\boldsymbol{\Omega}}_l) \} \sup_i\Vert m_2(\textbf{y}_i; \textbf{x}_i, \tilde{\boldsymbol{\Omega}}_l)-m_2(\textbf{y}_i; \textbf{x}_i, \tilde{\boldsymbol{\Omega}}_{l^\prime}) \Vert_\infty \\
 & \quad + \sup_i \{ L_l(\textbf{y}_i; \tx_i, \tilde{\boldsymbol{\Omega}}_l)- L_l(\textbf{y}_i; \tx_i, \tilde{\boldsymbol{\Omega}}_{l^\prime})\} \sup_i \Vert m_2(\textbf{y}_i; \textbf{x}_i, \tilde{\boldsymbol{\Omega}}_{l^\prime})\Vert_\infty, 
\end{split}
\]
where $m_2(\textbf{y}_i; \textbf{x}_i, \tilde{\boldsymbol{\Omega}}_l)=\frac{1}{2}\left[ \mbox{vec}(\tilde{\boldsymbol{\Theta}}_l)^{-1} - \mbox{vec}(\textbf{y}_i - \tilde{\boldsymbol{\Gamma}}_l\textbf{x}_i)(\textbf{y}_i - \tilde{\boldsymbol{\Gamma}}_l\textbf{x}_i)\right]$.  In addition, $m_2(\textbf{y}_i; \textbf{x}_i, \cdot)$ is continuous. Thus, we have: 
\[
\sup_l \Vert \textbf{D}_{2l} - \textbf{D}_{2l^\prime} \Vert_\infty = O(\phi_n).
\]
Further, we can obtain that:
\begin{equation}
    \mathcal{L}(\boldsymbol{\Omega}) - \mathcal{L}(\breve{\boldsymbol{\Omega}}) \leq \sum_{k=1}^{K_0}\sum_{\{l, l^\prime \in \mathcal{T}_k^*, l<l^\prime\}}  \left[ \phi_n |\mathcal{T}_{\min}|^{-1}- \lambda_3 \rho^\prime(4\phi_n) \right] \Vert \boldsymbol{\Omega}_l - \boldsymbol{\Omega}_{l^\prime}\Vert_2.
\end{equation}
Let $\phi_n = \epsilon_n$. Then $\rho^\prime(4\phi_n) \rightarrow 1$  according to Condition (C7). Since $\lambda_3 \gg \epsilon_n$ and $|\mathcal{T}_{\min}| \ge 1$,
we have $|\mathcal{T}_{\min}| \lambda_3 \gg \phi_n$, and then $\mathcal{L}(\boldsymbol{\Omega}) - \mathcal{L}(\breve{\boldsymbol{\Omega}}) \leq 0$. This completes the proof of result (ii) and thus  Result 2.

\end{proof}

\subsection{Additional Simulation Results}

\begin{table}[H]
\caption{Simulation results under S1 with $p=q=100$. In each cell, mean (sd).
} \label{tab:s1_p100}
\centering
\renewcommand\arraystretch{0.4}  
\resizebox{\linewidth}{!}{
\begin{tabular}{c c c c c c c c} 
\toprule
$n$ & Method  & &RMSE &TPR &FRP &RI &$\hat{K}_0$ \\
\midrule
\multirow{18}{*}{(200,200,200)} &\multirow{2}{*}{Proposed} & $\btheta$ &5.867(5.208) &0.952(0.054)	&0.184(0.084) &\multirow{2}{*}{0.644(0.291)} &\multirow{2}{*}{2.70(1.22)} \\
& &$\bgamma$ &7.297(2.298)	&0.792(0.091)	&0.005(0.009) & & \\
&\multirow{2}{*}{HeteroGGM} & $\btheta$ &6.194(0.490)	&0.980(0.009)	&0.888(0.014) &\multirow{2}{*}{0.128(0.203)} &\multirow{2}{*}{4.75(1.77)} \\
& &$\bgamma$ &- &- &- & & \\
&\multirow{2}{*}{CGLasso($K=6$)} & $\btheta$ &5.344(0.156)	&0.828(0.012)	&0.279(0.004) &\multirow{2}{*}{0.022(0.012)} &\multirow{2}{*}{6(0)} \\
& &$\bgamma$ &11.701(0.335)	&0.770(0.061)	&0.088(0.004) & & \\
&\multirow{2}{*}{CGLasso($K=4$)} & $\btheta$ &5.683(0.245)	&0.899(0.016)	&0.249(0.017) &\multirow{2}{*}{0.090(0.119)} &\multirow{2}{*}{4(0)} \\
& &$\bgamma$ &10.469(1.419)	&0.904(0.065)	&0.069(0.008) & & \\
&\multirow{2}{*}{CGLasso($K=3$)} & $\btheta$ &5.484(1.165)	&0.937(0.012)	&0.176(0.063) &\multirow{2}{*}{0.350(0.350)} &\multirow{2}{*}{3(0)} \\
& &$\bgamma$ &8.075(3.449)	&0.969(0.034)	&0.044(0.018) & & \\
&\multirow{2}{*}{MRCE($K=6$)} & $\btheta$
&5.938(0.176)	&0.846(0.013)	&0.360(0.015) &\multirow{2}{*}{0.022(0.012)} &\multirow{2}{*}{6(0)} \\
& &$\bgamma$ &12.537(0.099)	&0.284(0.048)	&0.026(0.003) & & \\
&\multirow{2}{*}{MRCE($K=4$)} & $\btheta$ &5.872(0.344)	&0.910(0.021)	&0.330(0.035) &\multirow{2}{*}{0.090(0.119)} &\multirow{2}{*}{4(0)} \\
& &$\bgamma$ &11.452(1.532)	&0.431(0.192)	&0.034(0.021) & & \\
&\multirow{2}{*}{MRCE($K=3$)} & $\btheta$ &5.415(1.436)	&0.955(0.026)	&0.252(0.059) &\multirow{2}{*}{0.350(0.350)} &\multirow{2}{*}{3(0)} \\
& &$\bgamma$ &8.535(4.008)	&0.782(0.242)	&0.061(0.031) & & \\
&\multirow{2}{*}{MCGGM($K=3$)} & $\btheta$ &6.671(1.435)	&0.750(0.109)	&0.086(0.085) &\multirow{2}{*}{0.666(0.173)} &\multirow{2}{*}{3(0)} \\
& &$\bgamma$ &- &- &- & & \\
\hline

\multirow{18}{*}{(150,200,250)} &\multirow{2}{*}{Proposed} & $\btheta$ &4.488(0.621)	&0.963(0.026)	&0.162(0.039) &\multirow{2}{*}{0.705(0.159)} &\multirow{2}{*}{2.20(0.41)} \\
& &$\bgamma$ &6.816(1.364)	&0.781(0.111)	&0.002(0.001) & & \\
&\multirow{2}{*}{HeteroGGM} & $\btheta$ &6.063(0.370)	&0.983(0.009)	&0.895(0.016) &\multirow{2}{*}{0.049(0.122)} &\multirow{2}{*}{5.30(1.49)} \\
& &$\bgamma$ &- &- &- & & \\
&\multirow{2}{*}{CGLasso($K=6$)} & $\btheta$ &5.727(0.156)	&0.859(0.012)	&0.278(0.005) &\multirow{2}{*}{0.030(0.014)} &\multirow{2}{*}{6(0)} \\
& &$\bgamma$ &11.542(0.407)	&0.806(0.056)	&0.087(0.003) & & \\
&\multirow{2}{*}{CGLasso($K=4$)} & $\btheta$ &6.111(0.348)	&0.921(0.016)	&0.253(0.018) &\multirow{2}{*}{0.049(0.100)} &\multirow{2}{*}{4(0)} \\
& &$\bgamma$ &10.980(1.108)	&0.878(0.058)	&0.071(0.007) & & \\
&\multirow{2}{*}{CGLasso($K=3$)} & $\btheta$ &5.745(0.881)	&0.956(0.014)	&0.195(0.047) &\multirow{2}{*}{0.239(0.265)} &\multirow{2}{*}{3(0)} \\
& &$\bgamma$ &8.961(2.687)	&0.951(0.055)	&0.047(0.013) & & \\
&\multirow{2}{*}{MRCE($K=6$)} & $\btheta$
&5.980(0.106)	&0.874(0.013)	&0.359(0.015) &\multirow{2}{*}{0.030(0.014)} &\multirow{2}{*}{6(0)} \\
& &$\bgamma$ &12.694(0.153)	&0.235(0.070)	&0.025(0.002) & & \\
&\multirow{2}{*}{MRCE($K=4$)} & $\btheta$ &6.065(0.328)	&0.928(0.016)	&0.333(0.050) &\multirow{2}{*}{0.049(0.100)} &\multirow{2}{*}{4(0)} \\
& &$\bgamma$ &11.885(1.281)	&0.357(0.208)	&0.033(0.020) & & \\
&\multirow{2}{*}{MRCE($K=3$)} & $\btheta$ &5.680(1.164)	&0.961(0.018)	&0.239(0.029) &\multirow{2}{*}{0.239(0.265)} &\multirow{2}{*}{3(0)} \\
& &$\bgamma$ &9.285(3.124)	&0.785(0.216)	&0.068(0.028) & & \\
&\multirow{2}{*}{MCGGM($K=3$)} & $\btheta$ &6.544(0.569)	&0.611(0.203)	&0.035(0.033) &\multirow{2}{*}{0.700(0.142)} &\multirow{2}{*}{3(0)}\\
& &$\bgamma$ &- &- &- & & \\
 \hline

 \multirow{18}{*}{(500,500,500)} &\multirow{2}{*}{Proposed} & $\btheta$ &1.165(0.037)	&0.998(0.002)	&0.036(0.001) &\multirow{2}{*}{1.000(0.000)} &\multirow{2}{*}{3.00(0.00)} \\
& &$\bgamma$ &0.456(0.022)	&1.000(0.000)	&0.001(0.000) & & \\
&\multirow{2}{*}{HeteroGGM} & $\btheta$ &5.793(0.150)	&0.993(0.004)	&0.851(0.010) &\multirow{2}{*}{0.730(0.112)} &\multirow{2}{*}{5.35(0.99)} \\
& &$\bgamma$ &- &- &- & & \\
&\multirow{2}{*}{CGLasso($K=6$)} & $\btheta$ &3.954(0.338)	&0.957(0.024)	&0.124(0.035) &\multirow{2}{*}{0.613(0.036)} &\multirow{2}{*}{6(0)} \\
& &$\bgamma$ &4.322(0.688)	&0.999(0.002)	&0.042(0.018) & & \\
&\multirow{2}{*}{CGLasso($K=4$)} & $\btheta$ &3.134(0.300)	&0.960(0.016)	&0.050(0.021) &\multirow{2}{*}{0.829(0.015)} &\multirow{2}{*}{4(0)} \\
& &$\bgamma$ &2.560(0.483)	&0.999(0.002)	&0.022(0.007) & & \\
&\multirow{2}{*}{CGLasso($K=3$)} & $\btheta$ &2.646(0.180)	&0.978(0.020)	&0.020(0.007) &\multirow{2}{*}{0.964(0.007)} &\multirow{2}{*}{3(0)} \\
& &$\bgamma$ &1.854(0.036)	&1.000(0.000)	&0.011(0.001) & & \\
&\multirow{2}{*}{MRCE($K=6$)} & $\btheta$
&3.627(0.311)	&0.987(0.007)	&0.172(0.041) &\multirow{2}{*}{0.613(0.036)} &\multirow{2}{*}{6(0)} \\
& &$\bgamma$ &4.289(0.828)	&0.979(0.047)	&0.080(0.011) & & \\
&\multirow{2}{*}{MRCE($K=4$)} & $\btheta$ &2.680(0.221)	&0.995(0.007)	&0.090(0.035) &\multirow{2}{*}{0.829(0.015)} &\multirow{2}{*}{4(0)} \\
& &$\bgamma$ &2.209(0.463)	&0.998(0.008)	&0.057(0.007) & & \\
&\multirow{2}{*}{MRCE($K=3$)} & $\btheta$ &2.216(0.104)	&0.999(0.002)	&0.050(0.013) &\multirow{2}{*}{0.964(0.007)} &\multirow{2}{*}{3(0)} \\
& &$\bgamma$ &1.449(0.039)	&1.000(0.000)	&0.050(0.006) & & \\
&\multirow{2}{*}{MCGGM($K=3$)} & $\btheta$ &6.234(0.680)	&0.750(0.249)	&0.036(0.038) &\multirow{2}{*}{0.729(0.185)} &\multirow{2}{*}{3(0)} \\
& &$\bgamma$ &- &- &- & & \\
\bottomrule	

\end{tabular}}
\end{table}

\begin{table}[H]
\caption{Simulation results under S2 with $p=q=50$. In each cell, mean (sd).} \label{tab:s2}
\centering
\renewcommand\arraystretch{0.4}  
\resizebox{\linewidth}{!}{
\begin{tabular}{c c c c c c c c} 
\toprule
$n$ & Method  & &RMSE &TPR &FRP &RI &$\hat{K}_0$ \\
\midrule
\multirow{18}{*}{(200,200,200)} &\multirow{2}{*}{Proposed} & $\btheta$ &2.792(0.697)	&0.874(0.046)	&0.115(0.044) &\multirow{2}{*}{0.785(0.220)} &\multirow{2}{*}{2.50(0.51)} \\
& &$\bgamma$ &4.066(1.536)	&0.878(0.066) &0.005(0.004) & & \\
&\multirow{2}{*}{HeteroGGM} & $\btheta$ &4.809(0.093)	&0.975(0.008)	&0.921(0.011) &\multirow{2}{*}{0.107(0.172)} &\multirow{2}{*}{5.35(1.31)} \\
& &$\bgamma$ &- &- &- & & \\
&\multirow{2}{*}{CGLasso($K=6$)} & $\btheta$ &4.334(0.044)	&0.701(0.016)	&0.319(0.008) &\multirow{2}{*}{0.023(0.014)} &\multirow{2}{*}{6(0)} \\
& &$\bgamma$ &8.715(0.290)	&0.853(0.037) &0.141(0.003) & & \\
&\multirow{2}{*}{CGLasso($K=4$)} & $\btheta$ &4.583(0.037)	&0.741(0.018)	&0.290(0.007) &\multirow{2}{*}{0.032(0.016)} &\multirow{2}{*}{4(0)} \\
& &$\bgamma$ &8.211(0.229)	&0.927(0.020) &0.116(0.003) & & \\
&\multirow{2}{*}{CGLasso($K=3$)} & $\btheta$ &4.711(0.038)	&0.762(0.019)	&0.263(0.009) &\multirow{2}{*}{0.029(0.023)} &\multirow{2}{*}{3(0)} \\
& &$\bgamma$ &8.166(0.428)	&0.941(0.032) &0.097(0.005) & & \\
&\multirow{2}{*}{MRCE($K=6$)} & $\btheta$
&4.667(0.064)	&0.736(0.020)	&0.396(0.025) &\multirow{2}{*}{0.023(0.014)} &\multirow{2}{*}{6(0)} \\
& &$\bgamma$ &9.489(0.196)	&0.454(0.069) &0.058(0.010) & & \\
&\multirow{2}{*}{MRCE($K=4$)} & $\btheta$ &4.741(0.103)	&0.777(0.025)	&0.375(0.053) &\multirow{2}{*}{0.032(0.016)} &\multirow{2}{*}{4(0)} \\
& &$\bgamma$ &8.921(0.346)	&0.580(0.120) &0.071(0.024) & & \\
&\multirow{2}{*}{MRCE($K=3$)} & $\btheta$ &4.823(0.118)	&0.784(0.024)	&0.322(0.052) &\multirow{2}{*}{0.029(0.023)} &\multirow{2}{*}{3(0)} \\
& &$\bgamma$ &8.804(0.619)	&0.617(0.167) &0.070(0.028) & & \\
&\multirow{2}{*}{MCGGM($K=3$)} & $\btheta$ 
&4.459(0.578)	&0.703(0.193)	&0.143(0.112)
&\multirow{2}{*}{0.473(0.141)} &\multirow{2}{*}{3(0)} \\
& &$\bgamma$ &- &- &- & & \\
\hline

\multirow{18}{*}{(150,200,250)} &\multirow{2}{*}{Proposed} & $\btheta$ &2.558(0.338)	&0.879(0.028)	&0.113(0.053) &\multirow{2}{*}{0.934(0.136)}  &\multirow{2}{*}{2.80(0.41)} \\
& &$\bgamma$ &3.482(0.751)	&0.844(0.079) &0.003(0.003) & & \\
&\multirow{2}{*}{HeteroGGM} & $\btheta$ &4.820(0.071)	&0.971(0.007)	&0.919(0.011) &\multirow{2}{*}{0.144(0.243)} &\multirow{2}{*}{5.20(1.54)} \\
& &$\bgamma$ &- &- &- & & \\
&\multirow{2}{*}{CGLasso($K=6$)} & $\btheta$ &4.301(0.045)	&0.702(0.022)	&0.314(0.007) &\multirow{2}{*}{0.029(0.012)} &\multirow{2}{*}{6(0)} \\
& &$\bgamma$ &8.350(0.235)	&0.879(0.029) &0.137(0.004) & & \\
&\multirow{2}{*}{CGLasso($K=4$)} & $\btheta$ &4.543(0.069)	&0.740(0.020)	&0.286(0.010) &\multirow{2}{*}{0.040(0.029)} &\multirow{2}{*}{4(0)} \\
& &$\bgamma$ &7.824(0.411)	&0.937(0.025) &0.109(0.006) & & \\
&\multirow{2}{*}{CGLasso($K=3$)} & $\btheta$ &4.663(0.101)	&0.765(0.028) &0.257(0.013) &\multirow{2}{*}{0.047(0.063)} &\multirow{2}{*}{3(0)} \\
& &$\bgamma$ &7.974(0.796)	&0.914(0.042) &0.092(0.007) & & \\
&\multirow{2}{*}{MRCE($K=6$)} & $\btheta$ &4.723(0.591)	&0.708(0.142)	&0.381(0.081) &\multirow{2}{*}{0.029(0.012)} &\multirow{2}{*}{6(0)} \\
& &$\bgamma$ &9.107(0.228)	&0.516(0.117) &0.067(0.019) & & \\
&\multirow{2}{*}{MRCE($K=4$)} & $\btheta$ &4.654(0.132)	&0.771(0.038)	&0.363(0.057) &\multirow{2}{*}{0.040(0.029)} &\multirow{2}{*}{4(0)} \\
& &$\bgamma$ &8.436(0.515)	&0.655(0.093) &0.082(0.023) & & \\
&\multirow{2}{*}{MRCE($K=3$)} & $\btheta$ &4.693(0.176)	&0.792(0.030)	&0.333(0.058) &\multirow{2}{*}{0.047(0.063)} &\multirow{2}{*}{3(0)} \\
& &$\bgamma$ &8.352(0.898)	&0.666(0.132) &0.088(0.023) & & \\
&\multirow{2}{*}{MCGGM($K=3$)} & $\btheta$  
&4.612(0.555)	&0.604(0.199)	&0.098(0.109)	&\multirow{2}{*}{0.473(0.086)} &\multirow{2}{*}{3(0)}
\\
& &$\bgamma$ &- &- &- & & \\
\hline

\multirow{18}{*}{(500,500,500)} &\multirow{2}{*}{Proposed} & $\btheta$ &0.930(0.040)	&0.980(0.006)	&0.038(0.003) &\multirow{2}{*}{1.000(0.000)} &\multirow{2}{*}{3.00(0.00)} \\
& &$\bgamma$ &0.369(0.062)	&1.000(0.001) &0.001(0.001) & & \\
&\multirow{2}{*}{HeteroGGM} & $\btheta$ &4.660(0.021)	&0.980(0.004) &0.895(0.009) &\multirow{2}{*}{0.642(0.047)} &\multirow{2}{*}{5.90(0.45)} \\
& &$\bgamma$ &- &- &- & & \\
&\multirow{2}{*}{CGLasso($K=6$)} & $\btheta$ &3.894(0.184)	&0.800(0.018)	&0.143(0.018) &\multirow{2}{*}{0.298(0.049)} &\multirow{2}{*}{6(0)} \\
& &$\bgamma$ &4.972(0.603)	&0.993(0.008) &0.057(0.019) & & \\
&\multirow{2}{*}{CGLasso($K=4$)} & $\btheta$ &3.679(0.303)	&0.822(0.022)	&0.090(0.022) &\multirow{2}{*}{0.476(0.120)} &\multirow{2}{*}{4(0)} \\
& &$\bgamma$ &3.890(0.901)	&0.999(0.002) &0.054(0.014) & & \\
&\multirow{2}{*}{CGLasso($K=3$)} & $\btheta$ &3.954(0.682)	&0.827(0.033)	&0.089(0.051) &\multirow{2}{*}{0.400(0.303)} &\multirow{2}{*}{3(0)} \\
& &$\bgamma$ &4.610(2.453)	&0.999(0.002) &0.052(0.026) & & \\
&\multirow{2}{*}{MRCE($K=6$)} & $\btheta$ &3.586(0.257)	&0.872(0.036)	&0.283(0.037) &\multirow{2}{*}{0.298(0.049)} &\multirow{2}{*}{6(0)} \\
& &$\bgamma$ &4.854(0.698)	&0.960(0.052) &0.126(0.013) & & \\
&\multirow{2}{*}{MRCE($K=4$)} & $\btheta$ &3.387(0.401)	&0.907(0.032) &0.235(0.031) &\multirow{2}{*}{0.476(0.120)} &\multirow{2}{*}{4(0)} \\
& &$\bgamma$ &3.910(1.060)	&0.985(0.034) &0.081(0.006) & & \\
&\multirow{2}{*}{MRCE($K=3$)} & $\btheta$ &3.766(0.905)	&0.899(0.067)	&0.215(0.044) &\multirow{2}{*}{0.400(0.303)} &\multirow{2}{*}{3(0)} \\
& &$\bgamma$ &4.825(2.778)	&0.939(0.108) &0.057(0.004) & & \\
&\multirow{2}{*}{MCGGM($K=3$)} & $\btheta$
&4.493(0.544)	&0.665(0.236)	&0.064(0.052)	
&\multirow{2}{*}{0.493(0.188)} &\multirow{2}{*}{3(0)}
\\
& &$\bgamma$ &- &- &- & & \\
\bottomrule
\end{tabular}}
\end{table}

\begin{table}[H]
\caption{Simulation results under S2 with $p=q=100$. In each cell, mean (sd).}  \label{tab:s2_p100}
\centering
\renewcommand\arraystretch{0.4}  
\resizebox{\linewidth}{!}{
\begin{tabular}{c c c c c c c c} 
\toprule
$n$ & Method  & &RMSE &TPR &FRP &RI &$\hat{K}_0$ \\
\midrule
\multirow{18}{*}{(200,200,200)} &\multirow{2}{*}{Proposed} & $\btheta$ &4.313(0.707)	&0.869(0.021)	&0.092(0.028) &\multirow{2}{*}{0.678(0.191)} &\multirow{2}{*}{2.25(0.44)} \\
& &$\bgamma$ &6.810(1.586)	&0.846(0.059) &0.003(0.002) & & \\
&\multirow{2}{*}{HeteroGGM} & $\btheta$ &6.661(0.181)	&0.966(0.007)	&0.893(0.012) &\multirow{2}{*}{0.019(0.013)} &\multirow{2}{*}{5.15(1.53)} \\
& &$\bgamma$ &- &- &- & & \\
&\multirow{2}{*}{CGLasso($K=6$)} & $\btheta$ &5.813(0.045)	&0.723(0.014)	&0.277(0.005) &\multirow{2}{*}{0.025(0.012)} &\multirow{2}{*}{6(0)} \\
& &$\bgamma$ &11.951(0.317)	&0.823(0.044)	&0.089(0.002) & & \\
&\multirow{2}{*}{CGLasso($K=4$)} & $\btheta$ &6.210(0.038)	&0.771(0.011)	&0.254(0.004) &\multirow{2}{*}{0.030(0.015)} &\multirow{2}{*}{4(0)} \\
& &$\bgamma$ &11.419(0.416)	&0.901(0.032) &0.073(0.002) & & \\
&\multirow{2}{*}{CGLasso($K=3$)} & $\btheta$ &6.285(0.287)	&0.799(0.011)	&0.218(0.020) &\multirow{2}{*}{0.095(0.127)} &\multirow{2}{*}{3(0)} \\
& &$\bgamma$ &10.535(1.519)	&0.941(0.037) &0.056(0.008) & & \\
&\multirow{2}{*}{MRCE($K=6$)} & $\btheta$ &6.613(0.076)	&0.756(0.018) &0.363(0.013) &\multirow{2}{*}{0.025(0.012)} &\multirow{2}{*}{6(0)} \\
& &$\bgamma$ &13.084(0.098)	&0.333(0.046) &0.025(0.002) & & \\
&\multirow{2}{*}{MRCE($K=4$)} & $\btheta$ &6.902(0.836)	&0.747(0.157) &0.321(0.071) &\multirow{2}{*}{0.030(0.015)} &\multirow{2}{*}{4(0)} \\
& &$\bgamma$ &12.713(0.296)	&0.357(0.115) &0.023(0.009) & & \\
&\multirow{2}{*}{MRCE($K=3$)} & $\btheta$ &6.458(0.518)	&0.815(0.024) &0.289(0.056) &\multirow{2}{*}{0.095(0.127)} &\multirow{2}{*}{3(0)} \\
& &$\bgamma$ &11.535(1.958)	&0.583(0.254) &0.041(0.030) & & \\
&\multirow{2}{*}{MCGGM($K=3$)} & $\btheta$  
&6.746(0.510)	&0.539(0.199)	&0.023(0.027)	&\multirow{2}{*}{0.729(0.362)} &\multirow{2}{*}{3(0)}
\\
& &$\bgamma$ &- &- &- & & \\
\hline

\multirow{18}{*}{(150,200,250)} &\multirow{2}{*}{Proposed} & $\btheta$ &4.234(0.401)	&0.865(0.020)	&0.090(0.019) &\multirow{2}{*}{0.723(0.168)} &\multirow{2}{*}{2.25(0.44)} \\
& &$\bgamma$ &5.675(0.855)	&0.842(0.091) &0.002(0.001) & & \\
&\multirow{2}{*}{HeteroGGM} & $\btheta$ &6.539(0.094)	&0.965(0.006)	&0.898(0.006) &\multirow{2}{*}{0.028(0.013)} &\multirow{2}{*}{5.70(0.66)} \\
& &$\bgamma$ &- &- &- & & \\
&\multirow{2}{*}{CGLasso($K=6$)} & $\btheta$ &5.781(0.038)	&0.722(0.012)	&0.274(0.003) &\multirow{2}{*}{0.023(0.011)} &\multirow{2}{*}{6(0)} \\
& &$\bgamma$ &11.656(0.250)	&0.833(0.026) &0.088(0.002) & & \\
&\multirow{2}{*}{CGLasso($K=4$)} & $\btheta$ &6.154(0.122)	&0.772(0.009)	&0.250(0.008) &\multirow{2}{*}{0.036(0.044)} &\multirow{2}{*}{4(0)} \\
& &$\bgamma$ &10.815(0.722)	&0.919(0.027) &0.071(0.004) & & \\
&\multirow{2}{*}{CGLasso($K=3$)} & $\btheta$ &6.058(0.566)	&0.805(0.021)	&0.204(0.033) &\multirow{2}{*}{0.161(0.210)} &\multirow{2}{*}{3(0)} \\
& &$\bgamma$ &10.017(2.068)	&0.903(0.056) &0.051(0.011) & & \\
&\multirow{2}{*}{MRCE($K=6$)} & $\btheta$ &6.800(0.963)	&0.717(0.169) &0.337(0.081) &\multirow{2}{*}{0.023(0.011)} &\multirow{2}{*}{6(0)} \\
& &$\bgamma$ &12.824(0.125)	&0.366(0.095) &0.024(0.006) & & \\
&\multirow{2}{*}{MRCE($K=4$)} & $\btheta$ &6.640(0.508)	&0.772(0.085) &0.311(0.049) &\multirow{2}{*}{0.036(0.044)} &\multirow{2}{*}{4(0)} \\
& &$\bgamma$ &12.111(0.854)	&0.495(0.120) &0.030(0.016) & & \\
&\multirow{2}{*}{MRCE($K=3$)} & $\btheta$ &6.085(0.823)	&0.825(0.036) &0.269(0.037) &\multirow{2}{*}{0.161(0.210)} &\multirow{2}{*}{3(0)} \\
& &$\bgamma$ &10.503(2.385)	&0.676(0.204) &0.057(0.028) & & \\
&\multirow{2}{*}{MCGGM($K=3$)} & $\btheta$ &6.528(0.628)	&0.700(0.144)	&0.088(0.075) 
&\multirow{2}{*}{0.633(0.132)} &\multirow{2}{*}{3(0)}\\
& &$\bgamma$ &- &- &- & & \\
 \hline

 \multirow{18}{*}{(500,500,500)} &\multirow{2}{*}{Proposed} & $\btheta$ &1.432(0.048)	&0.958(0.011)	&0.024(0.011) &\multirow{2}{*}{1.000(0.000)} &\multirow{2}{*}{3.00(0.00)} \\
& &$\bgamma$ &0.575(0.178)	&0.999(0.002) &0.001(0.001) & & \\
&\multirow{2}{*}{HeteroGGM} & $\btheta$ &6.393(0.046)	&0.981(0.006)	&0.859(0.009) &\multirow{2}{*}{0.707(0.120)} &\multirow{2}{*}{5.25(1.02)} \\
& &$\bgamma$ &- &- &- & & \\
&\multirow{2}{*}{CGLasso($K=6$)} & $\btheta$ &4.766(0.654)	&0.836(0.012)	&0.117(0.029) &\multirow{2}{*}{0.440(0.152)} &\multirow{2}{*}{6(0)} \\
& &$\bgamma$ &6.190(1.667)	&0.992(0.012) &0.045(0.018) & & \\
&\multirow{2}{*}{CGLasso($K=4$)} & $\btheta$ &3.541(0.279)	&0.892(0.026)	&0.055(0.021) &\multirow{2}{*}{0.794(0.018)} &\multirow{2}{*}{4(0)} \\
& &$\bgamma$ &3.156(0.556)	&0.999(0.009) &0.023(0.008) & & \\
&\multirow{2}{*}{CGLasso($K=3$)} & $\btheta$ &2.799(0.130)	&0.940(0.013) &0.025(0.005) &\multirow{2}{*}{0.956(0.007)} &\multirow{2}{*}{3(0)} \\
& &$\bgamma$ &1.916(0.042)	&1.000(0.000) &0.011(0.001) & & \\
&\multirow{2}{*}{MRCE($K=6$)} & $\btheta$
&4.424(0.766)	&0.884(0.025)	&0.222(0.031) &\multirow{2}{*}{0.440(0.152)} &\multirow{2}{*}{6(0)} \\
& &$\bgamma$ &6.336(1.803)	&0.932(0.082) &0.070(0.011) & & \\
&\multirow{2}{*}{MRCE($K=4$)} & $\btheta$ &2.908(0.169)	&0.952(0.014)	&0.162(0.019) &\multirow{2}{*}{0.794(0.018)} &\multirow{2}{*}{4(0)} \\
& &$\bgamma$ &2.850(0.542)	&0.999(0.004) &0.058(0.008) & & \\
&\multirow{2}{*}{MRCE($K=3$)} & $\btheta$ &2.242(0.102)	&0.971(0.008) &0.089(0.015) &\multirow{2}{*}{0.956(0.007)} &\multirow{2}{*}{3(0)} \\
& &$\bgamma$ &1.520(0.043)	&1.000(0.000) &0.053(0.006) & & \\
&\multirow{2}{*}{MCGGM($K=3$)} & $\btheta$ &6.303(0.641)	&0.676(0.230)	&0.042(0.039)	&\multirow{2}{*}{0.787(0.125)} &\multirow{2}{*}{3(0)} \\
& &$\bgamma$ &- &- &- & & \\
\bottomrule
\end{tabular}}
\end{table}

\begin{table}[H]
\caption{Simulation results under S3 with $p=q=50$. In each cell, mean (sd).
} \label{tab:s3}
\centering
\renewcommand\arraystretch{0.4}  
\resizebox{\linewidth}{!}{
\begin{tabular}{c c c c c c c c} 
\toprule
$n$ & Method  & &RMSE &TPR &FRP &RI &$\hat{K}_0$ \\
\midrule
\multirow{18}{*}{(200,200,200)} &\multirow{2}{*}{Proposed} & $\btheta$ &1.728(1.083)	&0.938(0.034)	&0.048(0.021) &\multirow{2}{*}{0.994(0.029)} &\multirow{2}{*}{3.05(0.22)} \\
& &$\bgamma$ &1.241(1.336)	&0.983(0.027)	&0.009(0.017) & & \\
&\multirow{2}{*}{HeteroGGM} & $\btheta$ &4.131(0.171)	&0.983(0.010)	&0.900(0.012) &\multirow{2}{*}{0.473(0.244)} &\multirow{2}{*}{5.15(1.38)} \\
& &$\bgamma$ &- &- &- & & \\
&\multirow{2}{*}{CGLasso($K=6$)} & $\btheta$ &3.496(0.181)	&0.869(0.018)	&0.249(0.018) &\multirow{2}{*}{0.387(0.089)} &\multirow{2}{*}{6(0)} \\
& &$\bgamma$ &5.589(0.541)	&0.964(0.034)	&0.102(0.006) & & \\
&\multirow{2}{*}{CGLasso($K=4$)} & $\btheta$ &3.284(0.292)	&0.908(0.018)	&0.184(0.025) &\multirow{2}{*}{0.572(0.076)} &\multirow{2}{*}{4(0)} \\
& &$\bgamma$ &4.013(0.676)	&0.992(0.007)	&0.064(0.009) & & \\
&\multirow{2}{*}{CGLasso($K=3$)} & $\btheta$ &2.982(0.429)	&0.919(0.021)	&0.112(0.038) &\multirow{2}{*}{0.738(0.145)} &\multirow{2}{*}{3(0)} \\
& &$\bgamma$ &2.853(1.138)	&0.998(0.004)	&0.037(0.012) & & \\
&\multirow{2}{*}{MRCE($K=6$)} & $\btheta$
&3.929(0.640)	&0.867(0.149)	&0.353(0.063) &\multirow{2}{*}{0.387(0.089)} &\multirow{2}{*}{6(0)} \\
& &$\bgamma$ &6.463(0.697)	&0.784(0.153)	&0.116(0.030) & & \\
&\multirow{2}{*}{MRCE($K=4$)} & $\btheta$ &3.035(0.320)	&0.959(0.013)	&0.293(0.044) &\multirow{2}{*}{0.572(0.076)} &\multirow{2}{*}{4(0)} \\
& &$\bgamma$ &4.066(0.717)	&0.977(0.028)	&0.148(0.022) & & \\
&\multirow{2}{*}{MRCE($K=3$)} & $\btheta$ &2.625(0.519)	&0.979(0.013)	&0.238(0.033) &\multirow{2}{*}{0.738(0.145)} &\multirow{2}{*}{3(0)} \\
& &$\bgamma$ &2.640(1.222)	&0.997(0.005)	&0.120(0.013) & & \\
&\multirow{2}{*}{MCGGM($K=3$)} & $\btheta$  &4.654(1.275)	&0.742(0.250)	&0.141(0.115)	&\multirow{2}{*}{0.528(0.219)} &\multirow{2}{*}{3(0)} \\
& &$\bgamma$ &- &- &- & & \\
\hline

\multirow{18}{*}{(150,200,250)} &\multirow{2}{*}{Proposed} & $\btheta$ &1.399(0.054)	&0.949(0.018)	&0.053(0.004) &\multirow{2}{*}{1.000(0.000)}  &\multirow{2}{*}{3.00(0.00)} \\
& &$\bgamma$ &0.636(0.173)	&0.998(0.005)	&0.005(0.002) & & \\
&\multirow{2}{*}{HeteroGGM} & $\btheta$ &4.142(0.143)	&0.988(0.007)	&0.905(0.013) &\multirow{2}{*}{0.467(0.270)} &\multirow{2}{*}{4.55(1.64)} \\
& &$\bgamma$ &- &- &- & & \\
&\multirow{2}{*}{CGLasso($K=6$)} & $\btheta$ &3.432(0.147)	&0.899(0.020)	&0.237(0.016) &\multirow{2}{*}{0.337(0.067)} &\multirow{2}{*}{6(0)} \\
& &$\bgamma$ &5.204(0.500)	&0.972(0.024)	&0.094(0.007) & & \\
&\multirow{2}{*}{CGLasso($K=4$)} & $\btheta$ &3.196(0.233)	&0.922(0.018)	&0.173(0.026) &\multirow{2}{*}{0.541(0.089)} &\multirow{2}{*}{4(0)} \\
& &$\bgamma$ &3.741(0.784)	&0.995(0.008)	&0.058(0.009) & & \\
&\multirow{2}{*}{CGLasso($K=3$)} & $\btheta$ &3.164(0.542)	&0.927(0.028)	&0.134(0.050) &\multirow{2}{*}{0.638(0.226)} &\multirow{2}{*}{3(0)} \\
& &$\bgamma$ &3.440(1.657)	&0.996(0.008)	&0.044(0.016) & & \\
&\multirow{2}{*}{MRCE($K=6$)} & $\btheta$
&3.814(0.686)	&0.927(0.019)	&0.342(0.035) &\multirow{2}{*}{0.337(0.067)} &\multirow{2}{*}{6(0)}\\
& &$\bgamma$ &5.767(0.563)	&0.884(0.054)	&0.146(0.029) & &\\
&\multirow{2}{*}{MRCE($K=4$)} & $\btheta$ &2.921(0.284)	&0.969(0.011)	&0.287(0.029) &\multirow{2}{*}{0.541(0.089)} &\multirow{2}{*}{4(0)} \\
& &$\bgamma$ &3.737(0.860)	&0.977(0.049)	&0.150(0.015) & & \\
&\multirow{2}{*}{MRCE($K=3$)} & $\btheta$ &2.807(0.653)	&0.970(0.013)	&0.277(0.035) &\multirow{2}{*}{0.638(0.226)} &\multirow{2}{*}{3(0)} \\
& &$\bgamma$ &3.278(1.770)	&0.996(0.009)	&0.133(0.017) & & \\
&\multirow{2}{*}{MCGGM($K=3$)} & $\btheta$ &4.515(0.381)	&0.714(0.150)	&0.064(0.061)	
&\multirow{2}{*}{0.589(0.112)} &\multirow{2}{*}{3(0)}
 \\
& &$\bgamma$ &- &- &- & & \\
 \hline

\multirow{18}{*}{(500,500,500)} &\multirow{2}{*}{Proposed} & $\btheta$ &0.755(0.035)	&0.997(0.005)	&0.032(0.007) &\multirow{2}{*}{1.000(0.000)}  &\multirow{2}{*}{3.00(0.00)} \\
& &$\bgamma$ &0.324(0.021)	&1.000(0.000)	&0.001(0.000) & & \\
&\multirow{2}{*}{HeteroGGM} & $\btheta$ &4.055(0.096)	&0.990(0.005)	&0.881(0.008) &\multirow{2}{*}{0.669(0.085)} &\multirow{2}{*}{5.80(0.70)} \\
& &$\bgamma$ &- &- &- & & \\
&\multirow{2}{*}{CGLasso($K=6$)} & $\btheta$ &2.802(0.175)	&0.922(0.019)	&0.080(0.021) &\multirow{2}{*}{0.517(0.041)} &\multirow{2}{*}{6(0)} \\
& &$\bgamma$ &2.951(0.404)	&0.999(0.003)	&0.043(0.016) & & \\
&\multirow{2}{*}{CGLasso($K=4$)} & $\btheta$ &2.581(0.173)	&0.958(0.024)	&0.065(0.026) &\multirow{2}{*}{0.751(0.014)} &\multirow{2}{*}{4(0)} \\
& &$\bgamma$ &2.243(0.236)	&1.000(0.000)	&0.037(0.018) & & \\
&\multirow{2}{*}{CGLasso($K=3$)} & $\btheta$ &2.315(0.153)	&0.970(0.026)	&0.039(0.015) &\multirow{2}{*}{0.877(0.015)} &\multirow{2}{*}{3(0)} \\
& &$\bgamma$ &1.555(0.052)	&1.000(0.000)	&0.022(0.001) & & \\
&\multirow{2}{*}{MRCE($K=6$)} & $\btheta$
&2.331(0.162)	&0.982(0.009)	&0.226(0.031) &\multirow{2}{*}{0.517(0.041)} &\multirow{2}{*}{6(0)}\\
& &$\bgamma$ &2.726(0.382)	&0.991(0.029)	&0.113(0.012) & &\\
&\multirow{2}{*}{MRCE($K=4$)} & $\btheta$ &2.080(0.083)	&0.997(0.004)	&0.201(0.025) &\multirow{2}{*}{0.751(0.014)} &\multirow{2}{*}{4(0)} \\
& &$\bgamma$ &1.950(0.117)	&1.000(0.000)	&0.093(0.012) & & \\
&\multirow{2}{*}{MRCE($K=3$)} & $\btheta$ &1.884(0.099)	&0.996(0.007)	&0.136(0.009) &\multirow{2}{*}{0.877(0.015)} &\multirow{2}{*}{3(0)} \\
& &$\bgamma$ &1.299(0.067)	&1.000(0.000)	&0.060(0.003) & & \\
&\multirow{2}{*}{MCGGM($K=3$)} & $\btheta$ &4.220(0.432)	&0.825(0.206)	&0.070(0.046)	&\multirow{2}{*}{0.487(0.180)} &\multirow{2}{*}{3(0)}
  \\
& &$\bgamma$ &- &- &- & & \\
\bottomrule
\end{tabular}}
\end{table}

\begin{table}[H]
\caption{Simulation results under S3 with $p=q=100$. In each cell, mean (sd). 
} \label{tab:s3_p100}
\centering
\renewcommand\arraystretch{0.4}  
\resizebox{\linewidth}{!}{
\begin{tabular}{c c c c c c c c} 
\toprule
$n$ & Method  & &RMSE &TPR &FRP &RI &$\hat{K}_0$ \\
\midrule
\multirow{18}{*}{(200,200,200)} &\multirow{2}{*}{Proposed} & $\btheta$ &3.901(0.844)	&0.952(0.024)	&0.081(0.014) &\multirow{2}{*}{0.721(0.210)} &\multirow{2}{*}{2.35(0.49)} \\
& &$\bgamma$ &5.865(1.936)	&0.885(0.070)	&0.003(0.002) & & \\
&\multirow{2}{*}{HeteroGGM} & $\btheta$ &6.175(0.570)	&0.979(0.007)	&0.895(0.013) &\multirow{2}{*}{0.024(0.016)} &\multirow{2}{*}{5.30(1.34)} \\
& &$\bgamma$ &- &- &- & & \\
&\multirow{2}{*}{CGLasso($K=6$)} & $\btheta$ &5.412(0.178)	&0.827(0.016)	&0.272(0.019) &\multirow{2}{*}{0.126(0.159)} &\multirow{2}{*}{6(0)} \\
& &$\bgamma$ &11.072(0.954)	&0.826(0.090)	&0.083(0.008) & & \\
&\multirow{2}{*}{CGLasso($K=4$)} & $\btheta$ &4.636(0.984)	&0.917(0.024)	&0.176(0.066) &\multirow{2}{*}{0.490(0.334)} &\multirow{2}{*}{4(0)} \\
& &$\bgamma$ &6.655(3.210)	&0.976(0.043)	&0.049(0.019) & & \\
&\multirow{2}{*}{CGLasso($K=3$)} & $\btheta$ &3.845(1.242)	&0.926(0.018)	&0.088(0.070) &\multirow{2}{*}{0.775(0.355)} &\multirow{2}{*}{3(0)} \\
& &$\bgamma$ &4.270(3.336)	&0.983(0.044)	&0.025(0.017) & & \\
&\multirow{2}{*}{MRCE($K=6$)} & $\btheta$
&5.933(0.538)	&0.838(0.099)	&0.359(0.036) &\multirow{2}{*}{0.126(0.159)} &\multirow{2}{*}{6(0)} \\
& &$\bgamma$ &12.153(0.608)	&0.267(0.068)	&0.026(0.006) & & \\
&\multirow{2}{*}{MRCE($K=4$)} & $\btheta$ &4.806(0.981)	&0.947(0.034)	&0.278(0.081) &\multirow{2}{*}{0.490(0.334)} &\multirow{2}{*}{4(0)} \\
& &$\bgamma$ &7.751(3.422)	&0.716(0.285)	&0.042(0.022) & &\\
&\multirow{2}{*}{MRCE($K=3$)} & $\btheta$ &3.635(1.485)	&0.962(0.026)	&0.182(0.092) &\multirow{2}{*}{0.775(0.355)} &\multirow{2}{*}{3(0)} \\
& &$\bgamma$ &4.597(3.837)	&0.903(0.196)	&0.040(0.018) & & \\
&\multirow{2}{*}{MCGGM($K=3$)} & $\btheta$ &6.464(0.863)	&0.633(0.269)	&0.067(0.094)	&\multirow{2}{*}{0.869(0.128)} &\multirow{2}{*}{3(0)}
 \\
& &$\bgamma$ &- &- &- & & \\
\hline

\multirow{18}{*}{(150,200,250)} &\multirow{2}{*}{Proposed} & $\btheta$ &3.515(1.097)	&0.948(0.017)	&0.076(0.017) &\multirow{2}{*}{0.809(0.203)}  &\multirow{2}{*}{2.50(0.51)} \\
& &$\bgamma$ &4.434(2.427)	&0.905(0.115)	
&0.002(0.002) & & \\
&\multirow{2}{*}{HeteroGGM} & $\btheta$ &5.977(0.172)	&0.985(0.007)	&0.897(0.011) &\multirow{2}{*}{0.020(0.015)} &\multirow{2}{*}{5.40(1.27)} \\
& &$\bgamma$ &- &- &- & & \\
&\multirow{2}{*}{CGLasso($K=6$)} & $\btheta$ &5.782(0.217)	&0.856(0.012)	&0.283(0.007) &\multirow{2}{*}{0.027(0.024)} &\multirow{2}{*}{6(0)} \\
& &$\bgamma$ &11.572(0.570)	&0.822(0.064)	&0.088(0.004) & & \\
&\multirow{2}{*}{CGLasso($K=4$)} & $\btheta$ &4.759(1.327)	&0.936(0.015)	&0.173(0.072) &\multirow{2}{*}{0.441(0.323)} &\multirow{2}{*}{4(0)} \\
& &$\bgamma$ &6.592(3.503)	&0.974(0.041)	&0.047(0.021) & & \\
&\multirow{2}{*}{CGLasso($K=3$)} & $\btheta$ &3.760(0.918)	&0.935(0.024)	&0.087(0.053) &\multirow{2}{*}{0.790(0.276)} &\multirow{2}{*}{3(0)} \\
& &$\bgamma$ &4.368(2.994)	&0.959(0.100)	&0.024(0.011) & & \\
&\multirow{2}{*}{MRCE($K=6$)} & $\btheta$
&6.036(0.160)	&0.875(0.010)	&0.371(0.022) &\multirow{2}{*}{0.027(0.024)} &\multirow{2}{*}{6(0)} \\
& &$\bgamma$ &12.669(0.412)	&0.217(0.108)	&0.025(0.007) & & \\
&\multirow{2}{*}{MRCE($K=4$)} & $\btheta$ &4.983(1.724)	&0.910(0.215)	&0.252(0.096) &\multirow{2}{*}{0.441(0.323)} &\multirow{2}{*}{4(0)}\\
& &$\bgamma$ &7.689(3.910)	&0.675(0.364)	&0.039(0.021) & & \\
&\multirow{2}{*}{MRCE($K=3$)} & $\btheta$ &3.620(0.991)	&0.972(0.011)	&0.186(0.063) &\multirow{2}{*}{0.790(0.276)} &\multirow{2}{*}{3(0)} \\
& &$\bgamma$ &4.493(3.227)	&0.926(0.152)	&0.050(0.019) & & \\
&\multirow{2}{*}{MCGGM($K=3$)} & $\btheta$ &6.876(0.609)	&0.512(0.177)	&0.023(0.033)	&\multirow{2}{*}{0.787(0.087)} &\multirow{2}{*}{3(0)}
 \\
& &$\bgamma$ &- &- &- & & \\
 \hline

\multirow{18}{*}{(500,500,500)} &\multirow{2}{*}{Proposed} & $\btheta$ &1.177(0.037)	&0.995(0.003)	&0.017(0.008) &\multirow{2}{*}{1.000(0.000)}  &\multirow{2}{*}{3.00(0.00)} \\
& &$\bgamma$ &0.460(0.015)	&1.000(0.000)	&0.000(0.001) & & \\
&\multirow{2}{*}{HeteroGGM} & $\btheta$ &5.851(0.230)	&0.991(0.006)	&0.855(0.009) &\multirow{2}{*}{0.728(0.134)} &\multirow{2}{*}{5.35(1.18)} \\
& &$\bgamma$ &- &- &- & & \\
&\multirow{2}{*}{CGLasso($K=6$)} & $\btheta$ &3.618(0.330)	&0.944(0.029)	&0.095(0.031) &\multirow{2}{*}{0.651(0.038)} &\multirow{2}{*}{6(0)} \\
& &$\bgamma$ &3.900(0.628)	&0.999(0.003)	&0.038(0.017) & & \\
&\multirow{2}{*}{CGLasso($K=4$)} & $\btheta$ &2.962(0.390)	&0.961(0.025)	&0.049(0.019) &\multirow{2}{*}{0.859(0.016)} &\multirow{2}{*}{4(0)} \\
& &$\bgamma$ &2.546(0.623)	&1.000(0.001)	&0.021(0.005) & & \\
&\multirow{2}{*}{CGLasso($K=3$)} & $\btheta$ &2.755(0.416)	&0.964(0.026)	&0.017(0.014) &\multirow{2}{*}{0.956(0.125)} &\multirow{2}{*}{3(0)} \\
& &$\bgamma$ &2.190(1.781)	&0.984(0.074)	&0.012(0.002) & & \\
&\multirow{2}{*}{MRCE($K=6$)} & $\btheta$
&3.376(0.473)	&0.974(0.039)	&0.157(0.032) &\multirow{2}{*}{0.651(0.038)} &\multirow{2}{*}{6(0)} \\
& &$\bgamma$ &4.122(0.865)	&0.957(0.070)	&0.063(0.018) & & \\
&\multirow{2}{*}{MRCE($K=4$)} & $\btheta$ &2.487(0.224)	&0.993(0.017)	&0.083(0.026) &\multirow{2}{*}{0.859(0.016)} &\multirow{2}{*}{4(0)} \\
& &$\bgamma$ &2.270(0.563)	&0.993(0.024)	&0.054(0.010) & & \\
&\multirow{2}{*}{MRCE($K=3$)} & $\btheta$ &2.147(0.428)	&0.998(0.002) &0.050(0.017) &\multirow{2}{*}{0.956(0.125)} &\multirow{2}{*}{3(0)} \\
& &$\bgamma$ &1.810(1.871)	&0.984(0.070)	&0.044(0.004) & & \\
&\multirow{2}{*}{MCGGM($K=3$)} & $\btheta$ &5.771(0.457)	&0.924(0.092)	&0.068(0.034)	&\multirow{2}{*}{0.716(0.132)} &\multirow{2}{*}{3(0)}
 \\
& &$\bgamma$ &- &- &- & & \\
\bottomrule
\end{tabular}}
\end{table}

\end{document}